\documentstyle[12pt,epsf]{article}
\newcommand{\mbf}[1]{\mbox{\boldmath $#1$}}

\begin{document}

\begin{center}

{\LARGE 
Flow equations for chiral problem in QCD\\}
\vskip 20pt
{\Large
Elena Gubankova\\}
\bigskip
{\it
Institute of Theoretical and Experimental Physics,
B. Cheremushkinskaya 25,\\ RU-117 218 Moscow, Russia} 

\bigskip
\end{center}

\date{\today}

\begin{abstract}
We analyze the chiral symmetry breaking of QCD in the Coulomb gauge.
Using flow equations, we derive the renormalized gap equation
and the Bethe-Salpeter equation and show that they are finite
in both UV and IR regions. No additional UV renormalization
is required in the chiral limit. We take into account the hyperfine
interaction as well as chiral symmetry breaking and obtain 
the $\pi-\rho$ mass splitting caused by the instantaneous and
dynamical interactions.

\end{abstract}

\section{Introduction}
\label{sec:1}

The connection between the fundamental quantum chromodynamics (QCD) and
the constituent quark model (CQM) is still a puzzling feature in hadron physics.
While the CQM appears to be based on a rather simple vacuum, the QCD has a 
complicate nonperturbative vacuum structure manifested 
by the color confinement and the
dynamical breaking of the chiral symmetry. In this respect, one of the most 
intriguing phenomena in hadron physics may be the mass splitting between
the pion (spin 0) and the $\rho$ meson (spin 1), which seems to exibit
an interplay between the spin-spin interactions and the chiral symmetry
breaking. 

In the CQM the spin hyperfine interaction, 
$\psi^{\dagger}T\mbf{\alpha}\psi\psi^{\dagger}T\mbf{\alpha}\psi$,
which is a nonrelativistic reduction of one gluon exchange
is believed to produce the $\pi-\rho$ mass splitting \cite{Isgur}.
However, in the nonrelativistic constituent quark model
one obtains the smaller value of mass splitting than the experimental observation. 
Also the typical behavior of the hyperfine interaction yields
the splitting as $\alpha_s/m_{q}m_{\bar{q}}$ and
restricts the applications of the nonrelativistic CQM 
to mesons containing at least one heavy quark.
Alternatively, the $\pi-\rho$ mass splitting is explained in QCD 
by the dynamical chiral symmetry breaking caused by the chiral noninvariant 
QCD vacuum. According to the 't Hooft anomaly condition \cite{'tHooft},
the chiral symmetry in QCD is broken in the Nambu-Goldstone mode and  
the pion is the Goldstone boson. However, this argument is kinematical and  
the detailed dynamical mechanism of chiral symmetry breaking in QCD 
remains unclear.

Even before the existence of QCD, Nambu and Jona-Lasinio \cite{NambuJona-Lasinio} 
successfully described the dynamical breaking of chiral symmetry by the fermion pair 
condensation. In QCD a number of approaches have been applied to the problem of 
chiral symmetry breaking. Lattice gauge simulations, though limited to light
quarks, predict a nonzero quark condensate and show a nonlinear dependence on
the light quark mass. In continuum a dynamical mass is obtained by solving
some sort of Dyson-Schwinger equation in QCD \cite{Cornwall}. 
Moreover the chiral perturbation theory provides the results deviated 
from the exact chiral limit where Goldstone mode is given by $M_{\pi}=0$. 
 
Our approach referred as the pairing or the BCS vacuum model 
is close to the original suggestion by Nambu and Jona-Lasinio, which
was exploited later by number of authors \cite{LeYaouanc}, \cite{AdlerDavis},
\cite{FingerMandula}, \cite{BicudoRibeiro}.   
Given a chiral-invariant interaction between massless quarks, 
$\psi^{\dagger}T\psi\psi^{\dagger}T\psi$, one obtains a gap equation by seeking 
a lowest-energy solution, i.e. the ground vacuum state which turns out to be
noninvariant under chiral symmetry. Subsequent many-body solution of 
the Bethe-Salpeter equation in this vacuum provides 
the information on the light-meson spectrum,
in particular the Goldstone boson--pion. However, it is difficult to solve
directly the original theory based on the strongly correlated chiral-noninvariant 
vacuum. The general strategy is to perform a Bogoliubov-Valatin transformation
and obtain a simple chiral-invariant vacuum at the expence of generating more 
complicated chiral-noninvariant quark interaction.
Introducing concept of a quasiparticle, the quark mass becomes an unknown 
parameter, which can be determined by solving the gap equation. 
This program has been completed by Le Yaouanc {\it et al.} \cite{LeYaouanc}
and Adler and Davis \cite{AdlerDavis} for the Coulomb gauge QCD Hamiltonian, 
without invoking explicit gluon degrees of freedom. 
As a result, the nonzero dynamical quark mass, $m(\mbf{k})\neq 0$, and 
the chiral condensates, $\langle\bar{\psi}\psi\rangle\neq 0$, as well as the 
$\pi-\rho$ mass splitting have been obtained. However, one encountered
a ultraviolet divergent problem in both gap and Bethe-Salpeter equations. 
Depending on the formulation, one may also face the infrared problem.
In this work, we include the dynamic gluon degrees of freedom utilizing
the flow equations and obtain the renormalized gap and Bethe-Salpeter equations
that are finite both in UV and IR regions. Introducing quasiparticles as elementary
degrees of freedom, one may provide a possible connection between 
the BCS (pairing) model and the CQM. In particular, we investigate
what is responsible for the $\pi-\rho$ mass splitting. The spontaneous chiral 
symmetry breaking is nonperturbative in nature and manifests for small 
(zero) quark masses, while the hyperfine interaction is the leading order 
of the perturbation theory and describes nonrelativistic
dynamics of large quark masses as dictated by the CQM. 

Our starting point in this approach is the QCD which has a complex 
chiral noninvariant vacuum and a strong chiral-invariant 
interaction between light particles. It is however 
an extremely difficult problem to solve the QCD at hand, because it is
a strongly coupled theory with a nonfixed number of particles.
Instead, we notice that the CQM provides reliable predictions
in hadron physics and the relatively heavy constituents are coupled weakly 
by the chiral-noninvariant interactions. In the CQM, the vacuum is simple 
and invariant under the chiral transformations. It is an open question whether 
the weak coupling CQM is dual to the strong coupling QCD. Perhaps, 
an ultimate goal in solving QCD is to find such a dual transformation 
that converts the strong interacting QCD to a weak interacting effective theory. 
The many body approach suggests the BCS type 
Bogoliubov-Valatin (BV) transformation to fill the gap between QCD and CQM.    
We therefore adopt that the BV transformed QCD that has a chiral-invariant vacuum,
but contains chiral-noninvariant interactions. The interactions can be decomposed
into the strong BCS interactions, that reflect the chiral symmetry breaking 
and the residual weak interactions. Namely, the QCD Hamiltonian may be represented
in the BV transformed massive basis as
\begin{eqnarray}
H_{QCD}=H_0+H_{QCD}^{I}=(H_0+H_{phen})+(H_{QCD}^{I}-H_{phen})
\equiv H_{NP}+H_{PT}
\,,\label{eq:1.1}\end{eqnarray}
where $H_0$ is a free Hamiltonian, $H_{QCD}^{I}$ contains all QCD
interactions, and $H_{phen}$ is a phenomenological QCD inspired
interaction Hamiltonian. We assume that after BV transformation,
i.e. in the massive quasiparticle basis, one can choose $H_{phen}$
in a way that it absorbes all stong interacting part of $H_{QCD}^{I}$.
Therefore, $(H_0+H_{phen})=H_{NP}$ is a strong interacting part
which corresponds to the zero'th order BCS approximation, and
$(H_{QCD}^{I}-H_{phen})=H_{PT}$ is a weak residual interaction part
which can be treated perturbatively.

In this work, we use the Coulomb gauge QCD Hamiltonian, 
as a starting point. As discussed above, we then reorganize the Hamiltonian
in such a way that the nonperturbative part includes the free Hamiltonian and
the instantaneous interactions, summing the Coulomb and 
linear confining potentials, $H_{NP}=H_0+V_{inst}$, and the perturbative
part corresponds to the dynamical interactions with propagating perturbative gluons,
$H_{PT}=V_{dyn}$. 

The success of the BCS (pairing) model \cite{AdlerDavis}
may be due to the explicit nonperturbative features such as dynamical chiral 
symmetry breaking and massive quasiparticle modes. Our aim here is to include  
the dynamical interactions perturbatively in the framework of BCS (pairing) model. 
Previously in Refs. \cite{LeYaouanc}--\cite{FingerMandula} the many-body relativistic 
Hamiltonian which contains the chiral-noninvariant interactions in a quasiparticle 
basis was solved directly for the lowest Fock sectors. 
In this way, only the particle-number-conserving interactions were taken into account. 
In principle, such a model \cite{AdlerDavis} incorporates an extensive Fock space, 
because of the residual dynamical interactions which change the particle number 
to describe transition amplitudes between different Fock sectors. 
However, these interactions invariant under chiral symmetry were neglected 
since they were assumed to be weak. We now use the method of flow equations
and include these supposedly weak interactions. 
 
The flow equations perform a sequence of unitary transformations to generate 
an effective Hamiltonian which is block-diagonal in the particle number space.
This defines a resulting quasiparticle basis obtained both by the 
Bogoluibov-Valatin transformation and the flow equations, which includes 
the effects from the dynamical interactions as well as the dynamical 
chiral symmetry breaking of vacuum. Since the dynamical interactions mix low 
and high Fock states, they are responsible for the high energy region. 
Including these interactions is a vital step to reproduce a correct ultraviolet 
behavior, since it accounts for the dynamical propagation of gluons.
This is missing when only the static chiral noninvariant interaction is taken
into account. The presence of both static and dynamic interactions in an effective
Hamiltonian makes it possible to compare our results with the covariant
calculations.
 
The paper is organized as followes. Section \ref{sec:2} concentrates on the 
formulation of an effective block-diagonal Hamiltonian in the Coulomb gauge QCD. 
In Section \ref{sec:3} this Hamiltonian is applied to one- and two- body sectors,
and analytic formulations of the gap and Tamm-Dancoff/Random Phase
Approximation equations are presented. In Section \ref{sec:4}, the gap and
TDA/RPA equations are solved numerically and the results are summarized.
The summary and conclusions follow in Section \ref{sec:5}.
The details of the QCD motivated Hamiltonian, the second order flow equations,
the gap equation with a double normal ordering, the analytic proof of nonzero
pion mass in TDA/RPA and the RPA for the S and D wave $\rho$ mesons are
presented in Appendices \ref{app:A}, \ref{app:B}, \ref{app:C}, \ref{app:D}
and \ref{app:E}, respectively.

\section{Effective QCD Hamiltonian in Coulomb gauge}
\label{sec:2}

As mentioned in the introduction, we proceed with an effective Coulomb gauge 
QCD Hamiltonian given by
\begin{eqnarray}
H &=& H_{0}+V_{inst}+V_{dyn} = H_{NP}+H_{PT}
\,,\label{eq:2.1}\end{eqnarray}
where $H_{0}$ is the free Hamiltonian, $V_{inst}$ is the instantaneous interaction
describing static properties, and $V_{dyn}$ is the dynamical interaction 
involving the gluon propagation. 
It may be considered in this Hamiltonian that the gluon field
is decomposed into the nonperturbative, $A^{NP}$, and perturbative, $A^{PT}$, 
components, i.e. $A_{\mu}^{tot}=A_{\mu}^{NP}+A_{\mu}^{PT}$.
Once the nonperturbative gluon configurations are averaged over after fixing
the gauge and performing the BV transformation, the nonperturbative component
may give rise to the strong instantaneous interactions both in quark
\cite{SzczepaniakSwanson1} and gluon \cite{SzczepaniakSwanson2} sectors.
The nonperturbative gluon field is then absorbed completely into $V_{inst}$ and 
the nonperturbative gluon condensate 
$\langle H_0\rangle=\langle F_{\mu\nu}F^{\mu\nu}\rangle$
(mixing terms with both $A^{NP}$ and $A^{PT}$ give either zero
or higher powers in fields).
On the other hand, the perturbative gluon field is included in the free Hamiltonian 
and the dynamical interactions. Both of them describe the propagating dynamical 
gluons. We combine the free Hamiltonian and the instantaneous interactions and
build a nonperturbative part of the Hamiltonian, $H_0+V_{inst}=H_{NP}$,
which gives the zero'th order approximation. The dynamical interactions
are included in the perturbative Hamiltonian, $V_{dyn}=H_{PT}$, and are
treated perturbatively.

In the Coulomb gauge, the free Hamiltonian is then given by
\begin{eqnarray}
 H_{0} &=& \int d \mbf{x}\bar{\psi}(\mbf{x}) 
\left( -i \mbf{\gamma}\cdot\mbf{\nabla} + m \right)\psi(\mbf{x})
\nonumber\\ 
&+& {\rm Tr}\int d \mbf{x} \left( \mbf{\Pi}^2(\mbf{x})
+ \mbf{B}^2_{A}(\mbf{x}) \right)
\,,\label{eq:2.2}\end{eqnarray}
where the non-abelian magnetic field is
$\mbf{B}=B_i=\nabla_j A_k-\nabla_k A_j+g[A_j,A_k]$, and 
its abelian part is represented by $\mbf{B}_{A}$.
The degrees of freedom are the transverse perturbative gluon field 
$\mbf{A}=\mbf{A}^{a}T^a$ ($A\equiv A^{PT}$), 
its conjugate momentum $\mbf{\Pi}$, and the quark
field in the Coulomb gauge. Motivated by the quark model phenomenology \cite{},
we introduce a confining potential to account for the nonperturbative physics,
and add it to the existing Coulomb potential in the instantaneous interaction
\cite{SzczepaniakSwanson1} (only for the quark component)
\footnote{Generally, the nonabelian Coulomb interaction has both quark
and gluon components 
\begin{eqnarray}
 V_{inst} &=& -\frac{1}{2}\int d\mbf{x}d\mbf{y}
 \rho^a(\mbf{x})V^{ab}(\mbf{x},\mbf{y})
\rho^b(\mbf{y})
\,,\end{eqnarray} 
since the density $\rho$ in the non-abelian Coulomb interaction
is the full QCD color charge, given as
$\rho^{a}(\mbf{x})=\bar{\psi}(\mbf{x})\gamma_0T^a\psi(\mbf{x})
+f^{abc}\mbf{A}^{b}(\mbf{x})\cdot\mbf{\Pi}^{c}(\mbf{x})$. 
}
\begin{eqnarray}
 V_{inst} &=& -\frac{1}{2}\int d\mbf{x}d\mbf{y}
\bar{\psi}(\mbf{x})\gamma_0 T^a\psi(\mbf{x})
\tilde{V}_{L+C}(|\mbf{x}-\mbf{y}|)
\bar{\psi}(\mbf{y})\gamma_0 T^a\psi(\mbf{y})
\,,\label{eq:2.3}\end{eqnarray}   
where the kernel is a sum of linear and Coulomb potentials
defined by 
\begin{eqnarray}
C_f \tilde{V}_{L+C}(r) &=& \sigma r - C_f\frac{\alpha_s}{r}
\,,\label{eq:2.4}\end{eqnarray}
with the string tension $\sigma=0.18 GeV^2$ in accordance 
with lattice predictions, $\alpha_s=g^2/4\pi$ and the fundamental
Casimir operator $C_f=(N_c^2-1)/2N_c=4/3$ for $N_c=3$.
The fourier transform of the instantaneous potential, Eq. (\ref{eq:2.4}),
gives
\begin{eqnarray}
C_f \tilde{V}_{L+C}(|\mbf{q}|) &=& -\frac{C_f 4\pi\alpha_s}{\mbf{q}^2}
-\frac{8\pi\sigma}{\mbf{q}^4}
\,.\label{eq:2.4a}\end{eqnarray}  
The dynamical interaction includes the minimal quark-gluon coupling,
$V_{qg}$, and the non-abelian three- and four-gluon interactions,
$V_{gg}$, i.e. $V_{dyn} = V_{qg} + V_{gg}$, where
\begin{eqnarray}
 V_{qg} &=& -g \int d \mbf{x}\bar{\psi}(\mbf{x})
\mbf{\gamma}\cdot\mbf{A}(\mbf{x}) \psi(\mbf{x})
\nonumber\\
 V_{gg} &=& {\rm Tr}\int d \mbf{x}
\left(\mbf{B}^2(\mbf{x})-\mbf{B}^2_{A}(\mbf{x})
\right)
\,,\label{eq:2.5}\end{eqnarray}
with the perturbative gluon field $A\equiv A^{PT}$ and the corresponding
perturbative component of the magnetic field.
In what follows, we focus on the first term; the quark-gluon
coupling. The goal is to use flow equations to scale the dynamical
interactions to lower energies by eliminating the quark-gluon coupling
and to generate an effective quark interaction which
can be diagonalized nonperturbatively for bound states 
when the generated terms are added to the instantaneous phenomenological
interaction. In this process the renormalization will also be achieved by introducing
second order $O(g^2)$ canonical counterterms which remove the UV divergences
from equations for physical quantities. Since the dynamical interactions
represent the canonical Coulomb gauge QCD interactions proportional
to the coupling constant, we assume that they are weak and the entire 
procedure may be conducted perturbatively. There is no over representation
or double counting, because the phenomenological interaction
determine the infrared (IR) behavior while the dynamical interactions
free from IR divergences describe the UV part. As shown below,  
the dynamical part gives rise to the hyperfine interaction
$(\psi^{\dagger}\mbf{\alpha}T^a\psi)(\psi^{\dagger}\mbf{\alpha}T^a\psi)$
in the quark sector, which should be added to the instantaneous terms in order 
to avoid the UV divergences and noncanonical renormalization in 
the gap and Bethe-Salpeter equations.

Along with the second quantization we choose basis states and expand the quark 
and gluon field operators in normal modes. Because of the phenomenological
interaction in the quark sector given by Eq. (\ref{eq:2.3}), the trivial 
perturbative vacuum $|0\rangle$ may not be the minimum ground state for our 
Hamiltonian given by Eq. (\ref{eq:2.1}). Thus, we introduce
a trial nonperturbative vacuum state $|\Omega\rangle$ containing quark condensate
which can be determined variationally.
This is analogous to the BCS type vacuum studies \cite{LeYaouanc}, \cite{AdlerDavis}. 
We refer the state $|\Omega\rangle$ as the BCS vacuum.
The Fock space is constructed from this vacuum using quasiparticle
operators $b^{\dagger}$ and $d^{\dagger}$ which appear in the field expansions
given by
\begin{eqnarray}
&& \psi(\mbf{x}) = \sum_{s}\int\frac{d\mbf{k}}{(2\pi)^3}
    [u(\mbf{k},s)b(\mbf{k},s)
    +v(-\mbf{k},s)d^{\dagger}(-\mbf{k},s)] {\rm e}^{i\mbf{k}\mbf{x}} 
    \nonumber\\
&& \mbf{A}(\mbf{x}) = \sum_{a}\int\frac{d\mbf{k}}{(2\pi)^3}
\frac{1}{\sqrt{2\omega(\mbf{k})}}
    [a(\mbf{k},a)+a^{\dagger}(-\mbf{k},a)] {\rm e}^{i\mbf{k}\mbf{x}} 
    \nonumber\\
&& \mbf{\Pi}(\mbf{x}) = -i\sum_{a}\int\frac{d\mbf{k}}{(2\pi)^3}
\sqrt{ \frac{\omega(\mbf{k})}{2} }
    [a(\mbf{k},a)-a^{\dagger}(-\mbf{k},a)]{\rm e}^{i\mbf{k}\mbf{x}} 
\,,\label{eq:2.6}\end{eqnarray}
where $b|\Omega\rangle = d|\Omega\rangle =0$.
Note that, since the phenomenological interaction appears only in quark sector,  
the 'gluon part' of vacuum state is considered as trivial, 
i.e. $a|\Omega\rangle =a|0\rangle =0$.
Hence, the gluon Fock space represents eigenstates of free gluon Hamiltonian
with single particle energy $\omega(\mbf{k})=|\mbf{k}|$.
In Eq. (\ref{eq:2.6}), the quark operators are given in the helicity basis 
and all descrete numbers (helicity, color, and flavor for the quarks and 
color for the gluons) are collectively denoted as $s$ and $a$, respectively. 
We use the spinors in the massive basis, i.e. the nonzero effective quark mass
is explicitly included in the spinor;
\begin{eqnarray}
 u(\mbf{k},s) &=& \sqrt{E(\mbf{k})+M(\mbf{k})}\left(
  \begin{array}{c}
         1 \\
  \mbf{\sigma}\cdot\mbf{k}/(E(\mbf{k})+M(\mbf{k})) 
  \end{array} \right)\chi_{s}
\nonumber\\
&=& \frac{1}{\sqrt{2}}\left(
  \begin{array}{c}
  \sqrt{1+s(\mbf{k})} \chi_{s}\\
  \sqrt{1-s(\mbf{k})}(\mbf{\sigma}\cdot\mbf{\hat{k}})\chi_{s}
  \end{array} \right)
\nonumber\\
 v(-\mbf{k},s) &=& \sqrt{E(\mbf{k})+M(\mbf{k})}\left(
  \begin{array}{c}
  -\mbf{\sigma}\cdot\mbf{k}/(E(\mbf{k})+M(\mbf{k})) \\
         1 
  \end{array} \right)(-i\sigma_2\chi_{s})
\nonumber\\
&=& \frac{1}{\sqrt{2}}\left(
  \begin{array}{c}
  -\sqrt{1-s(\mbf{k})}(\mbf{\sigma}\cdot\mbf{\hat{k}})
  (-i\sigma_2\chi_{s}) \\
  \sqrt{1+s(\mbf{k})}(-i\sigma_2\chi_{s})
  \end{array} \right)  
\,,\label{eq:2.7}\end{eqnarray}
where the sine and cosine of the Bogoliubov angle $\Phi(\mbf{k})$ denoted as 
$\sin(\Phi(\mbf{k}))=s(\mbf{k})$ and $\cos(\Phi(\mbf{k}))=c(\mbf{k})$,
respectively, are given by
\begin{eqnarray}
s(\mbf{k})&=&\frac{M(\mbf{k})}{\sqrt{\mbf{k}^2+M^2(\mbf{k})}}\,,\,
c(\mbf{k})=\frac{k}{\sqrt{\mbf{k}^2+M^2(\mbf{k})}}\nonumber\\
E(\mbf{k})&=& \sqrt{\mbf{k}^2+M^2(\mbf{k})}
\,.\label{eq:2.8}\end{eqnarray}   
Here, $E(\mbf{k})$ is a single-quark energy which we refer as a gap energy.  
The effective quark mass $M(\mbf{k})$ is kept as an unknown variational parameter 
in the calculations and found from the gap equation by minimizing
the ground state (vacuum) energy. This approach is equivalent to set up
the Coulomb gauge pairing model \cite{LeYaouanc}, \cite{AdlerDavis},  
that involves a Bogoliubov-Valatin transformation to a vacuum containing 
a $q\bar{q}$ condensate and uses optimization principle to give an equation for 
the condensate wave function $|\Omega(\mbf{k})\rangle$. 
Here $\chi_{s}$ and $\eta_{s}$ are the standard 
two-component Pauli spinors of a particle and an antiparticle, respectively,
and for an antiparticle $\eta_{-s}=-i\sigma_2\chi_{s}$, i.e.
$\chi_{1/2}=(1,0)$, $\chi_{-1/2}=(0,1)$ and
$\eta_{-1/2}=(0,1)$, $\eta_{1/2}=(-1,0)$. 
With the definition given by Eq. (\ref{eq:2.7}), the spinors satisfy the
nonrelativistic normalization and orthogonality relations;  
$u^{\dagger}(\mbf{k},s)u(\mbf{k},s)
=v^{\dagger}(-\mbf{k},s)v(-\mbf{k},s)=1$
and $u^{\dagger}(\mbf{k},s)v(-\mbf{k},s)
=v^{\dagger}(-\mbf{k},s)u(\mbf{k},s)=0$.
Canonical (anti)commutation relations are
\begin{eqnarray}
&& \{b(\mbf{k},s),b^{\dagger}(\mbf{k}^{\prime},s^{\prime})\}
=\{d(-\mbf{k},s),d^{\dagger}(-\mbf{k}^{\prime},s^{\prime})\}
=(2\pi)^3\delta(\mbf{k}-\mbf{k}^{\prime})\delta_{s,s^{\prime}}
\nonumber\\
&& [a_i(\mbf{k},a),a^{\dagger}_j(\mbf{k}^{\prime},a^{\prime})]
=(2\pi)^3\delta(\mbf{k}-\mbf{k}^{\prime})D_{ij}(\mbf{k})\delta_{a,a^{\prime}}
\,,\label{eq:2.9}\end{eqnarray}    
where the gluon operators $\mbf{a}=a_i(\mbf{k})^a=
\sum_{\lambda=1,2}\epsilon_i(\mbf{k},\lambda)a^a(\mbf{k},\lambda)$ 
are transverse, i.e. $\mbf{k}\cdot\mbf{a}^a(\mbf{k})=
\mbf{k}\cdot\mbf{a}^{a\dagger}(\mbf{k})=0$, and $D_{ij}(\mbf{k})$ is a
polarization sum
\begin{eqnarray}
D_{ij}(\mbf{k})=\sum_{\lambda=1,2}\epsilon_i(\mbf{k},\lambda)
\epsilon_j(\mbf{k},\lambda)=\delta_{ij}-\hat{k}_i\hat{k}_j
\,,\label{eq:2.10}\end{eqnarray} 
with unit vector component $\hat{k}_i=k_i/|\mbf{k}|$ and
$\hat{k}_i\cdot D_{ij}(\mbf{k})=0$.

The complete QCD motivated Hamiltonian, Eq. (\ref{eq:2.1}), can now be
expressed in second quantized form, using Eq. (\ref{eq:2.6}),
and normal ordered with respect to the trial vacuum $|\Omega\rangle$.
In addition to the two-body interactions, normal ordering leads to
the condensate (vacuum expectations) and one-body operators
which should be regulated in UV region with the cut-off 
$\Lambda\gg\Lambda_{QCD}$ since they include the fields at one point.
The final expressions are rather complicate as summarized in 
Appendix \ref{app:A}. 

Since our goal is to obtain
effective dynamical interactions in the quark sector
\footnote{
In this work only perturbative mixing between different Fock sectors
via $V_{qg}$ is considered. In order to consider the nonperturbative
mixing between quark and gluon sectors, one should
include the gluon component of the instantaneous phenomenological
interaction and the dynamical triple- and four-gluon terms.
Nonperturbative glueball calculations in a pure gluon sector
can be found in the first paper of \cite{Gubankova}.
},
we set up the flow equation scheme for the Hamiltonian
Eq. (\ref{eq:2.1}), ignoring pure gluon non-abelian contribution
$V_{gg}$, i.e. for $H =H_0+V_{inst}+V_{qg}$.
We restrict the nonperturbative part of the Hamiltonian to the diagonal
(particle number conserving) sector, since $H_{NP}$ is the zeroth approximation
that should not be eliminated by flow equations. 
The perturbative Hamiltonian which
includes the quark-gluon coupling and mixes different Fock sectors
is attributed to the rest (particle number changing) sector;
\begin{eqnarray}
H_d &=& H_{NP}=H_0+V_{inst}
\nonumber\\
H_r &=& V_{qg}
\,.\label{eq:2.11}\end{eqnarray}
Since the quark-gluon coupling is first order, $O(g)$, we eliminate it 
using flow equations perturbatively. All terms are detailed
in Appendix \ref{app:A} along with the free Hamiltonian operator containing
the kinetic $K$ and the condensate $O$ terms in the zero'th and second orders
in coupling, $H_0^{(0)}=K^{(0)}+O^{(0)}$ (the same for $H_0^{(2)}$).
The details of Wegner's flow equations \cite{Wegner} in QCD were presented
in the previous papers \cite{Gubankova}. 

Having identified the diagonal and rest parts of the Hamiltonian,
we now construct the generator, $\eta$, and solve flow equations
for the first two leading orders. The leading order $O(g)$
flow equation reads
\begin{eqnarray}
\frac{dV_{qg}^{(1)}(l)}{dl} &=& [\eta^{(1)}(l),K^{(0)}(l)]
\nonumber\\
\eta^{(1)}(l) &=& [K^{(0)}(l),V^{(1)}_{qg}(l)]
\,,\label{eq:2.12}\end{eqnarray} 
where the kinetic and quark-gluon vertex terms are given by 
(see also Eqs. (\ref{eq:a6}) and (\ref{eq:a18}) in Appendix \ref{app:A}); 
\begin{eqnarray}
K^{(0)} &=& \sum_{s}\int\frac{d\mbf{k}}{(2\pi)^3}
E(\mbf{k})
[b^{\dagger}_s(\mbf{k})b_s(\mbf{k}) + d^{\dagger}_s(\mbf{k})d_s(\mbf{k})]
+\sum_a \int\frac{d\mbf{k}}{(2\pi)^3}
 \omega(\mbf{k}) a_i^{a\dagger}(\mbf{k})a_i^a(\mbf{k})\,,
\nonumber\\ 
\label{eq:2.13} 
\end{eqnarray}
and 
\begin{eqnarray}
&& V_{qg}^{(1)} = -\sum_{s_1,s_2,a}\int\left(\prod_{n=1}^{3}
\frac{d\mbf{k}_n}{(2\pi)^3} \right)\nonumber\\
&& \left[\,
g_0(\mbf{k}_1,\mbf{k}_2,\mbf{k}_3;l)
d_{s_1}(-\mbf{k}_1)T^{a}b_{s_2}(\mbf{k}_2)
\frac{a_i^a(\mbf{k}_3)}{\sqrt{2\omega(\mbf{k}_3)}}
v^{\dagger}_{s_1}(-\mbf{k}_1)\alpha_i u_{s_2}(\mbf{k}_2)
(2\pi)^3\delta^{(3)}(\mbf{k}_1-\mbf{k}_2-\mbf{k}_3)\right.
\nonumber\\ 
&+&\left. g_1(\mbf{k}_1,\mbf{k}_2,\mbf{k}_3;l) 
[b^{\dagger}_{s_1}(\mbf{k}_1)T^{a}b_{s_2}(\mbf{k}_2)
\frac{a_i^a(\mbf{k}_3)}{\sqrt{2\omega(\mbf{k}_3)}}
u^{\dagger}_{s_1}(\mbf{k}_1)\alpha_i u_{s_2}(\mbf{k}_2)
(2\pi)^3\delta^{(3)}(\mbf{k}_1-\mbf{k}_2-\mbf{k}_3)\right.
\nonumber\\
 &-& \left. d^{\dagger}_{s_1}(-\mbf{k}_1)T^{a}d_{s_2}(-\mbf{k}_2)
\frac{a_i^a(\mbf{k}_3)}{\sqrt{2\omega(\mbf{k}_3)}}
v^{\dagger}_{s_2}(-\mbf{k}_2)\alpha_i v_{s_1}(-\mbf{k}_1)
(2\pi)^3\delta^{(3)}(\mbf{k}_2-\mbf{k}_1-\mbf{k}_3) ] \right.
\nonumber\\ 
 &+& \left. g_{1^{\prime}}(\mbf{k}_1,\mbf{k}_2,\mbf{k}_3;l)
\frac{a_i^{a\dagger}(\mbf{k}_1)}{\sqrt{2\omega(\mbf{k}_1)}}
d_{s_2}(-\mbf{k}_2)T^{a}b_{s_3}(\mbf{k}_3)
v^{\dagger}_{s_2}(-\mbf{k}_2)\alpha_i u_{s_3}(\mbf{k}_3)
(2\pi)^3\delta^{(3)}(\mbf{k}_3-\mbf{k}_1-\mbf{k}_2) \right.
\nonumber\\
&+&\left. \mbf{H.c.} \phantom{\frac{x^2}{x^2}}\hspace{-0.5cm}\,\right]
\,,\label{eq:2.14}\end{eqnarray}
respectively.
Here, $E(\mbf{k})=\sqrt{\mbf{k}^2+M^2(\mbf{k})}$ (Eq. (\ref{eq:2.8}))
and $\omega(\mbf{k})=k$. By implementing flow equations, 
the effective coupling constants are generated (see below) as 
$g_0(\mbf{k}_1,\mbf{k}_2,\mbf{k}_3;l)$,
$g_1(\mbf{k}_1,\mbf{k}_2,\mbf{k}_3;l)$ and
$g_{1^{\prime}}(\mbf{k}_1,\mbf{k}_2,\mbf{k}_3;l)$,
which are the functions of all three momenta corresponding 
to a given Fock sector 
(Note here that the different Fock sector operators flow 
differently in energy-momentum as indicated by $0$, $1$ and $1^{\prime}$
subscripts here and below) and depend upon the flow parameter $l$. 
From Eq. (\ref{eq:2.14}) the leading order generator is
\begin{eqnarray}
&& \eta^{(1)} = -\sum_{s_1,s_2,a}\int\left(\prod_{n=1}^{3}
\frac{d\mbf{k}_n}{(2\pi)^3} \right)\nonumber\\
&& \left[\,
\eta_0(\mbf{k}_1,\mbf{k}_2,\mbf{k}_3;l)
d_{s_1}(-\mbf{k}_1)T^{a}b_{s_2}(\mbf{k}_2)
\frac{a_i^a(\mbf{k}_3)}{\sqrt{2\omega(\mbf{k}_3)}}
v^{\dagger}_{s_1}(-\mbf{k}_1)\alpha_i u_{s_2}(\mbf{k}_2)
(2\pi)^3\delta^{(3)}(\mbf{k}_1-\mbf{k}_2-\mbf{k}_3)\right.
\nonumber\\ 
&+&\left. \eta_1(\mbf{k}_1,\mbf{k}_2,\mbf{k}_3;l) 
[ b^{\dagger}_{s_1}(\mbf{k}_1)T^{a}b_{s_2}(\mbf{k}_2)
\frac{a_i^a(\mbf{k}_3)}{\sqrt{2\omega(\mbf{k}_3)}}
u^{\dagger}_{s_1}(\mbf{k}_1)\alpha_i u_{s_2}(\mbf{k}_2)
(2\pi)^3\delta^{(3)}(\mbf{k}_1-\mbf{k}_2-\mbf{k}_3)\right.
\nonumber\\
 &-& \left. d^{\dagger}_{s_1}(-\mbf{k}_1)T^{a}d_{s_2}(-\mbf{k}_2)
\frac{a_i^a(\mbf{k}_3)}{\sqrt{2\omega(\mbf{k}_3)}}
v^{\dagger}_{s_2}(-\mbf{k}_2)\alpha_i v_{s_1}(-\mbf{k}_1)
(2\pi)^3\delta^{(3)}(\mbf{k}_2-\mbf{k}_1-\mbf{k}_3) ] \right.
\nonumber\\ 
 &+& \left. \eta_{1^{\prime}}(\mbf{k}_1,\mbf{k}_2,\mbf{k}_3;l)
\frac{a_i^{a\dagger}(\mbf{k}_1)}{\sqrt{2\omega(\mbf{k}_1)}}
d_{s_2}(-\mbf{k}_2)T^{a}b_{s_3}(\mbf{k}_3)
v^{\dagger}_{s_2}(-\mbf{k}_2)\alpha_i u_{s_3}(\mbf{k}_3)
(2\pi)^3\delta^{(3)}(\mbf{k}_3-\mbf{k}_1-\mbf{k}_2) \right.
\nonumber\\
&-&\left. \mbf{H.c.} \phantom{\frac{x^2}{x^2}}\hspace{-0.5cm}\, \right]
\,,\label{eq:2.15}\end{eqnarray}
where
\begin{eqnarray}
\eta_{i}(\mbf{k}_1,\mbf{k}_2,\mbf{k}_3;l) &=&
D_i(\mbf{k}_1,\mbf{k}_2,\mbf{k}_3)
g_i(\mbf{k}_1,\mbf{k}_2,\mbf{k}_3;l)
\,,\label{eq:2.16}\end{eqnarray}
for $i=0,1,1^{\prime}$, with energy terms
(i.e. energy denominators in the old fashioned perturbation theory)
\begin{eqnarray}
D_0(\mbf{k}_1,\mbf{k}_2,\mbf{k}_3) &=& 
-(E(\mbf{k}_1)+E(\mbf{k}_2)+\omega(\mbf{k}_3))
\nonumber\\
D_1(\mbf{k}_1,\mbf{k}_2,\mbf{k}_3) &=& 
E(\mbf{k}_1)-E(\mbf{k}_2)-\omega(\mbf{k}_3)
\nonumber\\
D_{1^{\prime}}(\mbf{k}_1,\mbf{k}_2,\mbf{k}_3) &=& 
\omega(\mbf{k}_1)-E(\mbf{k}_2)-E(\mbf{k}_3)
 \,.\label{eq:2.17}\end{eqnarray}
The solution of the flow equations, Eq. (\ref{eq:2.12}), 
 for the effective coupling constants is given by
\begin{eqnarray} 
g_i(l) &=& g(0){\rm exp}(-D_i^2l)
\,,\label{eq:2.18}\end{eqnarray} 
which eliminates the quark-gluon coupling,
Eq. (\ref{eq:2.14}), for $l\rightarrow\infty$, as anticipated. 
Correspondingly, the new operators in the particle number conserving sectors
are generated from the second order $O(g^2)$ flow equation
\begin{eqnarray} 
\frac{dH_d^{(2)}(l)}{dl} &=& [\eta^{(1)}(l),V_{qg}^{(1)}(l)] 
\,,\label{eq:2.19}\end{eqnarray}
which contribute to the effective block-diagonal Hamiltonian. One can always
eliminate the particle number changing off-diagonal terms, $H_r^{(2)}$,
appearing in second order by choosing the generator 
$\eta^{(2)}=[K^{(0)},H_r^{(2)}]$ in addition to $\eta^{(1)}$.
Solving the flow equations, Eq. (\ref{eq:2.12}) and Eq. (\ref{eq:2.19}),
the block-diagonal effective Hamiltonian, $H_{eff}$, renormalized to the
second order is obtained (See details in Appendix \ref{app:B}) as
\begin{eqnarray}
 H_{eff}(\Lambda) &=& H_0(\Lambda) + V_{inst}(\Lambda)
+ V_{gen}(\Lambda)+\delta X_{CT}(\Lambda)
\,,\label{eq:2.20}\end{eqnarray}
where $V_{gen}$ includes new operators generated via perturbative 
elimination of the dynamical interaction, the quark-gluon coupling
$V_{qg}$, present in the original Hamiltonian, Eq. (\ref{eq:2.1}).
The cut-off dependence on $\Lambda$ is introduced by regulating
operator products which appear at one point and contribute 
UV divergences in the momentum space loop integrals. Regularization
results in a large sensitivity of the effective Hamiltonian
to parameter $\Lambda$ which is set at the UV scale. In order to
eliminate this $\Lambda$ dependence, a renormalization procedure is used.
The renormalization is achieved through second order by adding the counterterm,
$\delta X_{CT}$, such that in the limit $\Lambda\rightarrow\infty$
the effective Hamiltonian, Eq. (\ref{eq:2.20}), and hence equations
derived using $H_{eff}$ are UV finite and do not depend on the cut-off
parameter $\Lambda$. Therefore we can omit the cut-off notation
in $H_{eff}$, Eq. (\ref{eq:2.20}); i.e. $H_{eff}(\Lambda)=H_{eff}$.
As shown below, the cut-off sensitivity is very weak 
in the perturbative renormalization. 

In what follows we focus on the quark sector and consider
only the quark renormalization which goes through the second order
and includes quark self-energy calculation. 
In the Hartree approximation the Dyson equations for the instantaneous
and generated self-energies are (See details in Appendix \ref{app:C})
\begin{eqnarray}
\Sigma_{inst}(\mbf{k}) &=& 2 \int\frac{d\mbf{q}}{(2\pi)^3}
C_f V_{L+C}(\mbf{k},\mbf{q})\gamma_0 S^{(3)}(\mbf{q})\gamma_0
\nonumber\\
\Sigma_{gen}(\mbf{k}) &=& 2 \int\frac{d\mbf{q}}{(2\pi)^3}
\frac{1}{2}C_f W(\mbf{k},\mbf{q}) \gamma_i S^{(3)}(\mbf{q})\gamma_j
D_{ij}(\mbf{k}-\mbf{q})
\,,\label{eq:2.21}\end{eqnarray}   
where, using Eqs. (\ref{eq:2.6}) and (\ref{eq:2.7}), the equal time $3$-d quark 
propagator, $S^{(3)}$, is related to the Feynman propagator, $S^{(4)}$,
\begin{eqnarray} 
&& \langle\Omega|\frac{1}{2}[\psi_{\alpha}(\mbf{x},0),
\bar{\psi}_{\beta}(\mbf{y},0)]|\Omega\rangle
= \int\frac{d\mbf{k}}{(2\pi)^3}{\rm e}^{i\mbf{k}(\mbf{x}-\mbf{y})}
\int\frac{d k_0}{2\pi} iS^{(4)}(\mbf{k},k_0)
\nonumber\\
&=& \int\frac{d\mbf{k}}{(2\pi)^3}{\rm e}^{i\mbf{k}(\mbf{x}-\mbf{y})}
\frac{1}{2}\left(\sum_s u_s(\mbf{k})u^{\dagger}_s(\mbf{k})\gamma_0
-\sum_s v_s(-\mbf{k})v^{\dagger}_s(-\mbf{k})\gamma_0
\right)_{\alpha\beta}
\nonumber\\
&=& \int\frac{d\mbf{k}}{(2\pi)^3}{\rm e}^{i\mbf{k}(\mbf{x}-\mbf{y})}
\frac{1}{2}\left(\frac{1}{2}(\gamma_0+s(\mbf{k})
-\mbf{\gamma}\cdot\mbf{\hat{k}}c(\mbf{k}))_{\alpha\beta}
-\frac{1}{2}(\gamma_0-s(\mbf{k})
+\mbf{\gamma}\cdot\mbf{\hat{k}}c(\mbf{k}))_{\alpha\beta} \right)
\nonumber\\
&=&\int\frac{d\mbf{k}}{(2\pi)^3}{\rm e}^{i\mbf{k}(\mbf{x}-\mbf{y})}
\frac{1}{2}\left( s(\mbf{k})
-\mbf{\gamma}\cdot\mbf{\hat{k}}c(\mbf{k}) \right)_{\alpha\beta}
\,,\label{eq:2.22}\end{eqnarray}   
and is given by
\begin{eqnarray}
S^{(3)}(\mbf{k}) &=& \int\frac{d k_0}{2\pi} iS^{(4)}(\mbf{k},k_0)
=\frac{1}{2}\left(s(\mbf{k})
-\mbf{\gamma}\cdot\mbf{\hat{k}}c(\mbf{k})\right)
\,.\label{eq:2.23}\end{eqnarray}
In Eq. (\ref{eq:2.21}), the potential functions are given by   
\begin{eqnarray}
C_f V_{L+C}(\mbf{k},\mbf{q}) &=&\frac{1}{2}\frac{C_fg^2}{(\mbf{k}-\mbf{q})^2}
+\frac{4\pi\sigma}{(\mbf{k}-\mbf{q})^4}
\nonumber\\
C_f W(\mbf{k},\mbf{q}) &=& \frac{C_fg^2}
{\omega(\mbf{k}-\mbf{q})(E(\mbf{q})+\omega(\mbf{k}-\mbf{q}))} 
\,.\label{eq:2.24}\end{eqnarray}
Note that this form of generated potential is valid 
for a large loop momentum $\mbf{q}$ and sufficient for the renormalization.
In general $W$ is more complicate (For details, see Appendices 
\ref{app:B} and \ref{app:C}). The complete proper self-energy reads    
\begin{eqnarray}
\Sigma &=& \Sigma_{inst}+\Sigma_{gen} 
\,.\label{eq:2.25}\end{eqnarray}   
Making a non-relativistic ansatz for the self-energy, $\Sigma$
can be represented as
\begin{eqnarray}
\Sigma &=& \Sigma(\mbf{k}) = mA(\mbf{k}) + 
\mbf{\gamma}\cdot\mbf{k}B(\mbf{k}) 
\,,\label{eq:2.26}\end{eqnarray}  
where $A$ and $B$ are some scalar functions.
In order to find $A$ we apply trace, ${\rm Tr}$,
to the l.h.s. and r.h.s. of Eq. (\ref{eq:2.21}).
Multiplying the l.h.s. and r.h.s. of Eq. (\ref{eq:2.21})
by $\mbf{\gamma}\cdot\mbf{\hat{k}}$ and applying trace after that,
one can find $B$. The following self-energies are obtained    
\begin{eqnarray}
\Sigma_{inst}(\mbf{k}) &=& \int\frac{d\mbf{q}}{(2\pi)^3}
C_f V_{L+C}(\mbf{k},\mbf{q})s(\mbf{q}){\rm e}^{-q^2/\Lambda^2}
\nonumber\\
&+&\mbf{\gamma}\cdot\mbf{\hat{k}}\int\frac{d\mbf{q}}{(2\pi)^3}
C_f V_{L+C}(\mbf{k},\mbf{q}) c(\mbf{q})
\mbf{\hat{k}}\cdot\mbf{\hat{q}}{\rm e}^{-q^2/\Lambda^2}
\nonumber\\
\Sigma_{gen}(\mbf{k}) &=& \int\frac{d\mbf{q}}{(2\pi)^3}
C_f W(\mbf{k},\mbf{q})s(\mbf{q}){\rm e}^{-4q^2/\Lambda^2}
\nonumber\\
&+&\mbf{\gamma}\cdot\mbf{\hat{k}}\int\frac{d\mbf{q}}{(2\pi)^3}
C_f W(\mbf{k},\mbf{q})c(\mbf{q})
\mbf{\hat{k}}\cdot\mbf{\hat{l}}\mbf{\hat{q}}\cdot\mbf{\hat{l}}
{\rm e}^{-4q^2/\Lambda^2}
\,,\label{eq:2.27}\end{eqnarray}
where $\mbf{l}=\mbf{k}-\mbf{q}$, 
and exponents are the regulating functions (Appendix \ref{app:B}).
We calculate the divergences associated with the self-energy $\Sigma$
(Eq. (\ref{eq:2.26})) given by
\begin{eqnarray}
\Sigma^{div} &=& mA^{div}+\mbf{\gamma}\cdot\mbf{k}B^{div}
\,,\label{eq:2.28}\end{eqnarray}
and define the quark mass and wave function corrections as
\begin{eqnarray} 
\delta m &=& mA^{div}
\nonumber\\
Z-1 &=& B^{div}
\,,\label{eq:2.29}\end{eqnarray}
respectively. 
\footnote{
In general,
\begin{eqnarray}
\delta m &=& \Sigma^{div}|_{k^2=\bar{m}^2} 
\nonumber\\
Z-1 &=& \left(d\Sigma^{div}/(\gamma_{\mu}k^{\mu})\right)|_{k^2=\bar{m}^2}
\,,\label{eq:2.29a}\end{eqnarray}
where $\bar{m}$ is the renormalization point (the renormalized mass). 
This definition is consistent with the one given above. }
Separating the divergences, one has
\begin{eqnarray}
\Sigma^{div}_{inst}(\mbf{k}) &=& m \int\frac{d\mbf{q}}{(2\pi)^3}
\frac{1}{q}\lim_{q\rightarrow\infty}
C_f V_{L+C}(\mbf{k},\mbf{q}){\rm e}^{-q^2/\Lambda^2}
\nonumber\\
&+& \mbf{\gamma}\cdot\mbf{k}\frac{1}{k}\int\frac{d\mbf{q}}{(2\pi)^3}
\lim_{q\rightarrow\infty}
C_f V_{L+C}(\mbf{k},\mbf{q}) \mbf{\hat{k}}\cdot\mbf{\hat{q}}
{\rm e}^{-q^2/\Lambda^2}
\nonumber\\
\Sigma^{div}_{gen}(\mbf{k}) &=& m \int\frac{d\mbf{q}}{(2\pi)^3}
\frac{1}{q}\lim_{q\rightarrow\infty}
C_f W(\mbf{k},\mbf{q}){\rm e}^{-4q^2/\Lambda^2}
\nonumber\\
&+& \mbf{\gamma}\cdot\mbf{k}\frac{1}{k}\int\frac{d\mbf{q}}{(2\pi)^3}
\lim_{q\rightarrow\infty}C_f W(\mbf{k}-\mbf{q})
\mbf{\hat{k}}\cdot\mbf{\hat{l}}\mbf{\hat{q}}\cdot\mbf{\hat{l}}
{\rm e}^{-4q^2/\Lambda^2}
\,,\label{eq:2.27a}\end{eqnarray}
since for the large momenta, $q\rightarrow\infty$, $s(\mbf{q})\rightarrow m/q$
and $c(\mbf{q})\rightarrow 1$ with the current quark mass $m$.
The integrals following the structures $Im$ and $\mbf{\gamma}\cdot\mbf{k}$
are identified as $A^{div}$ and $B^{div}$, respectively. 
In Eq. (\ref{eq:2.28}), $\Sigma^{div}$ includes the instantaneous and generated
divergent contributions, $\Sigma^{div}=\Sigma^{div}_{inst}+\Sigma^{div}_{gen}$,
while only a Coulomb potential in the instantaneous interaction
has UV divergent behavior. As shown below, the divergent terms
$A^{div}$ and $B^{div}$ and hence both corrections $\delta m$ and $Z$
reduce to constants. This means that the effective Hamiltonian Eq. (\ref{eq:2.20})
can be renormalized canonically by introducing a momentum-independent counterterm,
$\delta X_{CT}$, which corresponds to a local operator. 
The divergent self-energy operator, Eq. (\ref{eq:2.28}), 
appears as a correction to the quark free Hamiltonian, Eq. (\ref{eq:2.2}), 
\begin{eqnarray}
 H_0 + \Sigma^{div} &=& \int d \mbf{x}\bar{\psi}(\mbf{x})
[ m(1+A^{div})- i\mbf{\gamma}\cdot\mbf{\nabla}(1+B^{div})] 
\psi(\mbf{x})\nonumber\\
&=&  \int d \mbf{x}\bar{\psi}(\mbf{x})
[ (m+\delta m)- Z\,i\mbf{\gamma}\cdot\mbf{\nabla}]\psi(\mbf{x})
\,,\label{eq:2.30}\end{eqnarray}
that justifies the above definitions of the mass 
and wave function corrections, 
Eq. (\ref{eq:2.29}). 
Calculating integrals in Eq. (\ref{eq:2.27}), we find
\begin{eqnarray}
\Sigma^{div}_{inst} &=& m 
\left(\frac{C_fg^2}{(4\pi)^2}4\ln\Lambda \right)
+\mbf{\gamma}\cdot\mbf{k}
\left(\frac{C_fg^2}{(4\pi)^2}\frac{8}{3}\ln\Lambda \right)
\nonumber\\
\Sigma^{div}_{gen} &=& m
\left(\frac{C_fg^2}{(4\pi)^2}2\ln\Lambda \right)
+\mbf{\gamma}\cdot\mbf{k}
\left(\frac{C_fg^2}{(4\pi)^2}(-\frac{8}{3})\ln\Lambda \right)
\,,\label{eq:2.31}\end{eqnarray}
where, as expected, the leading $\Lambda^2$ divergence does not appear
in the quark sector. This results in   
\begin{eqnarray}
\delta m &=& \frac{C_fg^2}{(4\pi)^2}6m \ln\Lambda 
\nonumber\\
Z &=& 1
\,,\label{eq:2.32}\end{eqnarray}
for the sum of instantaneous and generated terms. Note that the combined 
instantaneous and dynamical terms together do not require the wave function 
renormalization, however each term alone requires this renormalization.
Therefore the only necessary counterterm is the quark mass counterterm
\begin{eqnarray}
\delta X_{CT}(\Lambda) &=& M_{CT}(\Lambda)\int d \mbf{x}
\bar{\psi}(\mbf{x})\psi(\mbf{x})
\,,\label{eq:2.33}\end{eqnarray}
which absorbs the UV divergences in the quark sector of $H_{eff}$ 
as $\Lambda\rightarrow\infty$. Here,
\begin{eqnarray}
M_{CT}(\Lambda) &=& -\delta m =  
-\frac{C_fg^2}{(4\pi)^2}6m \ln\Lambda 
\,.\label{eq:2.34}\end{eqnarray}
The mass counterterm is proportional to the bare quark mass, $m$, 
and thus vanishes in the chiral limit $m\rightarrow 0$. This means that, 
provided both instantaneous and dynamical terms are included,
the quark sector of the effective Hamiltonian, Eq. (\ref{eq:2.20}),
is UV finite in the chiral limit and does not require any renormalization at all.
As shown above, Eq. (\ref{eq:2.31}), this happens due to a complete
cancelation of divergent wave function corrections from the instantaneous
and generated terms. The renormalized free quark Hamiltonian reads 
\begin{eqnarray}
 H_0^{ren}(\Lambda) &=& H_0 + \delta X_{CT}(\Lambda) 
 = \int d \mbf{x}\bar{\psi}(\mbf{x}) m(\Lambda) \psi(\mbf{x})
\nonumber\\
m(\Lambda) &=& m+M_{CT}(\Lambda) = 
m\left(1-\frac{C_fg^2}{(4\pi)^2}6\ln\Lambda\right)
\,,\label{eq:2.35}\end{eqnarray} 
which insures that the effective Hamiltonian is UV finite
in the quark sector. 

Our results are in accordance with the canonical renormalization
of the Coulomb gauge QCD Hamiltonian \cite{TDLee}, $H_{QCD}$, since
both Hamiltonians, $H_{eff}$ and $H_{QCD}$, have the same UV behavior. 
Extensive discussion in the literature is devoted to 
a correct renormalization procedure of an effective Hamiltonian
based on the Coulomb gauge, and the corresponding formulation of UV finite 
equations \cite{AdlerDavis}. The authors in Ref. \cite{AdlerDavis}
claim that in the chiral limit one is unable to avoid UV divergences and 
therefore suggested to introduce a non-canonical momentum-dependent counterterm 
to absorb them. In related studies \cite{FingerMandula} the authors adopt 
a special prescription of double normal ordering with respect to the perturbative, 
$|0\rangle$, and the non-perturbative, $|\Omega\rangle$, vacua in order to cancel
the divergent term (see Appendix \ref{app:C}). However,
as shown in Appendix \ref{app:C}, this prescription fails for some potentials 
in the IR region. In all of these cases, the divergence is caused by 
the term \cite{AdlerDavis}
\begin{eqnarray}
\frac{1}{k}\int\frac{d\mbf{q}}{(2\pi)^3}
\lim_{q\rightarrow\infty}
C_f V_{L+C}(\mbf{k},\mbf{q}) \mbf{\hat{k}}\cdot\mbf{\hat{q}}
\,,\label{eq:2.36}\end{eqnarray}    
which appears from the instantaneous interaction in the chiral limit
(see Eq. (\ref{eq:2.27a})).
As shown above, the dynamical interaction which is missing
in the mentioned works \cite{AdlerDavis}, \cite{FingerMandula} 
has to be included in the Hamiltonian in order to
cancel the instantaneous divergent contribution.
It may be analogous to the standard time-ordered perturbation theory,
where all time-ordered diagrams in a given order should be added to
obtain a correct covariant result.

We summarize the matrix elements of $H_{eff}$, Eq. (\ref{eq:2.20}),
in the quark sectors of interest (up to two quark states)  
\begin{eqnarray}
\langle\Omega|H_{eff}|\Omega\rangle &=& O^{ren}(\Lambda)
+O_{inst}(\Lambda) + O_{gen}(\Lambda)
\nonumber\\
\langle 1|H_{eff}|1\rangle &=& K^{ren}(\Lambda)
+\Sigma_{inst}(\Lambda) + \Sigma_{gen}(\Lambda)
\nonumber\\
\langle 2|H_{eff}|2\rangle &=& V_{inst} + V_{gen}
\,,\label{eq:2.37}\end{eqnarray}
where $|\Omega\rangle$ is a shorthand notation for the zero-quark sector
(also vacuum state), $|1\rangle$ is the single-quark sector, etc.,
and all other terms are specified below.
One should distinguish between two types of $H_{eff}$ terms. The first one
arises from the normal ordering of the original Hamiltonian, Eq. (\ref{eq:2.1}):
the instantaneous interaction with linear plus Coulomb potentials, labeled
by $inst$. This leads to the self-energy operator $\Sigma_{inst}$ in the one-body
sector, and the condensate term $O_{inst}$ in the zero-body sector.
The energy of the quark ground state $O=O_q$ comes from the normal ordering 
of the free quark Hamiltonian $H_0$ with respect to the vacuum $|\Omega\rangle$.
The second type of terms are dynamical operators generated by flow equations
and labeled by $gen$. In Eq. (\ref{eq:2.37}) the renormalized condensate
$O^{ren}$ and the kinetic $K^{ren}$ terms are given by (Appendix \ref{app:A})
\begin{eqnarray}
O^{ren}(\Lambda) &=& O + \delta X_{CT}^{0body}(\Lambda)
= -4N_c\mbf{V}\int\frac{d\mbf{k}}{(2\pi)^3}
\left[\, kc(\mbf{k}) + m(\Lambda)s(\mbf{k})\,\right]
\nonumber\\
K^{ren}(\Lambda) &=& K + \delta X_{CT}^{1body}(\Lambda)
\nonumber\\
 &=& \sum_s \int\frac{d\mbf{k}}{(2\pi)^3}\left[\,
(kc(\mbf{k}) + m(\Lambda)s(\mbf{k})) 
[b_s^{\dagger}(\mbf{k})b_s(\mbf{k}) + d_s^{\dagger}(\mbf{k})d_s(\mbf{k})]
\right.
\nonumber\\
&+&\left. (ks(\mbf{k}) - m(\Lambda)c(\mbf{k}) ) 
[b_s^{\dagger}(\mbf{k})d_s^{\dagger}({-\bf k}) + d_s({-\bf k})b_s(\mbf{k})]
\,\right]
\,,\label{eq:2.38}\end{eqnarray}
where $m(\Lambda)$ is defined in Eq. (\ref{eq:2.35}). Here, $K=K_q$ and
$O=O_q$ are defined by Eqs. (\ref{eq:a3}) and (\ref{eq:a8}), respectively,
in the Appendix \ref{app:A} and $\delta X_{CT}^{0body}$ 
is the mass counterterm given by Eq. (\ref{eq:2.33}) in the zero-body sector 
(analogous for $\delta X_{CT}^{1body}$). The mass counterterms 
$\delta X_{CT}$ cancel the leading $\ln\Lambda$ behavior
of the radiative corrections to the vacuum and kinetic terms.
In Eq. (\ref{eq:2.37}) the corrections
to $O^{ren}$ and $K^{ren}$, regulated by the exponential
cut-off function, include the condensate terms (Appendix \ref{app:B})
\begin{eqnarray}
O_{inst}(\Lambda) &=& 2 N_c\mbf{V}\int 
\frac{d\mbf{k}d\mbf{q}}{(2\pi)^6} C_f V_{L+C}(\mbf{k},\mbf{q})
\left[\, 1-s(\mbf{k})s(\mbf{q})-c(\mbf{k})c(\mbf{q})
\mbf{\hat{k}}\cdot\mbf{\hat{q}} \,\right]
{\rm e}^{-(q+k)^2/\Lambda^2}
\label{eq:2.39} \\
O_{gen}(\Lambda) &=& -2 N_c\mbf{V}\int 
\frac{d\mbf{k}d\mbf{q}}{(2\pi)^6} C_f W(\mbf{k},\mbf{q})
\left[\, 1+s(\mbf{k})s(\mbf{q})+c(\mbf{k})c(\mbf{q})
\mbf{\hat{k}}\cdot\mbf{\hat{l}}\mbf{\hat{q}}\cdot\mbf{\hat{l}} \,\right]
{\rm e}^{-(q+k+l)^2/\Lambda^2} \nonumber
\,,\end{eqnarray}
and the polarization operators (Appendix \ref{app:B})
\begin{eqnarray}
\Sigma_{inst}(\Lambda) &=& 
\sum_s\int\frac{d\mbf{k}d\mbf{q}}{(2\pi)^6} C_f V_{L+C}(\mbf{k},\mbf{q})
\left[\, s(\mbf{k})s(\mbf{q})+c(\mbf{k})c(\mbf{q})
\mbf{\hat{k}}\cdot\mbf{\hat{q}} \,\right]
{\rm e}^{-q^2/\Lambda^2}
\nonumber\\ 
&\times& 
[b^{\dagger}_s(\mbf{k})b_s(\mbf{k}) + d^{\dagger}_s(-\mbf{k})d_s(-\mbf{k})]
\nonumber\\
&+& \sum_s\int\frac{d\mbf{k}d\mbf{q}}{(2\pi)^6} C_f V_{L+C}(\mbf{k},\mbf{q})
\left[\,-c(\mbf{k})s(\mbf{q})+s(\mbf{k})c(\mbf{q})
\mbf{\hat{k}}\cdot\mbf{\hat{q}}\,\right]
{\rm e}^{-q^2/\Lambda^2}
\nonumber\\
&\times&
[b^{\dagger}_s(\mbf{k})d^{\dagger}_s({-\bf k}) + d_s({-\bf k})b_s(\mbf{k})]
\nonumber\\
\Sigma_{gen}(\Lambda) &=& \sum_{s}\int 
\frac{d\mbf{k}d\mbf{q}}{(2\pi)^6} C_f W(\mbf{k},\mbf{q})
\left[\, s(\mbf{k})s(\mbf{q})+c(\mbf{k})c(\mbf{q})
\mbf{\hat{k}}\cdot\mbf{\hat{l}}\mbf{\hat{q}}\cdot\mbf{\hat{l}} \,\right]
{\rm e}^{-4q^2/\Lambda^2}
\nonumber\\
&\times& [b_s^{\dagger}(\mbf{k})b_s(\mbf{k})
+d_s^{\dagger}(-\mbf{k})d_s(-\mbf{k})]
\nonumber\\
&+&\sum_{s}\int \frac{d\mbf{k}d\mbf{q}}{(2\pi)^6} C_f W(\mbf{k},\mbf{q})
\left[\, -c(\mbf{k})s(\mbf{q})+s(\mbf{k})c(\mbf{q})
\mbf{\hat{k}}\cdot\mbf{\hat{l}}\mbf{\hat{q}}\cdot\mbf{\hat{l}} \,\right]
{\rm e}^{-4q^2/\Lambda^2}
\nonumber\\
&\times& [b_s^{\dagger}(\mbf{k})d_s^{\dagger}(-\mbf{k})
+d_s^{\dagger}(-\mbf{k})b_s(\mbf{k})]
\,,\label{eq:2.40}\end{eqnarray}
where the potential functions $V_{L+C}$ and $W$ are defined
in Eq. (\ref{eq:2.24}).
The effective quark interaction includes the two interactions,
$V_{inst}+V_{gen}$, that define the effective Hamiltonian Eq. (\ref{eq:2.20})
in the two-body sector, $\langle 2|H_{eff}|2 \rangle$.
We consider only two-quark interactions
which contribute to a meson bound state equation in
Tamm-Dancoff (TDA) and Random Phase (RPA) approximations.
In the c.m. frame the instantaneous and 
generated interactions contributing to TDA 
($X$ component of the RPA wave function) are (see Appendix \ref{app:B})   
\begin{eqnarray}
V_{inst} &=& \sum_{\alpha\beta\delta\gamma}\int\frac{d\mbf{k}d\mbf{q}}{(2\pi)^6}
2 V_{L+C}(\mbf{k},\mbf{q}) 
\label{eq:2.41} \\ 
&\times&\left[\, (u^{\dagger}_{\delta}(\mbf{q})u_{\alpha}(\mbf{k}))
(v^{\dagger}_{\beta}(-\mbf{k})v_{\gamma}(-\mbf{q}))
\colon
b_{\delta}^{\dagger}(\mbf{q})T^a b_{\alpha}(\mbf{k})
d_{\beta}(-\mbf{k})T^a d_{\gamma}^{\dagger}(-\mbf{q}) \colon\right.
\nonumber\\
&+& \left. (u^{\dagger}_{\alpha}(\mbf{k})u_{\delta}(\mbf{q}))
(v^{\dagger}_{\gamma}(-\mbf{q})v_{\beta}(-\mbf{k}))
\colon
b_{\alpha}^{\dagger}(\mbf{k})T^a b_{\delta}(\mbf{q})
d_{\gamma}(-\mbf{q})T^a d_{\beta}^{\dagger}(-\mbf{k}) \colon
\,\right] 
\nonumber\\
V_{gen} &=& 
\sum_{\alpha\beta\delta\gamma}\int\frac{d\mbf{k}d\mbf{q}}{(2\pi)^6}
2 W_1(\mbf{k},\mbf{q}) D_{ij}(\mbf{k}-\mbf{q})
\nonumber\\ 
&\times&\left[\, (u^{\dagger}_{\delta}(\mbf{q})\alpha_iu_{\alpha}(\mbf{k}))
(v^{\dagger}_{\beta}(-\mbf{k})\alpha_jv_{\gamma}(-\mbf{q}))
\colon
b_{\delta}^{\dagger}(\mbf{q})T^a b_{\alpha}(\mbf{k})
d_{\beta}(-\mbf{k})T^a d_{\gamma}^{\dagger}(-\mbf{q}) \colon\right.
\nonumber\\
&+& \left. (u^{\dagger}_{\alpha}(\mbf{k})\alpha_iu_{\delta}(\mbf{q}))
(v^{\dagger}_{\gamma}(-\mbf{q})\alpha_jv_{\beta}(-\mbf{k}))
\colon
b_{\alpha}^{\dagger}(\mbf{k})T^a b_{\delta}(\mbf{q})
d_{\gamma}(-\mbf{q})T^a d_{\beta}^{\dagger}(-\mbf{k}) \colon
\,\right] \nonumber
\,.\end{eqnarray}
In RPA ($Y$ component of the RPA wave function), they are given by
\begin{eqnarray}
V_{inst} &=& 
\sum_{\alpha\beta\delta\gamma}\int\frac{d\mbf{k}d\mbf{q}}{(2\pi)^6}
2 V_{L+C}(\mbf{k},\mbf{q})
\label{eq:2.42} \\ 
&\times&\left[\, (v^{\dagger}_{\gamma}(-\mbf{q})u_{\alpha}(\mbf{k}))
(v^{\dagger}_{\beta}(-\mbf{k})u_{\delta}(\mbf{q}))
\colon
d_{\gamma}(-\mbf{q})T^a b_{\alpha}(\mbf{k})
d_{\beta}(-\mbf{k})T^a b_{\delta}(\mbf{q}) \colon\right.
\nonumber\\
&+& \left. (u^{\dagger}_{\delta}(\mbf{q})v_{\beta}(-\mbf{k}))
(u^{\dagger}_{\alpha}(\mbf{k})v_{\gamma}(-\mbf{q}))
\colon
b_{\delta}^{\dagger}(\mbf{q})T^a d_{\beta}^{\dagger}(-\mbf{k})
b_{\alpha}^{\dagger}(\mbf{k})T^a d_{\gamma}^{\dagger}(-\mbf{q}) \colon
\,\right]
\nonumber\\
V_{gen} &=& 
\sum_{\alpha\beta\delta\gamma}\int\frac{d\mbf{k}d\mbf{q}}{(2\pi)^6}
2 W_2(\mbf{k},\mbf{q}) D_{ij}(\mbf{k}-\mbf{q})
\nonumber\\ 
&\times&\left[\, (v^{\dagger}_{\gamma}(-\mbf{q})\alpha_iu_{\alpha}(\mbf{k}))
(v^{\dagger}_{\beta}(-\mbf{k})\alpha_ju_{\delta}(\mbf{q}))
\colon
d_{\gamma}(-\mbf{q})T^a b_{\alpha}(\mbf{k})
d_{\beta}(-\mbf{k})T^a b_{\delta}(\mbf{q}) \colon\right.
\nonumber\\
&+& \left. (u^{\dagger}_{\delta}(\mbf{q})\alpha_iv_{\beta}(-\mbf{k}))
(u^{\dagger}_{\alpha}(\mbf{k})\alpha_jv_{\gamma}(-\mbf{q}))
\colon
b_{\delta}^{\dagger}(\mbf{q})T^a d_{\beta}^{\dagger}(-\mbf{k})
b_{\alpha}^{\dagger}(\mbf{k})T^a d_{\gamma}^{\dagger}(-\mbf{q}) \colon
\,\right] \nonumber
\,,\end{eqnarray}
where potential functions are given
\begin{eqnarray}
C_f W_1(\mbf{k},\mbf{q}) &=& -\frac{1}{2}
\frac{C_f g^2}{\omega^2(\mbf{k}-\mbf{q})+(E(\mbf{k})-E(\mbf{q}))^2}
\nonumber\\
C_f W_2(\mbf{k},\mbf{q}) &=& -\frac{1}{2}
\frac{C_f g^2}{\omega^2(\mbf{k}-\mbf{q})+(E(\mbf{k})+E(\mbf{q}))^2}
\,.\label{eq:2.43}\end{eqnarray}
and $V_{L+C}$ is defined in Eq. (\ref{eq:2.24}).
In the next section we utilize the obtained effective Hamiltonian,
with matrix elements given by Eq. (\ref{eq:2.37}), to derive
and solve the quark gap and meson bound state equations.

\section{Sector solution of the effective Hamiltonian: 
gap and bound--state equations}
\label{sec:3}

Now that we have eliminated the quark-gluon coupling, which mixes
different quark sectors, and obtained the effective quark Hamiltonian,
$H_{eff}$, valid up to the second order in the coupling constant,
we can nonperturbatively diagonalize each sector Hamiltonian,
Eq. (\ref{eq:2.37}). Further, because $H_{eff}$ is also renormalized,
the equations for physical observables are free from the UV divergences.
In subsection \ref{subsec:3.1} we first investigate the quark vacuum by formulating 
the quark gap equation and also calculate the quark condensate.
Then we address the meson spectrum in subsection \ref{subsec:3.2}.
Numerical solutions of the gap equation and the bound state equations
are discussed in the next section (Section \ref{sec:4}).

\subsection{Gap equation}
\label{subsec:3.1}

The gap equation allows the determination of a nontrivial vacuum
with quark condensates and propagating quasiparticles
(or quarks with a dynamical mass). There are several ways
to obtain this equation, the most common based upon a variational
principle to minimize the vacuum (ground state) energy. The 
variational parameter is the angle of transformation
from undressed to dressed particle (quasiparticle) operators,
$\Phi(\mbf{k})$, which defines a quasiparticle basis,
Eq. (\ref{eq:2.7}), with a dynamical quark mass $M(\mbf{k})$.
Therefore, minimizing the vacuum energy of the effective
Hamiltonian, i.e.
\begin{eqnarray}
\frac{\delta\langle\Omega|H_{eff}|\Omega\rangle}{\delta\Phi(\mbf{k})} &=& 0
\,,\label{eq:3.1}\end{eqnarray}
generates the gap equation for the unknown $\Phi({\mbf{k}})$
or $M(\mbf{k})$. Using Eq. (\ref{eq:2.39}) for the condensate terms 
\begin{eqnarray}
\frac{\delta}{\delta\Phi(\mbf{k})}\left(O_{ren}+O_{inst}+O_{gen}\right) &=& 0
\,,\label{eq:3.1a}\end{eqnarray}
the following gap equation is obtained:
\begin{eqnarray}
&& ks(\mbf{k})-m(\Lambda)c(\mbf{k})
 = \int\frac{d\mbf{q}}{(2\pi)^3} C_f V_{L+C}(\mbf{k},\mbf{q})
\left[\, c(\mbf{k})s(\mbf{q}) - s(\mbf{k})c(\mbf{q})
\mbf{\hat{k}}\cdot\mbf{\hat{q}} \,\right]
{\rm e}^{-q^2/\Lambda^2}
\nonumber\\
&+& \int\frac{d\mbf{q}}{(2\pi)^3} C_f W(\mbf{k},\mbf{q})
\left[\, c(\mbf{k})s(\mbf{q}) - s(\mbf{k})c(\mbf{q})
\mbf{\hat{k}}\cdot\mbf{\hat{l}}\mbf{\hat{q}}\cdot\mbf{\hat{l}} \,\right]
{\rm e}^{-4q^2/\Lambda^2}
\,,\label{eq:3.2}\end{eqnarray}
where $\mbf{l}=\mbf{k}-\mbf{q}$, potential functions
$V_{L+C}$ and $W$ are given in Eq. (\ref{eq:2.24}), and 
$m(\Lambda)$ includes the mass counterterm (Eq. (\ref{eq:2.35}))
defined by Eq. (\ref{eq:2.34}). 

The gap equation can also be obtained by demanding that the effective
Hamiltonian should not contain off-diagonal one-body terms
of the type $bd$ and $b^{\dagger}d^{\dagger}$. This means that 
the BCS vacuum $|\Omega\rangle$ is stable against quasiparticle 
pair creation. Therefore, the operator
\begin{eqnarray}
\sum_s \int\frac{d\mbf{k}}{(2\pi)^3}F(\mbf{k},\Phi)
[b^{\dagger}_s(\mbf{k})d^{\dagger}_s(-\mbf{k})+d_s(-\mbf{k})b_s(\mbf{k})]
\,\label{eq:3.3}\end{eqnarray}
of $H_{eff}$ vanishes by choosing
\begin{eqnarray}
F(\mbf{k},\Phi) &=& 0
\,.\label{eq:3.4}\end{eqnarray}
Imposing this condition on the nondiagonal matrix elements, labeled $nd$,
of the effective Hamiltonian in the single quark sector 
$\langle 1|H_{eff}|1\rangle$ given by
Eqs. (\ref{eq:2.37}) and (\ref{eq:2.40}), i.e.
\begin{eqnarray}
\bar{\psi}\left[\,H_0^{ren}+\Sigma\,\right]\psi|_{nondiag.}\rightarrow
K^{nd}_{ren}+\Sigma^{nd}_{inst}+\Sigma^{nd}_{gen}=0
\,,\label{eq:3.5}\end{eqnarray}
yields the same gap equation as above. 

The alternative way to obtain the quark gap equation without specifying
matrix elements of $H_{eff}$ is to use the Dyson equation for 
the self-energy operator $\Sigma$, Eq. (\ref{eq:2.21}).
With the propagator given by Eq. (\ref{eq:2.23}) the instantaneous and
generated self-energies have been found in section \ref{sec:2}, Eq. (\ref{eq:2.27}).
Therefore $A$ and $B$ functions, defined in Eq. (\ref{eq:2.26}), are
\begin{eqnarray}
mA(\mbf{k}) &=& \int\frac{d\mbf{q}}{(2\pi)^3}C_f V_{L+C}(\mbf{k},\mbf{q})
s(\mbf{q}){\rm e}^{-q^2/\Lambda^2}
+\int\frac{d\mbf{q}}{(2\pi)^3}C_f W(\mbf{k},\mbf{q})
s(\mbf{q}){\rm e}^{-4q^2/\Lambda^2}
\label{eq:3.6}\\
kB(\mbf{k}) &=& \int\frac{d\mbf{q}}{(2\pi)^3}C_f V_{L+C}(\mbf{k},\mbf{q})
c(\mbf{q})\mbf{\hat{k}}\cdot\mbf{\hat{q}}
{\rm e}^{-q^2/\Lambda^2}
+\int\frac{d\mbf{q}}{(2\pi)^3}C_f W(\mbf{k},\mbf{q})
c(\mbf{q})\mbf{\hat{k}}\cdot\mbf{\hat{l}}\mbf{\hat{q}}\cdot\mbf{\hat{l}}
{\rm e}^{-4q^2/\Lambda^2}\nonumber
\,.\end{eqnarray}
On the other hand, using the general expression for the self-energy $\Sigma$,
Eq. (\ref{eq:2.26}), the 4-d dressed Feynman propagator is
\begin{eqnarray}
S^{(4)}(\mbf{k},k_0) &=& \frac{1}{\gamma_0k_0-\mbf{\gamma}\cdot\mbf{k}
-m-\Sigma(\mbf{k})}\nonumber\\
&=&\frac{\gamma_0k_0-\mbf{\gamma}\cdot\mbf{k}(1+B(\mbf{k}))
+m(1+A(\mbf{k}))}{k_0^2-\Omega^2(\mbf{k})}
= \frac{R_{+}(\mbf{k})}{k_0-\Omega(\mbf{k})}
+ \frac{R_{-}(\mbf{k})}{k_0+\Omega(\mbf{k})}
\nonumber\\
\Omega(\mbf{k}) &=& \sqrt{\mbf{k}^2(1+B(\mbf{k}))^2+m^2(1+A(\mbf{k}))^2}
\,,\label{eq:3.7}\end{eqnarray}
with the residues $R_{\pm}(\mbf{k})$ given by
\begin{eqnarray}
R_{\pm}(\mbf{k}) &=&\frac{1}{2}\left[\gamma_0\pm
\frac{m(1+A(\mbf{k}))}{\Omega(\mbf{k})}\right]
\mp\frac{1}{2}\mbf{\gamma}\cdot\mbf{k}\frac{1+B(\mbf{k})}{\Omega(\mbf{k})}
\,.\label{eq:3.8}\end{eqnarray}
The integration over $k_0$, Eq. (\ref{eq:2.22}), with the residues 
$R_{\pm}(\mbf{k})$ gives the 3-d equal-time propagator
\begin{eqnarray}
S^{(3)}(\mbf{k}) &=& \int\frac{dk_0}{2\pi}iS^{(4)}(\mbf{k},k_0)
= \frac{m(1+A(\mbf{k}))-\mbf{\gamma}\cdot\mbf{k}(1+B(\mbf{k}))}
{2\Omega(\mbf{k})}
\,,\label{eq:3.9}\end{eqnarray}
and comparing Eq. (\ref{eq:3.9}) with Eq. (\ref{eq:2.23}) for $S^{(3)}$
we see that 
\begin{eqnarray}
s(\mbf{k}) = \frac{m(1+A(\mbf{k}))}{\Omega(\mbf{k})}\,,\,
c(\mbf{k}) = \frac{k(1+B(\mbf{k}))}{\Omega(\mbf{k})}
\,,\label{eq:3.10}\end{eqnarray}
where $\Omega(\mbf{k})$ is given in Eq. (\ref{eq:3.7}).
Essentially Eq. (\ref{eq:3.10}) is the gap equation.
Eliminating $\Omega$ in Eq. (\ref{eq:3.10}) 
\begin{eqnarray}
s(\mbf{k})k(1+B(\mbf{k}))-c(\mbf{k})m(1+A(\mbf{k})) &=& 0
\,,\label{eq:3.11}\end{eqnarray}
and substituting $A$ and $B$, Eq. (\ref{eq:3.6}),
we obtain again the same quark gap equation, Eq. (\ref{eq:3.1}).

Due to the gap equation, Eq. (\ref{eq:3.5}),
the single-quark operator is diagonal
\begin{eqnarray}
\sum_s\int\frac{d\mbf{k}}{(2\pi)^3}\varepsilon(\mbf{k})
[b_s^{\dagger}(\mbf{k})b_s(\mbf{k})+d_s^{\dagger}(\mbf{k})d_s(\mbf{k})]
\,,\label{eq:3.12}\end{eqnarray}
and therefore it can be associated with an effective quark energy,
$\varepsilon(\mbf{k})$,   
\begin{eqnarray}
\bar{\psi}\left[\,H_0^{ren}+\Sigma\,\right]\psi|_{diag.}\rightarrow
K^{d}_{ren}+\Sigma^{d}_{inst}+\Sigma^{d}_{gen}=\varepsilon(\mbf{k})
\,.\label{eq:3.13}\end{eqnarray}
The vanishing non-diagonal part of this operator, Eq. (\ref{eq:3.5}), 
is given by Eq. (\ref{eq:3.11}), then its diagonal part is
\begin{eqnarray}
\varepsilon(\mbf{k}) &=& c(\mbf{k})k(1+B(\mbf{k}))+s(\mbf{k})m(1+A(\mbf{k})) 
\,,\label{eq:3.14}\end{eqnarray}
and substituting $A$ and $B$ functions, Eq. (\ref{eq:3.6}), gives
\begin{eqnarray}
\varepsilon(\mbf{k}) &=& kc(\mbf{k})+m(\Lambda)s(\mbf{k})
 + \int\frac{d\mbf{q}}{(2\pi)^3} C_f V_{L+C}(\mbf{k},\mbf{q})
\left[\, s(\mbf{k})s(\mbf{q}) + c(\mbf{k})c(\mbf{q})
\mbf{\hat{k}}\cdot\mbf{\hat{q}} \,\right]
{\rm e}^{-q^2/\Lambda^2}
\nonumber\\
&+& \int\frac{d\mbf{q}}{(2\pi)^3} C_f W(\mbf{k},\mbf{q})
\left[\, s(\mbf{k})s(\mbf{q}) + c(\mbf{k})c(\mbf{q})
\mbf{\hat{k}}\cdot\mbf{\hat{l}}\mbf{\hat{q}}\cdot\mbf{\hat{l}} \,\right]
{\rm e}^{-4q^2/\Lambda^2}
\,.\label{eq:3.15}\end{eqnarray}
This equation can also be obtained using $\langle1|H_{eff}|1\rangle$,
Eqs. (\ref{eq:2.37}) and (\ref{eq:2.40}).
It is convenient to represent $\varepsilon(\mbf{k})$, using the gap equation, 
Eq. (\ref{eq:3.11}), \mbox{$k(1+B)=[c(\mbf{k})/s(\mbf{k})]m(1+A)$}, as
\begin{eqnarray}
&\varepsilon(\mbf{k})& = \frac{m(1+A(\mbf{k}))}{s(\mbf{k})}
\label{eq:3.15a}\\
&=& \frac{m}{s(\mbf{k})}
+ \int\frac{d\mbf{q}}{(2\pi)^3}C_fV_{L+C}(\mbf{k},\mbf{q})
\left[\,\frac{s(\mbf{q})}{s(\mbf{k})}\,\right]
{\rm e}^{-q^2/\Lambda^2}
+\int\frac{d\mbf{q}}{(2\pi)^3}C_fW(\mbf{k},\mbf{q})
\left[\,\frac{s(\mbf{q})}{s(\mbf{k})}\,\right]
{\rm e}^{-4q^2/\Lambda^2} \nonumber
\,.\end{eqnarray}
The effective energy $\varepsilon(\mbf{k})$, Eq. (\ref{eq:3.14}),
equals to the frequency $\Omega(\mbf{k})$, 
appearing as a pole in the propagator, Eq. (\ref{eq:3.7}),
\begin{eqnarray}
\varepsilon(\mbf{k}) &=& \Omega(\mbf{k})
\,.\label{eq:3.16}\end{eqnarray}
In order to introduce $\varepsilon(\mbf{k})$ the Bogoluibov-Valatin angle
should satisfy the gap equation, Eq. (\ref{eq:3.10}) or Eq. (\ref{eq:3.11}).  
Substituting Eq. (\ref{eq:3.10}) into $\varepsilon(\mbf{k})$, 
Eq. (\ref{eq:3.14}), gives Eq. (\ref{eq:3.16}). However, one should
distinguish between the effective energy $\varepsilon(\mbf{k})$ and
the gap energy $E(\mbf{k})$, Eq. (\ref{eq:2.8}). Comparing
Eq. (\ref{eq:2.8}) with Eq. (\ref{eq:3.10}) for the sine and cosine 
we find
\begin{eqnarray}
E(\mbf{k}) &=& \varepsilon(\mbf{k})/(1+B(\mbf{k}))
\,.\label{eq:3.16a}\end{eqnarray} 
A natural question arises, what value can be considered physical,
the effective quark energy $\varepsilon(\mbf{k})=\Omega(\mbf{k})$
or the mass gap $M(\mbf{k})$ (or related gap energy $E(\mbf{k})$). 
Consider first the propagator for a free massive Dirac
particle with mass $M(\mbf{k})$
\begin{eqnarray}
&& \frac{1}{\gamma_0k_0-\mbf{\gamma}\cdot\mbf{k}
-M(\mbf{k})}=\frac{\gamma_0k_0-\mbf{\gamma}\cdot\mbf{k}+M(\mbf{k})}
{k_0^2-E^2(\mbf{k})}
= \frac{R(\mbf{k})}{k_0-E(\mbf{k})}
+ \left[ analytic\, at\, k_0=E(\mbf{k}) \right]
\,,\label{eq:3.17}\end{eqnarray}
with $E(\mbf{k})=\sqrt{\mbf{k}^2+M^2(\mbf{k})}$ and
the residue $R(\mbf{k})$ at the positive frequency pole given by
\begin{eqnarray}
R(\mbf{k}) =\frac{1}{2}\left[\gamma_0+
\frac{M(\mbf{k})}{E(\mbf{k})}\right]
-\frac{1}{2}\mbf{\gamma}\cdot\mbf{\hat{k}}\frac{k}{E(\mbf{k})}
= \frac{1}{2}\left[\gamma_0+s(\mbf{k})\right]
-\frac{1}{2}\mbf{\gamma}\cdot\mbf{\hat{k}}c(\mbf{k})
\,,\label{eq:3.18}\end{eqnarray}
where we have used Eq. (\ref{eq:2.8}) for the sine and cosine.
Let us compare this with the propagator in our model, 
Eqs. (\ref{eq:3.7}) and (\ref{eq:3.8})
\begin{eqnarray}
&& \frac{1}{\gamma_0k_0-\mbf{\gamma}\cdot\mbf{k}
-m-\Sigma(\mbf{k})}= \frac{R_{+}(\mbf{k})}{k_0-\Omega(\mbf{k})}
+ \left[ analytic\, at\, k_0=\Omega(\mbf{k}) \right]
\,,\label{eq:3.19}\end{eqnarray}
where, using Eq. (\ref{eq:3.10}) for the sine and cosine,
the residue $R_{+}(\mbf{k})$ at the positive frequency pole
is given by 
\begin{eqnarray}
R_{+}(\mbf{k}) =\frac{1}{2}\left[\gamma_0+
\frac{m(1+A(\mbf{k}))}{\Omega(\mbf{k})}\right]
-\frac{1}{2}\mbf{\gamma}\cdot\mbf{k}\frac{1+B(\mbf{k})}{\Omega(\mbf{k})}
= \frac{1}{2}\left[\gamma_0+s(\mbf{k})\right]
-\frac{1}{2}\mbf{\gamma}\cdot\mbf{\hat{k}}c(\mbf{k})
\,,\label{eq:3.20}\end{eqnarray}
which is the same as Eq. (\ref{eq:3.18}).
Using Eq. (\ref{eq:2.8}) and Eq. (\ref{eq:3.10}) for the sine and cosine,
the propagator Eq. (\ref{eq:3.19}) can be written as
\begin{eqnarray}
&& \frac{1}{\gamma_0k_0-(\mbf{\gamma}\cdot\mbf{k}
+M(\mbf{k}))\Omega(\mbf{k})/E(\mbf{k})}
\,.\label{eq:3.21}\end{eqnarray}
Hence, though the two covariant $4$-d Feynman propagators, 
Eqs. (\ref{eq:3.17}) and (\ref{eq:3.19}) (or (\ref{eq:3.21})),
have different behavior with different poles, $E(\mbf{k})$ and 
$\Omega(\mbf{k})=\varepsilon(\mbf{k})$, respectively, they
have the same $3$-d image $S^{(3)}(\mbf{k})$, 
given by Eq. (\ref{eq:2.23}), which appears as an equal-time
quark propagator in the Coulomb gauge Hamiltonian,
Eq. (\ref{eq:2.1}). Therefore, $E(\mbf{k})$
can be considered as a physical pole, associated with 
a quasiparticle having an effective mass $M(\mbf{k})$.
This also justifies our definition of sine and cosine
through $M(\mbf{k})$ and $E(\mbf{k})$, Eq. (\ref{eq:2.8}).
We show below that the frequency pole $\Omega(\mbf{k})$
is not well defined, and thus can not represent a particle.

We investigate the UV and IR behavior of 
$\varepsilon(\mbf{k})=\Omega(\mbf{k})$, Eq. (\ref{eq:3.15}),
and $M(\mbf{k})$, given by the gap equation Eq. (\ref{eq:3.2}).
In the UV region both equations are finite, since based
on the renormalized $H_{eff}(\Lambda)$, the mass counterterm
$m(\Lambda)$ cancels exactly the only UV-divergence from
$A(\mbf{k})$, Eq. (\ref{eq:3.6}), while $B(\mbf{k})$ does not contribute
to the UV-divergence (as discussed in Section \ref{sec:2}). 
In the chiral limit $\varepsilon(\mbf{k})$
and $M(\mbf{k})$ are defined by the UV finite equations even without
renormalization (See Section \ref{sec:2}).

Infrared problems are caused by the linear potential in the instantaneous
interaction, $V_{L+C}(\mbf{k},\mbf{q})$, diverging as $|\mbf{k}-\mbf{q}|^{-4}$
as $\mbf{k}\rightarrow\mbf{q}$. 
Consider the limit $\mbf{k}\rightarrow\mbf{q}$ in the gap equation, 
Eq. (\ref{eq:3.2}),
\begin{eqnarray}
\int\frac{d\mbf{q}}{(2\pi)^3} C_f V_{L+C}(\mbf{k},\mbf{q})
\left[\, c(\mbf{k})s(\mbf{q}) - s(\mbf{k})c(\mbf{q})
\mbf{\hat{k}}\cdot\mbf{\hat{q}} \,\right]
\,.\label{eq:3.22}\end{eqnarray}
Expanding up to the second order
\begin{eqnarray}
s(\mbf{q}) &=& s(\mbf{k}) +\mbf{\hat{k}}\cdot\mbf{\delta}s^{\prime}(\mbf{k})
+ O(\mbf{\delta}^2)
\nonumber\\
c(\mbf{q}) &=& c(\mbf{k}) +\mbf{\hat{k}}\cdot\mbf{\delta}c^{\prime}(\mbf{k})
+ O(\mbf{\delta}^2)
\nonumber\\
\mbf{\hat{k}}\cdot\mbf{\hat{q}} &=& 1 + O(\mbf{\delta}^2) 
\,,\label{eq:3.23}\end{eqnarray}
where $\mbf{\delta}=\mbf{q}-\mbf{k}$ and $s^{\prime}(\mbf{k})$ denotes
the derivative of sine in $\mbf{k}$, we find for the term Eq. (\ref{eq:3.22}) 
\begin{eqnarray}
\int\frac{d\mbf{q}}{(2\pi)^3} C_f V_{L+C}(\mbf{k},\mbf{q})
\left[\,\mbf{\hat{k}}\cdot\mbf{\delta} 
[c(\mbf{k})s^{\prime}(\mbf{k}) - s(\mbf{k})c^{\prime}(\mbf{k})]
 + O(\mbf{\delta}^2) \,\right]
\,,\label{eq:3.24}\end{eqnarray}
which behaves, after angular averaging, as
\begin{eqnarray}
\int d\mbf{q}|\mbf{k}-\mbf{q}|^{-4}O((\mbf{q}-\mbf{k})^2) 
\,,\label{eq:3.25}\end{eqnarray}
that converges. Thus, the mass gap $M(\mbf{k})$, Eq. (\ref{eq:3.2}),
is well defined in the IR. 
In Appendix \ref{app:C} we discuss the gap equation obtained by
double normal ordering with respect to the perturbative $|0\rangle$
and nonperturbative $|\Omega\rangle$ vacuum states 
(See Eqs. (\ref{eq:c10}) and (\ref{eq:c11})) which was used  
in Ref. \cite{FingerMandula} as well as the first paper in 
Ref. \cite{SzczepaniakSwanson1}.
One of the motivations for this prescription was to avoid UV divergences
for the Coulomb potential in the chiral limit.  
However, the dangerous term reads 
\begin{eqnarray}
\int\frac{d\mbf{q}}{(2\pi)^3} C_f V_{L+C}(\mbf{k},\mbf{q})
\left[\, c(\mbf{k})s(\mbf{q}) - s(\mbf{k})(c(\mbf{q})-1)
\mbf{\hat{k}}\cdot\mbf{\hat{q}} \,\right]
\,,\label{eq:3.26}\end{eqnarray}
which, using Eq. (\ref{eq:3.23}), behaves as $\mbf{k}\rightarrow\mbf{q}$ as
\begin{eqnarray}
\int\frac{d\mbf{q}}{(2\pi)^3} C_f V_{L+C}(\mbf{k},\mbf{q})s(\mbf{k})
\rightarrow \int d\mbf{q}|\mbf{k}-\mbf{q}|^{-4} s(\mbf{k})
\,,\label{eq:3.27}\end{eqnarray}
that diverges. Thus, the double normal ordering gap equation is IR singular,
and a mass gap $M(\mbf{k})$ does not exist for a confining potential.
On the contrary, our gap equation, Eq. (\ref{eq:3.2}), is well defined
for the Coulomb plus linear potentials in the UV and IR, and provides
a finite mass gap $M(\mbf{k})$ (or a finite gap energy $E(\mbf{k})$).

However, the effective quark energy
$\varepsilon(\mbf{k})$, Eq. (\ref{eq:3.15}),
\begin{eqnarray}
\int\frac{d\mbf{q}}{(2\pi)^3} C_f V_{L+C}(\mbf{k},\mbf{q})
\left[\, s(\mbf{k})s(\mbf{q}) + c(\mbf{k})c(\mbf{q})
\mbf{\hat{k}}\cdot\mbf{\hat{q}} \,\right]
\,,\label{eq:3.28}\end{eqnarray}
and the frequency pole $\Omega(\mbf{k})$, Eq. (\ref{eq:3.16}),
behave for a confining potential as $\mbf{k}\rightarrow\mbf{q}$ as
\begin{eqnarray}
\int\frac{d\mbf{q}}{(2\pi)^3} C_f V_{L+C}(\mbf{k},\mbf{q})
\rightarrow \int d\mbf{q}|\mbf{k}-\mbf{q}|^{-4}
\,,\label{eq:3.29}\end{eqnarray}
that diverges. Hence, contrary to the findings in Ref. \cite{FingerMandula},
the effective energy is not a physical observable and is clearly not IR finite.
Instead the excitation energy 
$\varepsilon(\mbf{k})-\varepsilon(0)$ is IR finite.
Using Eq. (\ref{eq:3.15}) and 
$V_{L+C}(\mbf{k},\mbf{q})\rightarrow V_{L+C}(\mbf{k}-\mbf{q})$, 
\begin{eqnarray}
\varepsilon(\mbf{k}) &=& \int\frac{d\mbf{q}}{(2\pi)^3} 
C_f V_{L+C}(\mbf{k},\mbf{q})
\left[\, s(\mbf{k})s(\mbf{q}) + c(\mbf{k})c(\mbf{q})
\mbf{\hat{k}}\cdot\mbf{\hat{q}} \,\right] + IR\,finite
\label{eq:3.30} \\
&\rightarrow & \int\frac{d\mbf{q}}{(2\pi)^3} 
C_f V_{L+C}(\mbf{q})
\left[\, s(\mbf{k})s(\mbf{k}-\mbf{q}) + c(\mbf{k})c(\mbf{k}-\mbf{q})
\mbf{\hat{k}}\cdot(\mbf{\hat{k-q}}) \,\right] + IR\, finite \nonumber 
\,,\end{eqnarray}
the excitation energy is given by the IR finite formula 
\begin{eqnarray}
\varepsilon(\mbf{k})-\varepsilon(0) &=& \int\frac{d\mbf{q}}{(2\pi)^3} 
C_f V_{L+C}(\mbf{q})
\left[\, s(\mbf{k})s(\mbf{k}-\mbf{q})-s(\mbf{q})
+ c(\mbf{k})c(\mbf{k}-\mbf{q})
\mbf{\hat{k}}\cdot(\mbf{\hat{k-q}}) \,\right]
\nonumber\\
&+& IR\, finite
\rightarrow \int d\mbf{q}|\mbf{q}|^{-4}O(\mbf{q}^2) 
\,,\label{eq:3.31}\end{eqnarray}
as $\mbf{q}\rightarrow 0$. The same holds for
$\Omega(\mbf{k})-\Omega(0)$. Using Eqs. (\ref{eq:3.10}) and 
(\ref{eq:3.6}),
\begin{eqnarray}
\Omega(\mbf{k}) &=& \frac{m(1+A(\mbf{k}))}{s(\mbf{k})} 
=\left(m+\int\frac{d\mbf{q}}{(2\pi)^3} 
C_f V_{L+C}(\mbf{k},\mbf{q})s(\mbf{q})\right)/s(\mbf{k})
+IR\, finite
\nonumber\\
&\rightarrow & \left(m+\int\frac{d\mbf{q}}{(2\pi)^3} 
C_f V_{L+C}(\mbf{q})s(\mbf{k}-\mbf{q})\right)/s(\mbf{k})
+IR\, finite
\,,\label{eq:3.32}\end{eqnarray}
the frequency pole difference is given by
\begin{eqnarray}
\Omega(\mbf{k})-\Omega(0) &=& \int\frac{d\mbf{q}}{(2\pi)^3} 
C_f V_{L+C}(\mbf{q})\left[\,\frac{s(\mbf{k}-\mbf{q})}{s(\mbf{k})} 
-s(\mbf{q})\,\right] +IR\, finite
\nonumber\\
&\rightarrow &\int d\mbf{q}|\mbf{q}|^{-4}O(\mbf{q}^2)
\,,\label{eq:3.33}\end{eqnarray}
that converges as $\mbf{q}\rightarrow 0$. However, by shifting the pole, 
the Feynman propagator, Eq. (\ref{eq:3.7}), still has the IR divergent 
structure
\begin{eqnarray}
S^{(4)}(\mbf{k},k_0) &=& 
\frac{R_{+}(\mbf{k})}{k_0-\Omega(0)-(\Omega(\mbf{k})-\Omega(0))}
+ \frac{R_{-}(\mbf{k})}{k_0+\Omega(0)+(\Omega(\mbf{k})-\Omega(0))}
\,,\label{eq:3.34}\end{eqnarray}
containing the IR divergent term $\Omega(0)$ in the denominators
but with the residues $R_{\pm}$ given by the IR finite expressions,
Eq. (\ref{eq:3.20}),
\begin{eqnarray}
R_{\pm}(\mbf{k}) &=& \frac{1}{2}\left[\gamma_0\pm s(\mbf{k})\right]
\mp \frac{1}{2}\mbf{\gamma}\cdot\mbf{\hat{k}}c(\mbf{k})
\,.\label{eq:3.35}\end{eqnarray}
This structure is a reflection of the fact that as a result of confinement,
an infinite amount of energy is required to create a single quasiparticle
state from the vacuum. This means that a color singlet state does not exist
and cannot be a physical state in our model. This is generally true 
in hadron physics. At the same time the energy gap, given by the gap equation, 
Eq. (\ref{eq:3.2}), does exist and corresponds to a physical quantity, 
defining the gap between a vacuum $|\Omega\rangle$ and the hadron scale.
This interpretation makes possible to map our model on the constituent quark model 
by associating the energy gap with the elementary degrees of freedom, quasiparticles. 
Quasiparticles with effective mass $M(\mbf{k})$ correspond to the valence quarks. 
The dynamics of quasiparticles is described by the Feynman propagator, Eq. (\ref{eq:3.17}), 
which is IR finite with a physical pole at $M(\mbf{k})$. In $3$-d a pole of 
the equal-time propagator is given by the effective energy of a quasiparticle,
$E(\mbf{k})=\sqrt{\mbf{k}^2+M^2(\mbf{k})}$.

Next we consider the quark condensate $\langle\Omega|\bar{\psi}\psi|\Omega\rangle$
for a single quark flavor. Using Eqs. (\ref{eq:2.6}) and (\ref{eq:2.7}), we obtain
\begin{eqnarray}
&& \langle\Omega|\bar{\psi}\psi|\Omega\rangle = \int\frac{d\mbf{k}}{(2\pi)^3}
\sum_s v^{\dagger}_s(-\mbf{k})\gamma_0 v_s(-\mbf{k})
= -2N_c \int\frac{d\mbf{k}}{(2\pi)^3}s(\mbf{k})
\,.\label{eq:3.36}\end{eqnarray}
This can also be evaluated in terms of the equal-time propagator 
$S^{(3)}(\mbf{k})$, 
\begin{eqnarray}
\langle\Omega|\bar{\psi}\psi|\Omega\rangle = N_c\delta_{\alpha\beta}
\langle\Omega| \frac{1}{2}[\bar{\psi}_{\beta}(0),\psi_{\alpha}(0)]
+\frac{1}{2}\{\bar{\psi}_{\beta}(0),\psi_{\alpha}(0)\} |\Omega\rangle
= -N_c \int\frac{d\mbf{k}}{(2\pi)^3} {\rm Tr}S^{(3)}(\mbf{k})
\,,\label{eq:3.37}\end{eqnarray}
where the minus sign arises because $1/2[\psi,\bar{\psi}]\rightarrow S^{(3)}$.
Substituting Eq. (\ref{eq:2.23}) for $S^{(3)}(\mbf{k})$ into 
Eq. (\ref{eq:3.37}), we get the quark condensate of Eq. (\ref{eq:3.36}).
We regulate the quark condensate, Eq. (\ref{eq:3.36}), by subtracting
the perturbative contribution 
\begin{eqnarray}
\langle\Omega|\bar{\psi}\psi|\Omega\rangle 
- \langle 0|\bar{\psi}\psi|0\rangle 
&=& -N_c \int\frac{d\mbf{k}}{(2\pi)^3} \left( {\rm Tr}S^{(3)}(\mbf{k})
-{\rm Tr}S_0^{(3)}(\mbf{k}) \right)
\nonumber\\
&=& -2N_c \int\frac{d\mbf{k}}{(2\pi)^3}\left( s(\mbf{k})
-\frac{m}{\sqrt{k^2+m^2}} \right)
\,,\label{eq:3.38}\end{eqnarray}
where $m$ is the bare quark mass.
As $|\mbf{k}|\rightarrow\infty$ the mass gap $M(\mbf{k})\rightarrow m$
and the nonperturbative sine behaves as 
$s(\mbf{k})\rightarrow m/\sqrt{k^2+m^2}$. Thus this subtraction improves 
the convergence of the quark condensate integral in the UV.

For further investigations it is convenient to introduce the scalar,
$\sigma_s$, and vector, $\sigma_v$, parts of the equal-time quark propagator, 
$S^{(3)}$ in Eq. (\ref{eq:3.9}), i.e.
\begin{eqnarray} 
S^{(3)}(\mbf{k})=\sigma_s-\mbf{\gamma}\cdot\mbf{k}\sigma_v
\,,\label{eq:3.39}\end{eqnarray}
and express all quantities of interest using them.
In particular, in the chiral limit $m\rightarrow 0$, 
one has from Eqs. (\ref{eq:3.9}) and (\ref{eq:3.10})
\begin{eqnarray} 
\sigma_s^{0}(\mbf{k}) &=& \frac{mA(\mbf{k})}{2\Omega(\mbf{k})}
=\frac{1}{2}s(\mbf{k}) \nonumber\\
\sigma_v^{0}(\mbf{k}) &=& \frac{1+B(\mbf{k})}{2\Omega(\mbf{k})}
=\frac{1}{2}\frac{c(\mbf{k})}{k} \nonumber\\
M_0(\mbf{k}) &=& \frac{mA(\mbf{k})}{1+B(\mbf{k})}
=\frac{\sigma_s(\mbf{k})}{\sigma_v(\mbf{k})} \nonumber\\
\langle\bar{\psi}\psi\rangle_0 &=& -2N_c\int\frac{d\mbf{k}}{(2\pi)^3}
\frac{mA(\mbf{k})}{\Omega(\mbf{k})} = -4N_c\int\frac{d\mbf{k}}{(2\pi)^3}
\sigma_s(\mbf{k}) \nonumber\\
E_0(\mbf{k}) &=& \frac{\Omega(\mbf{k})}{1+B(\mbf{k})}
=\frac{1}{2\sigma_v(\mbf{k})} \nonumber\\
\Omega_0(\mbf{k}) &=& \varepsilon(\mbf{k})
=\sqrt{m^2A^2(\mbf{k})+k^2(1+B(\mbf{k}))^2}
\,,\label{eq:3.40}\end{eqnarray}
where $A$ and $B$ functions are defined in Eq. (\ref{eq:3.6}),
and subscript $0$ denotes the chiral limit case.
When the scalar part \mbox{$mA\neq 0$}, 
the nonzero mass gap \mbox{$M_0=mA/(1+B)$} and chiral condensate
\mbox{$\langle\bar{\psi}\psi\rangle_0\sim\int d\mbf{k}mA/\Omega$}
are generated, and there is a dynamical chiral symmetry breaking 
($m=0$ corresponds to no explicit chiral symmetry breaking). 
An obvious property following from 
Eq. (\ref{eq:3.40})
\begin{eqnarray}
\sigma_s^2(\mbf{k})+k^2\sigma_v^2(\mbf{k}) = \frac{1}{4}
\,\label{eq:3.41}\end{eqnarray}
shows that low momentum behavior is governed by the scalar part 
of the propagator, $\sigma_s$.

One can generalize the above expressions for a nonzero current quark mass $m$
by substituting $mA(\mbf{k})\rightarrow m(1+A(\mbf{k}))$ in Eq. (\ref{eq:3.40}), 
and regulating the condensate by subtracting the perturbative value.

\subsection{Meson bound state equations: TDA and RPA}
\label{subsec:3.2}

As discussed in Section \ref{sec:2}, we map our effective model on 
the constituent quark model with the quasiparticles of mass $M(\mbf{k})$
playing the role of the valence quarks. In this Section, we represent
mesons as bound-states consisting of quasiparticles and seek
approximate eigensolution of our effective Hamiltonian. 
We do not solve here the Bethe-Salpeter or Salpeter equations.
Instead, we formulate bound state problem in the Tamm-Dancoff 
approximation (TDA) and subsequently in the Random Phase (RPA) 
approximation \cite{Walecka}. 
We show below that the flavour octet
family can be described well within these approximations, separating
the chiral symmetry and spin effects. However, flavour singlet states
need more elaborate approach, which we discuss in a separate publication.

In terms of the quasiparticle operators used in Section \ref{sec:2}, the TDA
meson creation operator reads
\begin{eqnarray}  
R_n^{\dagger} &=& \int\frac{d\mbf{q}}{(2\pi)^3}\sum_{\delta\gamma}
b_{\delta}^{\dagger}(\mbf{q})d_{\gamma}^{\dagger}(-\mbf{q})
\psi_n^{\delta\gamma}(\mbf{q})
\,,\label{eq:4.1}\end{eqnarray}   
which, acting on the vacuum, creates a meson with a wavefunction $|\psi_n\rangle$
with the quantum number $n$, and annihilates into the nonperturbative vacuum,
\begin{eqnarray}
R_n^{\dagger}|\Omega\rangle &=& |\psi_n\rangle
\nonumber\\
R_n|\Omega\rangle &=& 0 
\,.\label{eq:4.2}\end{eqnarray} 
Eq. (\ref{eq:4.2}) can be considered as a definition of the TDA vacuum, 
consisting of a condensate of only quasiparticle-quasihole pairs.
The commutation relation of the meson operators 
\begin{eqnarray} 
\langle\Omega|[R_{n^{\prime}},R_{n}^{\dagger}]|\Omega\rangle
=N\delta_{nn^{\prime}}
\,\label{eq:4.3}\end{eqnarray}  
leads to a normalization condition for the wave functions
\begin{eqnarray} 
\int\frac{d\mbf{q}}{(2\pi)^3} \sum_{\delta\gamma}
\psi_{n^{\prime}}^{\delta\gamma\ast}(\mbf{q})\psi_{n}^{\delta\gamma}(\mbf{q}) 
= N\delta_{nn^{\prime}}
\,,\label{eq:4.4}\end{eqnarray}
where $N$ is the normalization constant. Projecting the Schr{\"o}dinger 
equation $H_{eff}|\psi_{n}\rangle=E_n|\psi_{n}\rangle$
onto one-particle-one-hole truncated Fock sector, we get the TDA equation
\begin{eqnarray} 
\langle\Omega|[R_n,[H_{eff},b_{\alpha}^{\dagger}(\mbf{k})
d_{\beta}^{\dagger}(-\mbf{k})]]|\Omega\rangle &=& M_n \psi_n^{\alpha\beta}(\mbf{k})
\,,\label{eq:4.5}\end{eqnarray}  
where the binding energy is defined as $M_n=E_n-E_0$ with the vacuum energy $E_0$ 
subtracted ($H_{eff}|\Omega\rangle=E_0|\Omega\rangle$) and in the r.h.s.
of the TDA equation the operator $b^{\dagger}d^{\dagger}$ picks up 
the wave function component $\psi_n^{\alpha\beta}$, 
\begin{eqnarray} 
\langle\Omega|[R_n,b_{\alpha}^{\dagger}(\mbf{k})
d_{\beta}^{\dagger}(-\mbf{k})]|\Omega\rangle &=& 
\psi_n^{\alpha\beta}(\mbf{k})
\,.\label{eq:4.6}\end{eqnarray}  
The TDA is improved by extending the quark vacuum to contain 
also four quasiparticle condensates in addition
to two quasiparticle condensates.
Including up to four quasiparticle correlations beyond the BCS is known 
as the RPA approach.
Generalization of the operator of Eq. (\ref{eq:4.1}) containes meson
creation and annihilation terms
\begin{eqnarray}  
Q_n^{\dagger} &=& \int\frac{d\mbf{q}}{(2\pi)^3}\sum_{\delta\gamma}
\left[\, b_{\delta}^{\dagger}(\mbf{q})d_{\gamma}^{\dagger}(-\mbf{q})
X_n^{\delta\gamma}(\mbf{q})
-b_{\delta}(\mbf{q})d_{\gamma}(-\mbf{q})
Y_n^{\delta\gamma}(\mbf{q}) \,\right]
\,.\label{eq:4.7}\end{eqnarray} 
The RPA wavefunction and the RPA vacuum are
\begin{eqnarray}
Q_n^{\dagger}|\Omega\rangle &=& |\psi_n\rangle
\nonumber\\
Q_n|\Omega\rangle &=& 0 
\,,\label{eq:4.8}\end{eqnarray} 
where, though the same notations were used as above,
they should not be confused with the TDA wave function and TDA
vacuum. 
From the meson commutation relation  
\begin{eqnarray} 
\langle\Omega|[Q_{n^{\prime}},Q_{n}^{\dagger}]|\Omega\rangle
=N\delta_{nn^{\prime}}
\,,\label{eq:4.9}\end{eqnarray}  
the following normalization condition for the wave function
components, $X$  and $Y$, is obtained
\begin{eqnarray} 
\int\frac{d\mbf{q}}{(2\pi)^3} \sum_{\delta\gamma}\left[\,
X_{n^{\prime}}^{\delta\gamma\ast}(\mbf{q})X_{n}^{\delta\gamma}(\mbf{q})
-Y_{n^{\prime}}^{\delta\gamma\ast}(\mbf{q})Y_{n}^{\delta\gamma}(\mbf{q})
\,\right]
= N\delta_{nn^{\prime}}
\,,\label{eq:4.10}\end{eqnarray}
with the normalization constant $N$.
To derive the RPA equations of motion we calculate the commutators  
\begin{eqnarray} 
\langle\Omega|[Q_n,[H_{eff},b_{\alpha}^{\dagger}(\mbf{k})
d_{\beta}^{\dagger}(-\mbf{k})]]|\Omega\rangle &=& M_n X_n^{\alpha\beta}(\mbf{k})
\nonumber\\
\langle\Omega|[Q_n,[H_{eff},b_{\alpha}(\mbf{k})
d_{\beta}(-\mbf{k})]]|\Omega\rangle &=& M_n 
Y_n^{\alpha\beta}(\mbf{k})
\,,\label{eq:4.11}\end{eqnarray}  
which pick up the $X$ and $Y$ components in the r.h.s. of equations 
\begin{eqnarray} 
\langle\Omega|[Q_n,b_{\alpha}^{\dagger}(\mbf{k})
d_{\beta}^{\dagger}(-\mbf{k})]|\Omega\rangle &=& 
X_n^{\alpha\beta}(\mbf{k})
\nonumber\\
\langle\Omega|[Q_n,b_{\alpha}(\mbf{k})
d_{\beta}(-\mbf{k})]|\Omega\rangle &=& 
Y_n^{\alpha\beta}(\mbf{k})
\,.\label{eq:4.12}\end{eqnarray}
The RPA system of equations, Eq. (\ref{eq:4.11}), reduces to the TDA,
Eq. (\ref{eq:4.5}), by putting $Y=0$. In what follows, 
the RPA equations for $H_{eff}$, Eq. (\ref{eq:2.20}), are obtained. 
One- and two-body sectors of $H_{eff}$, Eq. (\ref{eq:2.37}), specified in 
Eqs. (\ref{eq:2.41}), (\ref{eq:2.42}) and (\ref{eq:3.15}), 
contribute to the RPA. Calculating the commutators of Eq. (\ref{eq:4.11}) with 
the effective Hamiltonian $H_{eff}$, the RPA equations are obtained as  
\begin{eqnarray} 
M_nX^{\alpha\beta}(\mbf{k}) &=& 2\varepsilon(\mbf{k})X^{\alpha\beta}(\mbf{k})
-2\int\frac{d\mbf{q}}{(2\pi)^3}I_{xx}^{\alpha\beta\delta\gamma}(\mbf{k},\mbf{q})
X^{\delta\gamma}(\mbf{q})
-2\int\frac{d\mbf{q}}{(2\pi)^3}I_{xy}^{\alpha\beta\delta\gamma}(\mbf{k},\mbf{q})
Y^{\delta\gamma}(\mbf{q})
\nonumber\\
&-&2\int\frac{d\mbf{q}}{(2\pi)^3}G_{xx}^{\alpha\beta\delta\gamma}(\mbf{k},\mbf{q})
X^{\delta\gamma}(\mbf{q})
-2\int\frac{d\mbf{q}}{(2\pi)^3}G_{xy}^{\alpha\beta\delta\gamma}(\mbf{k},\mbf{q})
Y^{\delta\gamma}(\mbf{q})
\nonumber\\
M_nY^{\alpha\beta}(\mbf{k}) &=& -2\varepsilon(\mbf{k})Y^{\alpha\beta}(\mbf{k})
+2\int\frac{d\mbf{q}}{(2\pi)^3}I_{yy}^{\alpha\beta\delta\gamma}(\mbf{k},\mbf{q})
Y^{\delta\gamma}(\mbf{q})
+2\int\frac{d\mbf{q}}{(2\pi)^3}I_{yx}^{\alpha\beta\delta\gamma}(\mbf{k},\mbf{q})
X^{\delta\gamma}(\mbf{q})
\nonumber\\
&+&2\int\frac{d\mbf{q}}{(2\pi)^3}G_{yy}^{\alpha\beta\delta\gamma}(\mbf{k},\mbf{q})
Y^{\delta\gamma}(\mbf{q})
+2\int\frac{d\mbf{q}}{(2\pi)^3}G_{yx}^{\alpha\beta\delta\gamma}(\mbf{k},\mbf{q})
X^{\delta\gamma}(\mbf{q})
\,,\label{eq:4.13}\end{eqnarray}
where an effective single particle energy $\varepsilon(\mbf{k})$ 
is defined in Eq. (\ref{eq:3.15}). Here, the instantaneous tensor terms 
$I(\mbf{k},\mbf{q})$ are
\begin{eqnarray} 
I_{xx}^{\alpha\beta\delta\gamma}(\mbf{k},\mbf{q}) &=& 
C_fV_{L+C}(\mbf{k},\mbf{q})(v^{\dagger}_{\beta}(-\mbf{k})v_{\gamma}(-\mbf{q}))
(u^{\dagger}_{\delta}(\mbf{q})u_{\alpha}(\mbf{k}))
\nonumber\\
I_{yy}^{\alpha\beta\delta\gamma}(\mbf{k},\mbf{q}) &=& 
C_fV_{L+C}(\mbf{k},\mbf{q})(u^{\dagger}_{\alpha}(\mbf{k})u_{\delta}(\mbf{q}))
(v^{\dagger}_{\gamma}(-\mbf{q})v_{\beta}(-\mbf{k}))=I_{xx}^{\dagger}
\nonumber\\
I_{xy}^{\alpha\beta\delta\gamma}(\mbf{k},\mbf{q})&=&
-C_fV_{L+C}(\mbf{k},\mbf{q})(v^{\dagger}_{\beta}(-\mbf{k})u_{\delta}(\mbf{q}))
(v^{\dagger}_{\gamma}(-\mbf{q})u_{\alpha}(\mbf{k}))
\label{eq:4.14} \\ 
I_{yx}^{\alpha\beta\delta\gamma}(\mbf{k},\mbf{q}) &=&
-C_fV_{L+C}(\mbf{k},\mbf{q})(u^{\dagger}_{\alpha}(\mbf{k})v_{\gamma}(-\mbf{q}))
(u^{\dagger}_{\delta}(\mbf{q})v_{\beta}(-\mbf{k}))=I_{xy}^{\dagger}\nonumber
\,,\end{eqnarray}
and the generated terms $G(\mbf{k},\mbf{q})$ are 
\begin{eqnarray} 
G_{xx}^{\alpha\beta\delta\gamma}(\mbf{k},\mbf{q}) &=& C_fW_1(\mbf{k},\mbf{q})
(v^{\dagger}_{\beta}(-\mbf{k})\alpha_iv_{\gamma}(-\mbf{q}))
(u^{\dagger}_{\delta}(\mbf{q})\alpha_ju_{\alpha}(\mbf{k}))
D_{ij}(\mbf{k}-\mbf{q})
\nonumber\\
G_{yy}^{\alpha\beta\delta\gamma}(\mbf{k},\mbf{q}) &=& C_fW_1(\mbf{k},\mbf{q})
(u^{\dagger}_{\alpha}(\mbf{k})\alpha_iu_{\delta}(\mbf{q}))
(v^{\dagger}_{\gamma}(-\mbf{q})\alpha_jv_{\beta}(-\mbf{k}))
D_{ij}(\mbf{k}-\mbf{q})=G_{xx}^{\dagger}
\nonumber\\
G_{xy}^{\alpha\beta\delta\gamma}(\mbf{k},\mbf{q})&=& -C_fW_2(\mbf{k},\mbf{q})
(v^{\dagger}_{\beta}(-\mbf{k})\alpha_iu_{\delta}(\mbf{q}))
(v^{\dagger}_{\gamma}(-\mbf{q})\alpha_ju_{\alpha}(\mbf{k}))
D_{ij}(\mbf{k}-\mbf{q})
\label{eq:4.15} \\
G_{yx}^{\alpha\beta\delta\gamma}(\mbf{k},\mbf{q}) &=& -C_fW_2(\mbf{k},\mbf{q})
(u^{\dagger}_{\alpha}(\mbf{k})\alpha_iv_{\gamma}(-\mbf{q}))
(u^{\dagger}_{\delta}(\mbf{q})\alpha_jv_{\beta}(-\mbf{k}))
D_{ij}(\mbf{k}-\mbf{q})=G_{xy}^{\dagger} \nonumber
\,,\end{eqnarray} 
with potential functions given by Eqs. (\ref{eq:2.24}) and (\ref{eq:2.43}).
Using Eq. (\ref{eq:2.7}), we represent the instantaneous terms,
Eq. (\ref{eq:4.14}), as
\begin{eqnarray} 
I_{xx}^{\alpha\beta\delta\gamma}(\mbf{k},\mbf{q}) &=& 
I_{yy}^{\alpha\beta\delta\gamma\dagger}(\mbf{k},\mbf{q})\nonumber\\
&=& C_fV_{L+C}(\mbf{k},\mbf{q})
\frac{1}{4}\left[\,\phantom{\mbf{\hat{k}}}\hspace{-0.3cm}
(1+s(\mbf{k}))(1+s(\mbf{q}))\delta_{\beta\gamma}
\delta_{\delta\alpha}\right.
\nonumber\\
&+&\left.(1-s(\mbf{k}))(1-s(\mbf{q}))\chi^{\dagger}_{\beta}\sigma_2
\mbf{\sigma}\cdot\mbf{\hat{k}}\mbf{\sigma}\cdot\mbf{\hat{q}}
\sigma_2\chi_{\gamma}\chi^{\dagger}_{\delta}
\mbf{\sigma}\cdot\mbf{\hat{q}}\mbf{\sigma}\cdot\mbf{\hat{k}}
\chi_{\alpha}\right.
\nonumber\\
&+&\left. c(\mbf{k})c(\mbf{q})(\delta_{\delta\alpha}
\chi^{\dagger}_{\beta}\sigma_2
\mbf{\sigma}\cdot\mbf{\hat{k}}\mbf{\sigma}\cdot\mbf{\hat{q}}
\sigma_2\chi_{\gamma}+\delta_{\beta\gamma}\chi^{\dagger}_{\delta}
\mbf{\sigma}\cdot\mbf{\hat{q}}\mbf{\sigma}\cdot\mbf{\hat{k}}
\chi_{\alpha})\, \right]
\label{eq:4.14a}\\
I_{xy}^{\alpha\beta\delta\gamma}(\mbf{k},\mbf{q}) &=& 
I_{yx}^{\alpha\beta\delta\gamma\dagger}(\mbf{k},\mbf{q})\nonumber\\
&=& C_fV_{L+C}(\mbf{k},\mbf{q})
\frac{1}{4}\left[\,(1+s(\mbf{k}))(1-s(\mbf{q}))
\chi^{\dagger}_{\beta}\sigma_2\mbf{\sigma}\cdot\mbf{\hat{q}}\chi_{\delta}
\chi^{\dagger}_{\gamma}\sigma_2\mbf{\sigma}\cdot\mbf{\hat{q}}\chi_{\alpha}
\right. \nonumber\\
&+&\left. (1-s(\mbf{k}))(1+s(\mbf{q}))
\chi^{\dagger}_{\beta}\sigma_2\mbf{\sigma}\cdot\mbf{\hat{k}}\chi_{\delta}
\chi^{\dagger}_{\gamma}\sigma_2\mbf{\sigma}\cdot\mbf{\hat{k}}\chi_{\alpha}
\right. \nonumber\\
&-&\left. c(\mbf{k})c(\mbf{q})
(\chi^{\dagger}_{\beta}\sigma_2\mbf{\sigma}\cdot\mbf{\hat{q}}\chi_{\delta}
\chi^{\dagger}_{\gamma}\sigma_2\mbf{\sigma}\cdot\mbf{\hat{k}}\chi_{\alpha}
+\chi^{\dagger}_{\beta}\sigma_2\mbf{\sigma}\cdot\mbf{\hat{k}}\chi_{\delta}
\chi^{\dagger}_{\gamma}\sigma_2\mbf{\sigma}\cdot\mbf{\hat{q}}\chi_{\alpha})
\,\right]\nonumber
\,,\end{eqnarray}
and use them in further calculations. 
Analogous expressions can also be found
for the generated terms $G(\mbf{k},\mbf{q})$, Eq. (\ref{eq:4.15}).
A crucial test of any approach dealing with chiral symmetry 
is the ability to describe the pseudoscalar meson channel. We therefore
consider an application of Eq. (\ref{eq:4.13}) to $\pi$,
$J^{PC}=0^{++}$ $L=S=1,J=0$ pseudoscalar, and $\rho$, $J^{PC}=0^{-+}$ $L=S=J=0$ 
vector, states. Based on the quantum numbers of these states, 
the tensor structure of $\pi$- and $\rho$-wave functions can be identified as 
\begin{eqnarray} 
X^{\alpha\beta}_{\pi}(\mbf{k}) &=& (i\sigma_2)^{\alpha\beta}X_{\pi}(\mbf{k})
\,,\,
Y^{\alpha\beta}_{\pi}(\mbf{k}) = (-i\sigma_2)^{\alpha\beta}Y_{\pi}(\mbf{k})
\nonumber\\
X^{\alpha\beta}_{\rho}(\mbf{k}) &=& (\mbf{\sigma}i\sigma_2)^{\alpha\beta}
X_{\rho}(\mbf{k})
\,,\,
Y^{\alpha\beta}_{\rho}(\mbf{k}) = 
(-i\sigma_2\mbf{\sigma})^{\alpha\beta}Y_{\rho}(\mbf{k})
\,.\label{eq:4.16}\end{eqnarray}
The normalization is chosen
\begin{eqnarray} 
\int\frac{d\mbf{k}}{(2\pi)^3}\left(X^*(\mbf{k})X(\mbf{k})
-Y^*(\mbf{k})Y(\mbf{k})\right)=1
\,,\label{eq:4.16a}\end{eqnarray}
reducing the normalization costant of the full wave function,
Eq. (\ref{eq:4.10}), to
\begin{eqnarray}
N=\langle\psi_n|\psi_n\rangle=2N_c
\,,\label{eq:4.16b}\end{eqnarray}
where the factor $2$ comes from the trace in the spinor space, and $N_c=3$.
The RPA equations for the momentum wave function components 
$X(\mbf{k}), Y(\mbf{k})$ have the same form for $\pi$ and $\rho$ states
\begin{eqnarray} 
M_nX(\mbf{k}) &=& 2\varepsilon(\mbf{k})X(\mbf{k})
-\int\frac{q^2dqdx}{4\pi^2}I_{xx}(\mbf{k},\mbf{q})X(\mbf{q})
-\int\frac{q^2dqdx}{4\pi^2}I_{xy}(\mbf{k},\mbf{q})Y(\mbf{q})
\nonumber\\
&-&\int\frac{q^2dqdx}{4\pi^2}G_{xx}(\mbf{k},\mbf{q})X(\mbf{q})
-\int\frac{q^2dqdx}{4\pi^2}G_{xy}(\mbf{k},\mbf{q})Y(\mbf{q})
\nonumber\\
-M_nY(\mbf{k}) &=& 2\varepsilon(\mbf{k})Y(\mbf{k})
-\int\frac{q^2dqdx}{4\pi^2}I_{yy}(\mbf{k},\mbf{q})Y(\mbf{q})
-\int\frac{q^2dqdx}{4\pi^2}I_{yx}(\mbf{k},\mbf{q})X(\mbf{q})
\nonumber\\
&-&\int\frac{q^2dqdx}{4\pi^2}G_{yy}(\mbf{k},\mbf{q})Y(\mbf{q})
-\int\frac{q^2dqdx}{4\pi^2}G_{yx}(\mbf{k},\mbf{q})X(\mbf{q})
\,,\label{eq:4.17}\end{eqnarray}
where the kernels $I$ and $G$ for $\pi$ are
\begin{eqnarray} 
I_{xx}^{\pi}(\mbf{k},\mbf{q}) &=& I_{yy}^{\pi}(\mbf{k},\mbf{q})
= C_fV_{L+C}(\mbf{k},\mbf{q})
\frac{1}{2}\left[\,
(1+s(\mbf{k}))(1+s(\mbf{q}))\right.\nonumber\\
&+&\left.(1-s(\mbf{k}))(1-s(\mbf{q}))
+2c(\mbf{k})c(\mbf{q})x\, \right]
\nonumber\\
I_{xy}^{\pi}(\mbf{k},\mbf{q}) &=& I_{yx}^{\pi}(\mbf{k},\mbf{q})
= C_fV_{L+C}(\mbf{k},\mbf{q})
\frac{1}{2}\left[\,
-(1+s(\mbf{k}))(1-s(\mbf{q}))\right.\nonumber\\
&-&\left.(1-s(\mbf{k}))(1+s(\mbf{q}))
+2c(\mbf{k})c(\mbf{q})x\, \right]
\nonumber\\
G_{xx}^{\pi}(\mbf{k},\mbf{q}) &=& G_{yy}^{\pi}(\mbf{k},\mbf{q})=
2C_fW_1(\mbf{k},\mbf{q})
\frac{1}{2}\left[\,\phantom{\frac{x^2}{x^2}}\hspace{-0.5cm}
-(1+s(\mbf{k}))(1-s(\mbf{q}))\right.\nonumber\\
&-&\left.(1-s(\mbf{k}))(1+s(\mbf{q}))
-2c(\mbf{k})c(\mbf{q})
\frac{(1+x^2)kq-x(k^2+q^2)}{(\mbf{k}-\mbf{q})^2}\,\right]
\nonumber\\
G_{xy}^{\pi}(\mbf{k},\mbf{q}) &=& G_{yx}^{\pi}(\mbf{k},\mbf{q})=
2C_fW_2(\mbf{k},\mbf{q})
\frac{1}{2}\left[\,\phantom{\frac{x^2}{x^2}}\hspace{-0.5cm}
(1+s(\mbf{k}))(1+s(\mbf{q}))\right.\nonumber\\
&+&\left.(1-s(\mbf{k}))(1-s(\mbf{q}))
- 2c(\mbf{k})c(\mbf{q})
\frac{(1+x^2)kq-x(k^2+q^2)}{(\mbf{k}-\mbf{q})^2}\, \right]
\,,\label{eq:4.18}\end{eqnarray}
and for $\rho$ are
\begin{eqnarray} 
I_{xx}^{\rho}(\mbf{k},\mbf{q}) &=& I_{yy}^{\rho}(\mbf{k},\mbf{q})=
C_fV_{L+C}(\mbf{k},\mbf{q})\frac{1}{2}\left[\,
\phantom{\frac{x^2}{x^2}}\hspace{-0.5cm}
(1+s(\mbf{k}))(1+s(\mbf{q}))\right.\nonumber\\
&+&\left.\frac{1}{3}(1-s(\mbf{k}))(1-s(\mbf{q}))
(4x^2-1)+2c(\mbf{k})c(\mbf{q})x\, \right]
\nonumber\\
I_{xy}^{\rho}(\mbf{k},\mbf{q}) &=& I_{yx}^{\rho}(\mbf{k},\mbf{q})=
C_fV_{L+C}(\mbf{k},\mbf{q})\frac{1}{2}\left[\,
-\frac{1}{3}(1+s(\mbf{k}))(1-s(\mbf{q}))\right.\nonumber\\
&-&\left.\frac{1}{3}(1-s(\mbf{k}))(1+s(\mbf{q})) 
+\frac{2}{3}c(\mbf{k})c(\mbf{q})x\, \right]
\nonumber\\
G_{xx}^{\rho}(\mbf{k},\mbf{q}) &=& G_{yy}^{\rho}(\mbf{k},\mbf{q})=
C_fW_1(\mbf{k},\mbf{q})\frac{1}{2}\left[\,
\frac{1}{3}(1+s(\mbf{k}))(1-s(\mbf{q}))
\left(1-\frac{2(1-x^2)k^2}{(\mbf{k}-\mbf{q})^2}\right)\right.\nonumber\\
&+&\left.\frac{1}{3}(1-s(\mbf{k}))(1+s(\mbf{q}))
\left(1-\frac{2(1-x^2)q^2}{(\mbf{k}-\mbf{q})^2}\right)
-\frac{2}{3}c(\mbf{k})c(\mbf{q})
\left(x+\frac{(1-x^2)kq}{(\mbf{k}-\mbf{q})^2}\right)\, \right]
\nonumber\\
G_{xy}^{\rho}(\mbf{k},\mbf{q}) &=& G_{yx}^{\rho}(\mbf{k},\mbf{q})=
C_fW_2(\mbf{k},\mbf{q})
\frac{1}{2}\left[\,
\frac{1}{3}(1+s(\mbf{k}))(1+s(\mbf{q}))\right.\nonumber\\
&+&\left.\frac{1}{3}(1-s(\mbf{k}))(1-s(\mbf{q}))(2x^2-1)
+\frac{2}{3}c(\mbf{k})c(\mbf{q})
\left(x-\frac{(1-x^2)kq}{(\mbf{k}-\mbf{q})^2}\right)\, \right]
\,,\label{eq:4.19}\end{eqnarray}
where we introduced $x=\mbf{\hat{k}}\cdot\mbf{\hat{q}}$, and used
\begin{eqnarray}
1-(\mbf{\hat{k}}\cdot\mbf{\hat{l}})^2 &=&\frac{(1-x^2)q^2}{(\mbf{k}-\mbf{q})^2}
\,,\,
1-(\mbf{\hat{q}}\cdot\mbf{\hat{l}})^2 =\frac{(1-x^2)k^2}{(\mbf{k}-\mbf{q})^2}
\nonumber\\
\mbf{\hat{k}}\cdot\mbf{\hat{l}}\mbf{\hat{q}}\cdot\mbf{\hat{l}} &=&
\frac{x(k^2+q^2)-(1+x^2)kq}{(\mbf{k}-\mbf{q})^2}
\,,\label{eq:4.20}\end{eqnarray}
with $\mbf{l}=\mbf{k}-\mbf{q}$.
The obtained $\pi$ and $\rho$ RPA equations are IR finite
as discussed in the following. Consider the collinear limit 
$\mbf{k}\rightarrow\mbf{q}$ for the instantaneous
terms, $I(\mbf{k},\mbf{q})$, since the confining potential in $V_{L+C}$
causes the IR problem. We get
\begin{eqnarray}
&& I_{xx}^{\pi}\rightarrow V_{L+C}(\mbf{k},\mbf{q})
\left[\,1+s(\mbf{k})s(\mbf{q})+c(\mbf{k})c(\mbf{q})\,\right]
\rightarrow 2 (\mbf{k}-\mbf{q})^{-4}
\nonumber\\
&& I_{xx}^{\rho}\rightarrow V_{L+C}(\mbf{k},\mbf{q})
\left[\,1+s(\mbf{k})s(\mbf{q})+c(\mbf{k})c(\mbf{q})\,\right]
\rightarrow 2 (\mbf{k}-\mbf{q})^{-4}
\,,\label{eq:4.21}\end{eqnarray}
and
\begin{eqnarray}
&& I_{xy}^{\pi}\rightarrow V_{L+C}(\mbf{k},\mbf{q})
\left[\,-(1-s(\mbf{k})s(\mbf{q}))+c(\mbf{k})c(\mbf{q})\,\right]\nonumber\\
&& \rightarrow V_{L+C}(\mbf{k},\mbf{q})O(\mbf{k}-\mbf{q})
+O((\mbf{k}-\mbf{q})^{-2})
\nonumber\\
&& I_{xy}^{\rho}\rightarrow V_{L+C}(\mbf{k},\mbf{q})
\frac{1}{3}\left[\,-(1-s(\mbf{k})s(\mbf{q}))+c(\mbf{k})c(\mbf{q})\,\right]
\nonumber\\
&& \rightarrow V_{L+C}(\mbf{k},\mbf{q})O(\mbf{k}-\mbf{q})
+O((\mbf{k}-\mbf{q})^{-2})
\,.\label{eq:4.22}\end{eqnarray}
The TDA kernel $I_{xx}$ has the same IR behavior for both channels
and is IR singular. However the effective energy $\varepsilon(\mbf{k})$
behaves the same way in the IR, Eq. (\ref{eq:3.29}), 
and comes with an opposite sign in the TDA/RPA equation, 
cancelling exactly the IR divergence.
The IR behavior of the TDA kernel should be same for other channels,
and can be used to check calculations.
In the RPA kernel, $I_{xy}$, the first term $O(\mbf{k}-\mbf{q})$
disappears after the angular integration, and the second term
$O((\mbf{k}-\mbf{q})^{-2})$ converges in the intergral. Thus,
the TDA and RPA equations are IR finite for the confining potential.
In the UV the potential part of interaction,
which contains intergrals with kernels $I$ and $G$, is regulated
by the wave functions $X$ and $Y$, vanishing for large momenta.
The kinetic part contains the UV finite effective energy,
$\varepsilon(\mbf{k})$, Eq. (\ref{eq:3.15}), which has been renormalized
by adding the mass counterterm.

The RPA equations are the eigenvalue problem for $M_n$ which can be
diagonalized in the twice size space of $(X,Y)$, compare to the TDA
requiring the size of only $X$. In the matrix form, Eq. (\ref{eq:4.17}) 
is written as
\begin{eqnarray}
A(\mbf{k},\mbf{q})X(\mbf{q})+B(\mbf{k},\mbf{q})Y(\mbf{q})
&=& MX(\mbf{k})
\nonumber\\
-B(\mbf{k},\mbf{q})X(\mbf{q})-A(\mbf{k},\mbf{q})Y(\mbf{q})
&=& MY(\mbf{k})
\,,\label{eq:4.23}\end{eqnarray}
with
\begin{eqnarray}
A(\mbf{k},\mbf{q}) &=& (2\varepsilon(\mbf{k})\delta_{\mbf{k},\mbf{q}}
-F_{xx}(\mbf{k},\mbf{q}))d\mbf{q}
\rightarrow
2\varepsilon(\mbf{k})-d\mbf{q}F_{xx}(\mbf{k},\mbf{q})
\nonumber\\
B(\mbf{k},\mbf{q}) &=& -d\mbf{q}F_{xy}(\mbf{k},\mbf{q})
\,,\label{eq:4.24}\end{eqnarray}
where $F$ includes the instantaneous and generated terms,
Eqs. (\ref{eq:4.18}) and (\ref{eq:4.19});
$F_{xx}(\mbf{k},\mbf{q})=I_{xx}(\mbf{k},\mbf{q})+G_{xx}(\mbf{k},\mbf{q})$,
and the same for $xy$ component. In Eq. (\ref{eq:4.23}) the integration
over $\mbf{q}$ is implied, and we omited factors $(2\pi)^3$.
The RPA matrix size can be reduced by a factor of $2$. Using variables
\begin{eqnarray}
\psi_{\pm} &=& X\pm Y
\,,\label{eq:4.25}\end{eqnarray}
the RPA equations are given by
\begin{eqnarray}
(A+B)\psi_{+} &=& M\psi_{-}
\nonumber\\
(A-B)\psi_{-} &=& M\psi_{+}
\,,\label{eq:4.26}\end{eqnarray}
which can be decoupled, at the expence of more complicated kernel, as
\begin{eqnarray}
\left[\,(A-B)(A+B)\,\right](\mbf{k},\mbf{q})\psi_{+}(\mbf{q}) &=& 
M^2\psi_{+}(\mbf{k})
\nonumber\\
\left[\,(A+B)(A-B)\,\right](\mbf{k},\mbf{q})\psi_{-}(\mbf{q}) &=& 
M^2\psi_{-}(\mbf{k})
\,.\label{eq:4.27}\end{eqnarray}
The diagonalization of either $(A-B)(A+B)$ or $(A+B)(A-B)$
gives the eigenvalue $M$ for the RPA problem.
Taking $B=0$ we come back to the TDA equation.

Finally, we calculate pion decay constant
in the TDA and RPA schemes. Using Thouless' theorem applied
to the chiral charge in one flavor case  
\begin{eqnarray}
Q_5 &=& \int d\mbf{x}\psi^{\dagger}(\mbf{x})\gamma_5\psi(\mbf{x})
\,,\label{eq:4.28}\end{eqnarray}
and the effective Hamiltonian, one gets
\begin{eqnarray}
\langle\Omega|[Q_5,[Q_5,H_{eff}]]|\Omega\rangle &=&
4m_{const}\langle\Omega|\bar{\psi}\psi|\Omega\rangle
\,.\label{eq:4.29}\end{eqnarray}
Using $\hat{I}=\sum_n|\psi_n\rangle\langle\psi_n|/N$,
the l.h.s. can be written as
\begin{eqnarray}
\langle\Omega|[Q_5,[Q_5,H_{eff}]]|\Omega\rangle &=&
-2\sum_n\frac{1}{N}|\langle\Omega|Q_5|\psi_n\rangle|^2M_n
\,,\label{eq:4.30}\end{eqnarray}
where $M_n=E_n-E_0$, $H_{eff}|\psi_n\rangle=E_n|\psi_n\rangle$
and $H_{eff}|\Omega\rangle=E_0|\Omega\rangle$.
Here, the normalization constant is $N=\sum_n|\psi_n|^2$, 
and the volume is set to unity, $V=1$. 
Defining the weak decay constant as
\begin{eqnarray}
f_n &=& \frac{1}{\sqrt{NM_n}}\langle\Omega|Q_5|\psi_n\rangle
\,,\label{eq:4.31}\end{eqnarray}
we obtain the Gell'Mann-Oakes-Renner relation
\begin{eqnarray}
\sum_nf^2_nM^2_n &=&-2m_{const}\langle\bar{\psi}\psi\rangle
\,,\label{eq:4.32}\end{eqnarray}
where $m_{const}$ is the quark constituent mass, and
$\langle\bar{\psi}\psi\rangle$ is the quark condensate. 
Using Eq. (\ref{eq:2.6}), the chiral charge is given by
\begin{eqnarray}
Q_5 &=& \sum_s\int\frac{d\mbf{k}}{(2\pi)^3}\left(\,
c(\mbf{k})\mbf{\sigma}\cdot\hat{k}\,
[b_s^{\dagger}(\mbf{k})b_s(\mbf{k})+d_s^{\dagger}(\mbf{k})d_s(\mbf{k})]
\right.\nonumber\\
&+&\left.s(\mbf{k})\,
[b_s^{\dagger}(\mbf{k})d_s^{\dagger}(-\mbf{k})+d_s(-\mbf{k})b_s(\mbf{k})]
\,\right)
\,.\label{eq:4.33}\end{eqnarray}
Taking the matrix element of the chiral charge, Eq. (\ref{eq:4.33}),
between the vacuum state and the RPA pion wave function, Eqs. (\ref{eq:4.7})
and (\ref{eq:4.8}), we have
\begin{eqnarray}
f_{\pi} &=& \frac{2N_c}{\sqrt{M_{\pi}N}}\int\frac{d\mbf{k}}{(2\pi)^3}
s(\mbf{k})\left(X_{\pi}(\mbf{k})-Y_{\pi}(\mbf{k})\right)
\,,\label{eq:4.341}\end{eqnarray}
where from Eq. (\ref{eq:4.16b}) $N=2N_c$.
For $N_c=3$, the pion decay constant is given 
\begin{eqnarray}
f_{\pi} &=& \frac{\sqrt{6}}{\sqrt{M_{\pi}}}\int\frac{k^2dk}{2\pi^2}
s(\mbf{k})\left(X_{\pi}(\mbf{k})-Y_{\pi}(\mbf{k})\right)
\,,\label{eq:4.351}\end{eqnarray}
in the RPA. In the TDA, $Y=0$ and $M_{\pi}$ is the eigenvalue
of the TDA equation.

\section{Numerical results} 
\label{sec:4}

In this section we obtain the numerical solutions of the quark gap equation
(subsection \ref{subsec:4.1}) and the TDA/RPA bound state equations 
(subsection \ref{subsec:4.2}), and discuss the results.

For the numerical calculations we have used
the routines from the SLATEC linear algebra archive, part of 
the Netlib database maintained by UTK and ORNL. The routines are found at
$www.netlib.org/slatec/lin/$ and a description of the entire SLATEC archive
can be found at $www.netlib.org/slatec/toc$.

\subsection{Energy gap, quark propagator and chiral condensate}
\label{subsec:4.1}

\begin{figure}[!htb]
\begin{center}
\input{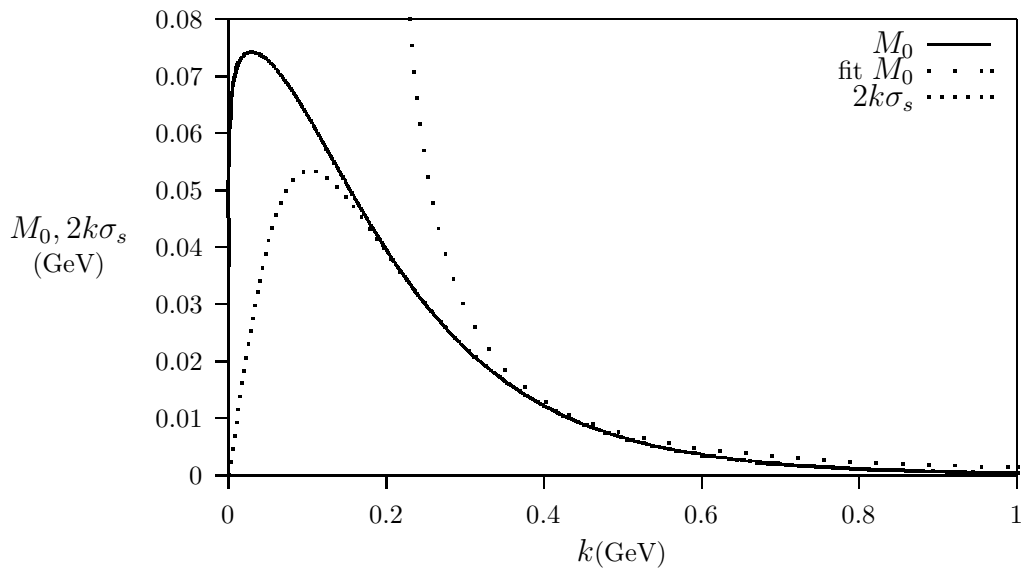}
\end{center}
\caption{The numerical solution of dynamical quark mass, 
$M_0(k)$, and the scalar part of the propagator, 
$2k\sigma_s$, in the chiral limit with confinement. 
The parameters for the numerical solution of 
the gap equation are $\sigma=0.18GeV^2$, $\Lambda=1GeV$.
The results are compared with the fit function given by 
$M_0(k)=0.0024/(k^2[\ln(k^2/0.04)]^{0.43})$ 
(parameters are in powers of $GeV$).}
\label{fig.1}
\end{figure}

\begin{figure}[!htb]
\begin{center}
\input{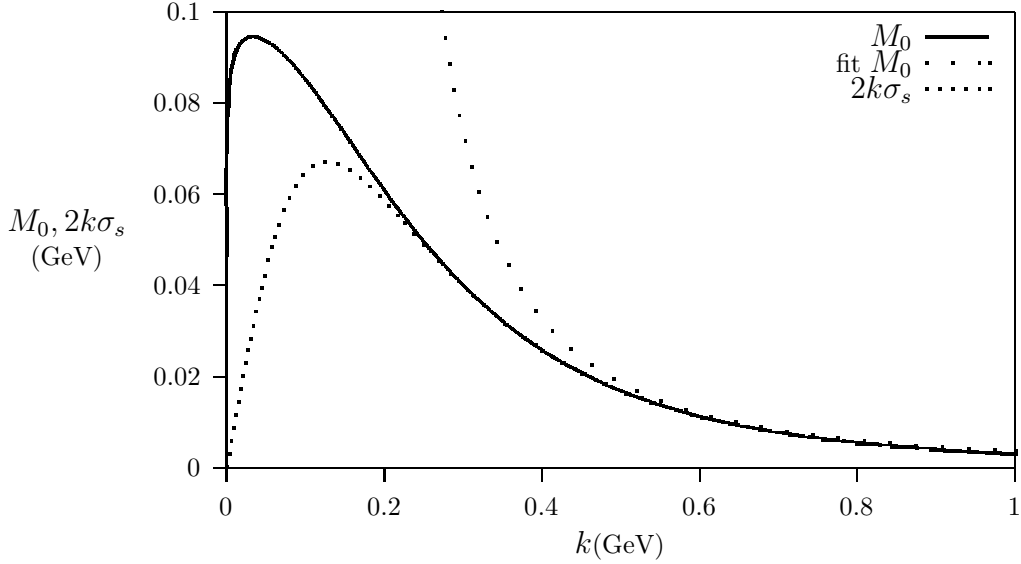}
\end{center}
\caption{The numerical solution of dynamical quark mass, $M_0(k)$, 
and the scalar part of the propagator, $2k\sigma_s$, in the chiral limit
when Coulomb and generated potentials are added with the running coupling
$\alpha_s(k^2)$ (same parameters as in Fig.1). 
The results are compared with the fit function given by 
$M_0(k)=0.0060/(k^2[\ln(k^2/0.04)]^{0.43})$.}
\label{fig.2}
\end{figure}

We have numerically solved the linearized gap equation, 
Eq. (\ref{eq:3.5}), for 
\begin{eqnarray} 
M(\mbf{k}) &=& M^{(0)}(\mbf{k})+\delta M(\mbf{k})
\,,\label{eq:3.42}\end{eqnarray}
where the mass correction is denoted as $\delta M(\mbf{k})$,
while the mass gap $M^{(0)}(\mbf{k})$ is the first iteration 
obtained by the gauss algorithm \cite{AdlerDavis}, 
providing an initial guess.
The linearized gap equation has the form
\begin{eqnarray}
\int d\mbf{q}A(\mbf{k},\mbf{q})\delta M(\mbf{q}) &=& B(\mbf{k})
\,,\label{eq:3.43}\end{eqnarray}
and, using Eq. (\ref{eq:2.8}) for sine and cosine, it reads
\begin{eqnarray}
&& \delta M(\mbf{k})\left[\,k+\int\frac{q^2dqdx}{4\pi^2}
\left(\frac{C_fV_{L+C}(\mbf{k},\mbf{q})qx}{\sqrt{q^2+(M^{(0)}(\mbf{q}))^{2}}}
{\rm e}^{-q^2/\Lambda^2}\right.\right.\nonumber\\
&+&\left.\left. \frac{C_fW(\mbf{k},\mbf{q})q}{\sqrt{q^2+(M^{(0)}(\mbf{q}))^{2}}}
\frac{(k^2+q^2)x-kq(1+x^2)}{(\mbf{k}-\mbf{q})^2}
{\rm e}^{-4q^2/\Lambda^2}\right)\,\right]
\nonumber\\
&+&\int\frac{q^2dqdx}{4\pi^2}\delta M(\mbf{q})\left[\,
\frac{C_fV_{L+C}(\mbf{k},\mbf{q})}{\sqrt{q^2+(M^{(0)}(\mbf{q}))^{2}}}
\left(\frac{(M^{(0)}(\mbf{q}))^{2}k}{q^2+(M^{(0)}(\mbf{q}))^{2}}
-\frac{M^{(0)}(\mbf{q})M^{(0)}(\mbf{k})qx}{q^2+(M^{(0)}(\mbf{q}))^{2}}-k\right)
{\rm e}^{-q^2/\Lambda^2}\right.\nonumber\\
&+&\left.\frac{C_fW(\mbf{k},\mbf{q})}{\sqrt{q^2+(M^{(0)}(\mbf{q}))^{2}}}
\left(\frac{(M^{(0)}(\mbf{q}))^{2}k}{q^2+(M^{(0)}(\mbf{q}))^{2}}
-\frac{M^{(0)}(\mbf{q})M^{(0)}(\mbf{k})q}{q^2+(M^{(0)}(\mbf{q}))^{2}}
\, \frac{(k^2+q^2)x-kq(1+x^2)}{(\mbf{k}-\mbf{q})^2}-k\right)
{\rm e}^{-4q^2/\Lambda^2}\,\right]
\nonumber\\
&=& -k(M^{(0)}(\mbf{k})-m(\Lambda))
+\int\frac{q^2dqdx}{4\pi^2}\left[\,
\frac{C_fV_{L+C}(\mbf{k},\mbf{q})}{\sqrt{q^2+(M^{(0)}(\mbf{q}))^{2}}}
\left(M^{(0)}(\mbf{q})k-M^{(0)}(\mbf{k})qx\right){\rm e}^{-q^2/\Lambda^2}
\right.\nonumber\\
&+&\left.\frac{C_fW(\mbf{k},\mbf{q})}{\sqrt{q^2+(M^{(0)}(\mbf{q}))^{2}}}
\left(M^{(0)}(\mbf{q})k-M^{(0)}(\mbf{k})q
\frac{(k^2+q^2)x-kq(1+x^2)}{(\mbf{k}-\mbf{q})^2}\right)
{\rm e}^{-4q^2/\Lambda^2}\,\right]
\,,\label{eq:3.44}\end{eqnarray}
where $x=\mbf{\hat{k}}\cdot\mbf{\hat{q}}$, the potential functions
$V_{L+C}$ and $W$ are defined in Eq. (\ref{eq:2.24}), and 
the running mass $m(\Lambda)$ is given by Eq. (\ref{eq:2.35}).

\begin{figure}[!htb]
\begin{center}
\input{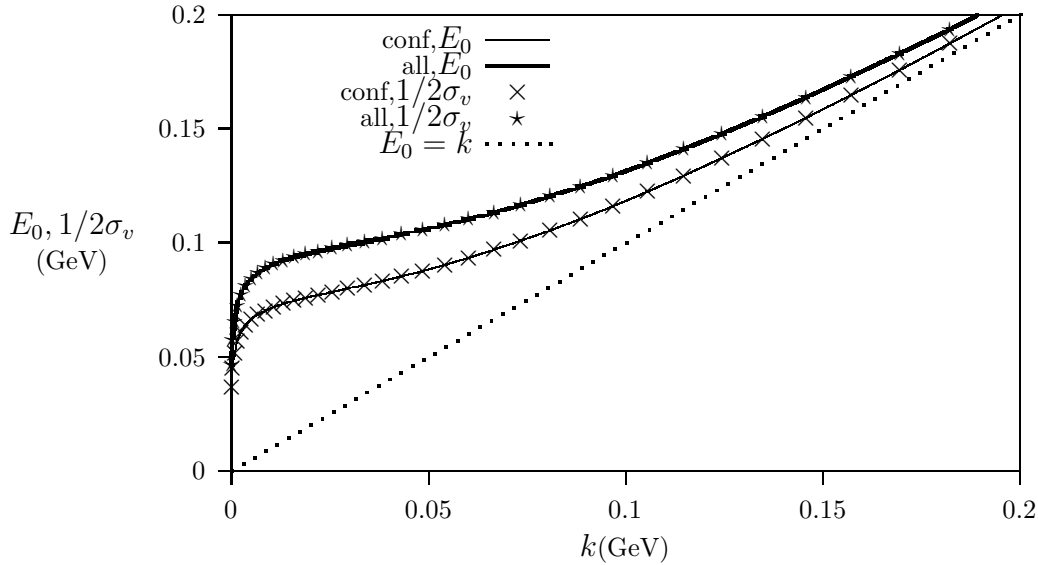}
\end{center}
\caption{One particle dispersion relation, $E_0(k)=\sqrt{k^2+M_0(k)^2}$,
free dispersion, $E_0(k)=k$, 
and the vector part of the propagator, $1/2\sigma_v$, in the chiral limit
with confinement and with confinement plus perturbative potentials (i.e.
when Coulomb and generated potentials are added).
The parameters are same as in Fig. \ref{fig.2}.}
\label{fig.3}
\end{figure}

\begin{figure}[!htb]
\begin{center}
\input{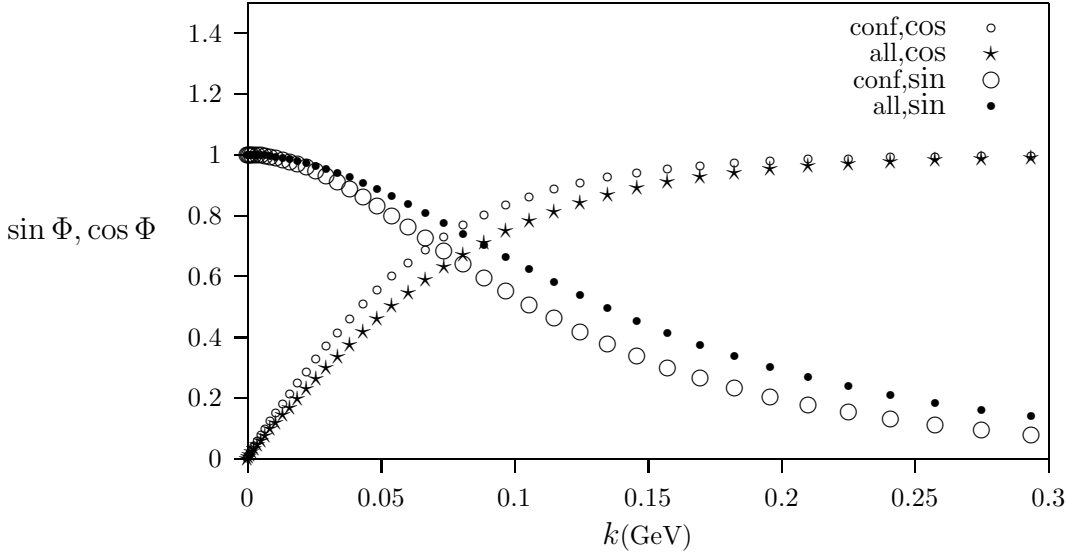}
\end{center}
\caption{Sine and cosine of the Bogoliubov-Valatin angle
in the chiral limit with confinement and 
with confinement and with confinement plus perturbative potentials (i.e.
when Coulomb and generated potentials are added). 
The parameters are same as in Fig. \ref{fig.2}.}
\label{fig.4}
\end{figure}

\begin{figure}[!htb]
\begin{center}
\input{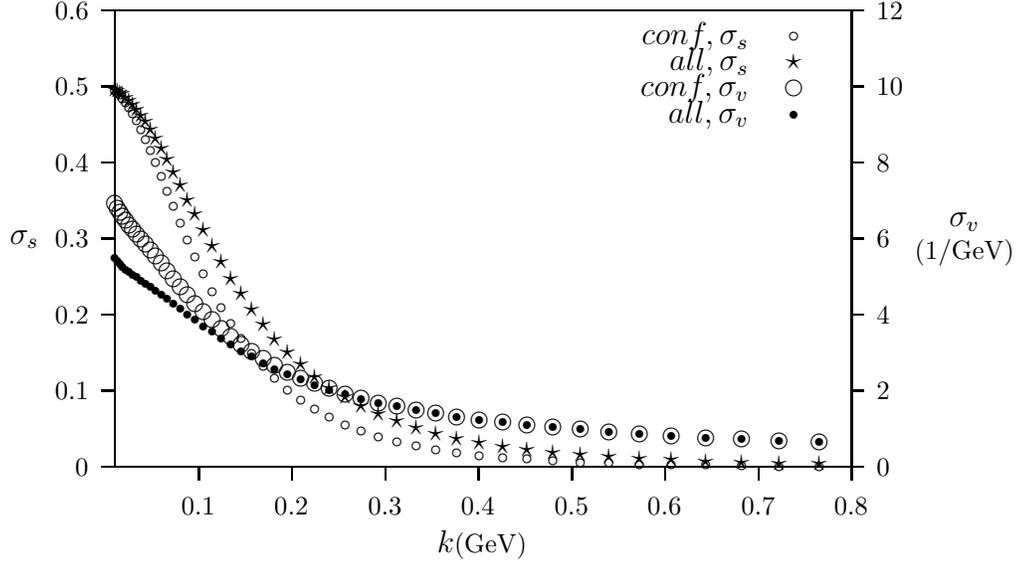}
\end{center}
\caption{Scalar, $\sigma_s$, and vector, $\sigma_v$, 
parts of the propagator, 
$S^{(3)}$, in the chiral limit with confinement and
with confinement and with confinement plus perturbative potentials (i.e. 
when Coulomb and generated potentials are added).
The parameters are same as in Fig. \ref{fig.2}.}
\label{fig.5}
\end{figure}

As discussed before the obtained gap equation is free 
from IR and UV divergences, 
and the mass gap, $M(\mbf{k})$, exists for a confining and Coulomb
potentials provided both instantaneous and generated by flow equations terms
are taken into account.

In numerical calculations of Eq. (\ref{eq:3.44}) we have used 
the file \texttt{dgeco.f} followed by \texttt{dgesl.f} 
from the SLATEC linear algebra archive,
which solves $A\times X=B$.

For a fixed cut-off $\Lambda=1\,GeV$, 
the mass gap, $M_0(\mbf{k})$, and the product of momentum and 
scalar part of propagator, $k\sigma_s^{0}(\mbf{k})$, 
in the chiral limit are displayed in Fig.\ref{fig.1} (confining potential) 
and in Fig.\ref{fig.2} (Coulomb and generated terms are added).
Hereafter, $M_0$ denotes the solution of the gap equation in the chiral
limit, not be confused with the first iteration in our calculations, $M^{(0)}$. 
Since $mA\neq 0 $, the nonzero mass gap $M_0=mA/(1+B)$ is generated,
and the chiral symmetry is broken dynamically. We denote
the maximum value of the gap as $M_0(0)$ and call it constituent quark mass.
In the chiral limit constituent quark mass is generated as  
$M_0(0)\approx 74\,MeV$ with confining potential, and slightly bigger
$M_0(0)\approx 95\,MeV$ when perturbative terms are included. 
At high momenta, from Eq. (\ref{eq:3.40}), 
\mbox{$2k\sigma_s^{0}(\mbf{k})\rightarrow M_0(\mbf{k})$},
and since BCS pairing is a low-momentum effect, 
the mass gap function $M_0(\mbf{k})$ vanishes rapidly as 
\mbox{$M_0(\mbf{k}\rightarrow \infty)\rightarrow 1/k^2 \rightarrow 0$}.
The behavior of the $A(\mbf{k})$ and $B(\mbf{k})$ amplitudes 
in the perturbative or UV asymptotic
region is well known from the QCD renormalization group and 
operator product expansion and QCD sum rules \cite{GasserLeutwyler}. 
The behavior has been summarized in the course of 
explicit numerical solutions and model building \cite{Roberts}.
For $k^2\gg\Lambda_{QCD}^2$ the leading-log result for the chiral mass gap,
$M_0(\mbf{k})=mA(\mbf{k})/(1+B(\mbf{k}))$, is
\begin{eqnarray}
M_0(\mbf{k}\rightarrow \infty) = 
\frac{\kappa}{k^2[\ln(k^2/\Lambda_{QCD}^2)]^{1-d}}
\,,\label{eq:3.45}\end{eqnarray}
where $d=12/(33-2N_f)$ is the anomalous dimension of the mass, 
$N_f$ is the number of quark flavours, $\Lambda_{QCD}\approx 0.20\,GeV$ 
is the scale parameter of QCD, and $\kappa$ is a constant given by
\begin{eqnarray}
\kappa\simeq -\frac{4\pi^2d}{3}\frac{\langle\bar{\psi}\psi\rangle_0}
{[\ln(\mu^2/\Lambda_{QCD}^2)]^d}
\,,\label{eq:3.46}\end{eqnarray}
with the scale $\mu^2=1\,GeV^2$. We fitted our numerically obtained solution,
$M_0(\mbf{k})$, with the function Eq. (\ref{eq:3.45}) at high momenta.
For $N_f=6$, $1-d=0.43$, we obtain $\kappa=2.4\,MeV$ and 
$\kappa=6.0\,MeV$, which corresponds to the chiral condensates
\mbox{$\langle\bar{\psi}\psi\rangle_0\approx -(85.3\,MeV)^3$} 
and \mbox{$\langle\bar{\psi}\psi\rangle_0\approx -(115.9\,MeV)^3$}
for confining and confining$+$perturbative potentials, respectively.
These estimates for condensates are lower than calculated directly from
Eq. (\ref{eq:3.40}) (see below). 
Though the condensate is a measure of chiral symmetry breaking, 
it is sensitive to the UV region, and therefore
it is enhanced by generated terms from the flow equation.
Note that we reproduce a correct high momentum behavior predicted from
the perturbative renormalization group analyses. 
It happens only if the terms from the flow equation 
together with the Coulomb interaction are added 
(Coulomb interaction alone does not work).

Similar behavior of the mass gap $M(\mbf{k})$
and scalar part of propagator is seen 
for the light $u,d$ quarks, with the current mass $m=8\,MeV$.
The predicted leading-log behavior for high momenta 
\begin{eqnarray}
M(\mbf{k}\rightarrow\infty)=\frac{\hat{m}}{[1/2\ln(k^2/\Lambda_{QCD}^2)]^d}
\,,\label{eq:3.47}\end{eqnarray}
provides a fitting function, where $\hat{m}$ is the renormalization point
independent current quark mass.
For $N_f=6$, $d=0.57$, we obtain \mbox{$\hat{m}\approx 20\,MeV$}
and \mbox{$\hat{m}\approx 32\,MeV$} for confining and confining$+$perturbative
potentials, respectively. 
Since \mbox{$\hat{m}\neq 0$}, there is an explicit chiral symmetry breaking.
The fitting functions, Eqs. (\ref{eq:3.45}) and (\ref{eq:3.47}),
were obtained in Ref. \cite{Roberts} using covariant calculations.
In our calculations, we obtain the same fitting functions.  

\begin{figure}[!htb]
\begin{center}
\input{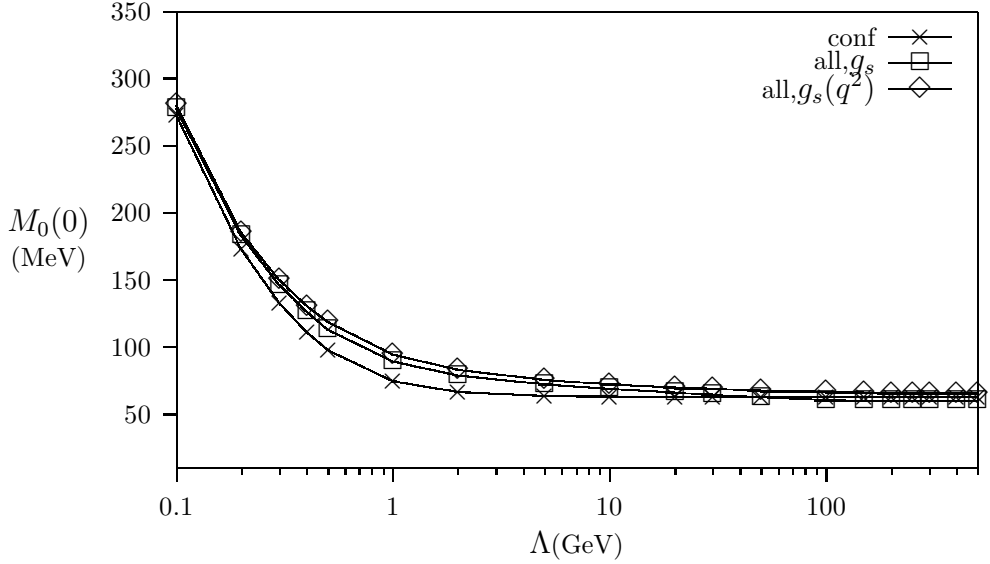}
\end{center}
\caption{Cut-off dependence of the constituent quark mass
in the chiral limit (same parameters as in Fig.1).
Crosses represent solution with confinement.
Boxes [diamonds] represent solution when Coulomb and generated
potentials are added with the constant value of coupling $g_s$
[with the running coupling $g_s(q^2)$].}
\label{fig.6}
\end{figure}

\begin{figure}[!htb]
\begin{center}
\input{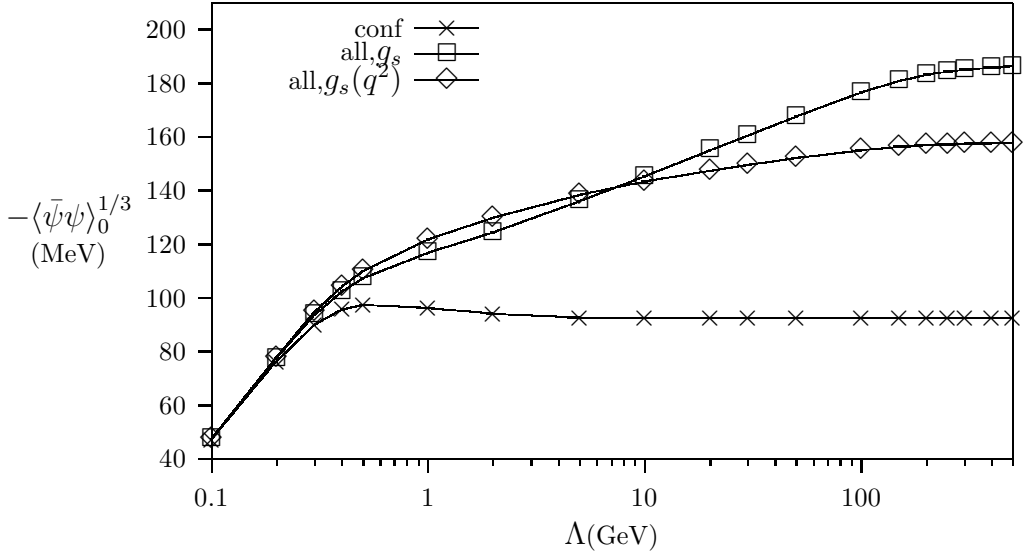}
\end{center}
\caption{Quark condensate cut-off dependence in the chiral limit 
(same parameters as in Fig.1).
Crosses represent solution with confinement.
Boxes [diamonds] represent solution when Coulomb and generated
potentials are added with the constant value of coupling $g_s$
[with the running coupling $g_s(q^2)$].}
\label{fig.7}
\end{figure}

\begin{figure}[!htb]
\begin{center}
\input{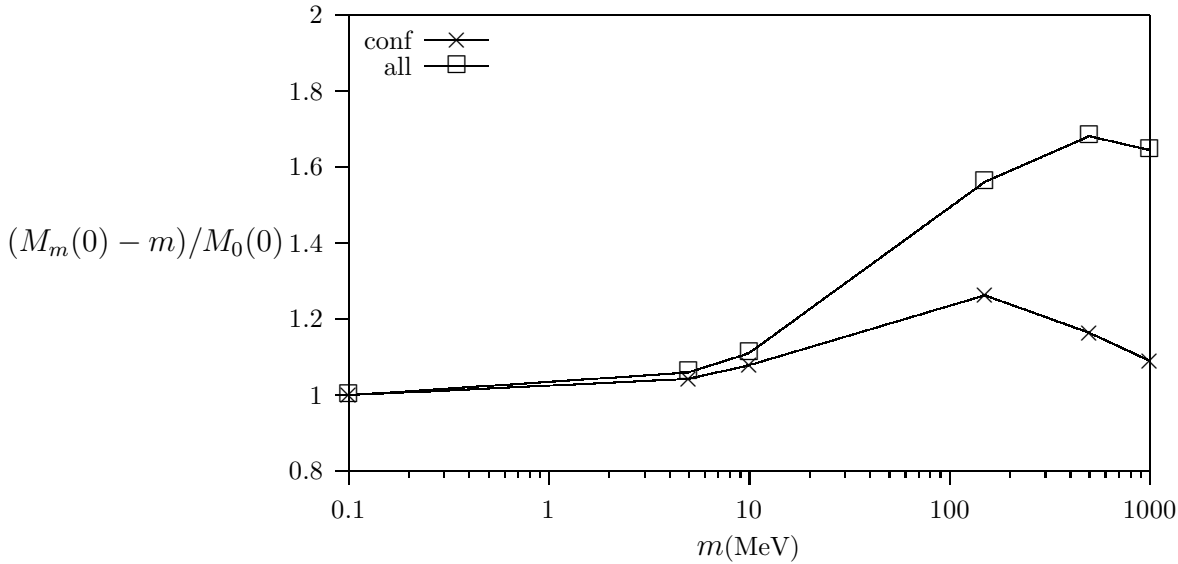}
\end{center}
\caption{Dependence of the constituent quark mass
on the current quark mass, \mbox{$(M_m(0)-m)/M_0(0)$},
where $M_m(0)$ and $M_0(0)$ are the constituent quark masses
for the current quark mass $m$ and in the chiral limit $m\rightarrow 0$,
respectively (same parameters as in Fig.1).
Crosses represent solution with confinement.
Boxes represent solution when Coulomb and generated
potentials are added.}
\label{fig.8}
\end{figure}

\begin{figure}[!htb]
\begin{center}
\input{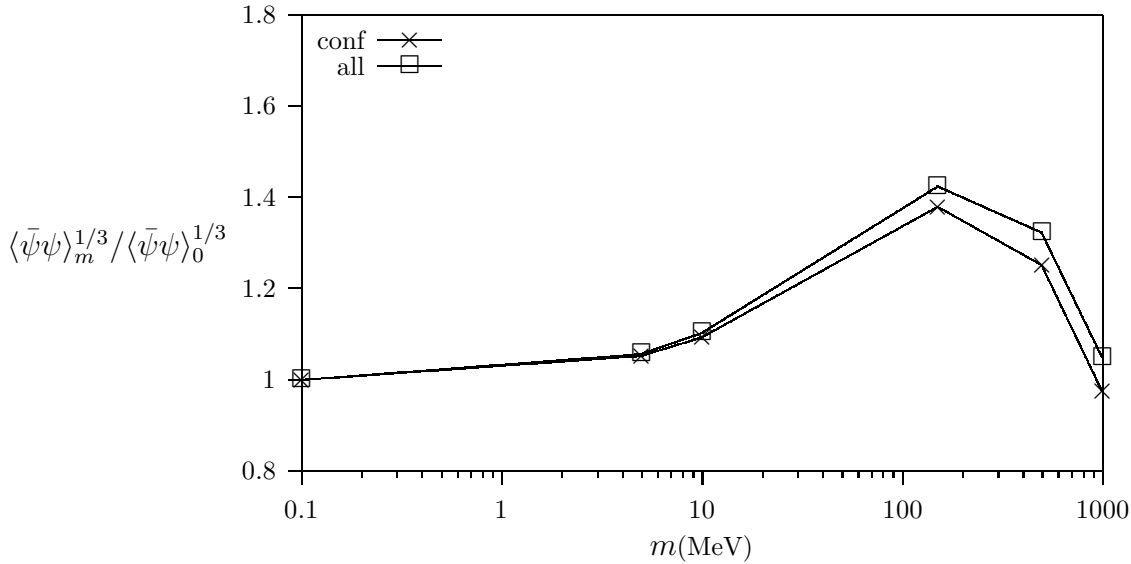}
\end{center}
\caption{Dependence of the quark condensate on the current quark mass, 
\mbox{$\langle\bar{\psi}\psi\rangle_m^{1/3}/
\langle\bar{\psi}\psi\rangle_0^{1/3}$},
where $\langle\bar{\psi}\psi\rangle_m$ and 
$\langle\bar{\psi}\psi\rangle_0$ are the quark condensates
for the current quark mass $m$ and in the chiral limit $m\rightarrow 0$,
respectively (same parameters as in Fig.1).
Crosses represent solution with confinement.
Boxes represent solution when Coulomb and generated
potentials are added.}
\label{fig.9}
\end{figure}

The quark energy dispersion, \mbox{$E_0(\mbf{k})=\sqrt{k^2+M_0^2(\mbf{k})}$},
where $M_0$ is the numerical solution of the gap equation, and the inverse
of vector part of propagator, 
\mbox{$1/2\sigma_v^{0}(\mbf{k})=\Omega(\mbf{k})/(1+B(\mbf{k}))$},
in the chiral limit at $\Lambda=1\,GeV$ are depicted in Fig.\ref{fig.3}. 
The free behavior, \mbox{$E_0(\mbf{k})=k$}, is recovered at high energies while for 
low energies constituent quark masses at $k=0$, roughly 
$36.8\,MeV$ and $46.7\,MeV$ are obtained for confining and 
confining$+$perturbative potentials,
respectively. The inverse of vector part
of propagator exactly reproduces the energy dispersion as expected
from Eq. (\ref{eq:3.40}), \mbox{$1/2\sigma_v^{0}(\mbf{k})=E_0(\mbf{k})$}.

The sine and cosine of the Bogoliubov-Valatin angle in the chiral limit
are depicted in Fig.\ref{fig.4}. At low momenta the cosine behaves linearly,
with the slope $1/M(0)$, 
\begin{eqnarray}
c(\mbf{k}\rightarrow 0)\rightarrow \frac{1}{M(0)}k
\,.\label{eq:3.48a}\end{eqnarray}
The obtained values of $M_0(0)$ in this way are the same 
as extracted from the dispersion relation. 
From Eq.(\ref{eq:3.40}), the relations 
\mbox{$s(\mbf{k})=2\sigma_s$} and \mbox{$c(\mbf{k})=2k\sigma_v$}
give additional insight into the behavior of scalar and vector parts of propagator. 

The scalar, $\sigma_s^{0}$, and vector, $\sigma_v^{0}$, parts of the propagator
in the chiral limit are presented in Fig.\ref{fig.5}. From Eq.(\ref{eq:3.40}),
at high momenta both amplitudes vanish rapidly,
\begin{eqnarray}
&& \sigma_s(\mbf{k}\rightarrow\infty)\rightarrow\frac{1}{2}\frac{M(\mbf{k})}{k}
\rightarrow\frac{1}{2}\frac{1}{k^3} \nonumber\\
&& \sigma_v(\mbf{k}\rightarrow\infty)\rightarrow\frac{1}{2}\frac{1}{k}
\,,\label{eq:3.48}\end{eqnarray}
while for low momenta both tend to be constants,
\begin{eqnarray}
&& \sigma_s(\mbf{k}\rightarrow 0)\rightarrow\frac{1}{2} \nonumber\\
&& \sigma_v(\mbf{k}\rightarrow 0)\rightarrow\frac{1}{2}\frac{1}{M(0)}
\,.\label{eq:3.49}\end{eqnarray}
As expected, adding Coulomb and flow equation terms increases the scalar
amplitude and decreases vector amplitude at low momenta, that amplifies
dynamical chiral symmetry breaking with larger $M_0(0)$ and 
\mbox{$\langle\bar{\psi}\psi\rangle_0$}.

The sensitivity of the constituent quark mass (maximum mass gap), $M_0(0)$,
and the quark condensate in the chiral limit to the cut-off $\Lambda$ 
is displayed in Fig.\ref{fig.6} and \ref{fig.7}, respectively. 
Remarkably, after roughly $\Lambda=4\,GeV$
the mass gap saturates to a constant value even 
when Coulomb and generated
potentials are added 
(adding only Coulomb causes slow logarithmic dependence).  
This proves that the obtained gap equation is
renormalized completely  and does not need counterterms in the chiral limit.
The constituent masses tend to values $63\,MeV$ for confining and 
\begin{eqnarray}
M_0(0)\approx 70\,MeV
\,,\label{eq:3.50a}\end{eqnarray}
for confining$+$perturbative potentials.
Taking a running coupling does not change the result for $M(0)$.
The chiral condensate, calculated using cut-off independent mass gap 
$M_0(\mbf{k})$, Eq.(\ref{eq:3.40}), rises logarithmically with cut-off 
$\Lambda$ when perturbative potentials are added. By including the leading-log
nonperturbative running coupling constant \cite{Brodsky}  
in the gap equation (in the perturbative kernel)
\begin{eqnarray}
\alpha_s(\vec{q})=\frac{d\pi}{\ln((q^2+\mu^2)/\kappa_{\Lambda}^2)}
\,,\label{eq:3.50}\end{eqnarray}
with scales $\mu\approx 0.87\,GeV$ (accounts for freezing of $\alpha_s$ since
confinement is present) and $\kappa_{\Lambda}\approx 0.16\,GeV$, 
found in \cite{Brodsky} 
by a fit to the non-relativistic heavy-quark lattice data, and
$d=12\pi/(33-2N_f)$, we damp the growth of the chiral condensate.
Since combination \mbox{$m(\Lambda)\langle\bar{\psi}\psi\rangle(\Lambda)$} 
is renormalization group invariant (it appears in Hamiltonian), 
the leading-log bahavior is given from Eq.(\ref{eq:2.35})       
\begin{eqnarray}
&& m(\Lambda)=m(\Lambda_0)
\left(1-\frac{C_fg^2}{(4\pi)^2}6\ln\Lambda/\Lambda_0\right) \nonumber\\
&& \langle\bar{\psi}\psi\rangle(\Lambda)=
\langle\bar{\psi}\psi\rangle(\Lambda_0)
\left(1-\frac{C_fg^2}{(4\pi)^2}6\ln\Lambda/\Lambda_0\right)
\,,\label{eq:3.51}\end{eqnarray}
where $\Lambda_0$ is the renormalization point. Hence the cut-off dependent
correction to condensate $\sim g^2\ln\Lambda$ can be absorbed 
by introducing running coupling $g^2\rightarrow g^2(\Lambda)\sim 1/\ln\Lambda$.
Relations between nonperturbative scales $\Lambda_0$, 
$\mu$ and $\kappa_{\Lambda}$ can be found. 
This procedure renormalizes quark condensate to leading-log order
and freezes its value at 
\begin{eqnarray}
\langle\bar{\psi}\psi\rangle_0\approx -(155\,MeV)^3 
\,,\label{eq:3.51a}\end{eqnarray}
while only with confinement condensate the value of $-(92\,MeV)^3$ 
(roughly $-(100\,MeV)^3$ reported in \cite{LeYaouanc}, \cite{AdlerDavis})
is obtained. Flow equations improve the chiral condensate by $\sim 68\%$, 
although the obtained value is still low (which is a common feature 
for most Hamiltonian methods).

Constituent masses and quark condensates for different current masses are shown
in Fig.\ref{fig.8} and \ref{fig.9}, respectively. 
In light quark sector ($u,d$ quarks)
there is practically no deviation from the chiral result. However,
constituent masses and condensates increase slowly when approaching 
the strange quark mass, and subsequently decrease. 
Generally, this behavior is very slow,
and depends on particular model used for calculations. 
In the instanton liquid model, it has been reported \cite{Musakhanov}
that the constitituent mass and the condensate decrease as
the current quark mass increases.
We recover the relation between quark and gluon condensates 
\cite{GorbarNatale}
\begin{eqnarray} 
m\langle-\bar{\psi}\psi\rangle = 
\frac{1}{\widetilde{\kappa}}\langle\frac{\alpha_s}{\pi}F_{\mu\nu}F^{\mu\nu}\rangle
\,,\label{eq:3.52}\end{eqnarray}
in heavy quark sector. Here the expectation value with respect to the quark
vacuum state is implied in 
$\hspace{-0.2cm}\phantom{1}_{f}\langle\Omega|\bar{\psi}\psi|\Omega\rangle_{f}$, and
the average over the gluon vacuum configurations stands in
$\hspace{-0.2cm}\phantom{1}_{g}\langle\Omega|F_{\mu\nu}F^{\mu\nu}|\Omega\rangle_{g}$,
assuming that the quark and gluon vacuum states factorize 
as the direct product
$|\Omega\rangle=|\Omega\rangle_{f}\otimes|\Omega\rangle_{g}$.   
We obtain \mbox{$\langle\frac{\alpha_s}{\pi}F_{\mu\nu}F^{\mu\nu}\rangle
\approx 0.010\,GeV^4$} for strange quark, with \mbox{$\widetilde{\kappa}=13.2$} 
\cite{GorbarNatale}, in agreement with the QCD sum rules $0.012\,GeV^4$
\cite{ShifmanVainshteinZakharov}. 
In the light quark sector we obtain too low value for the gluon condensate, 
suggesting that this relation holds only for heavy quarks.

\subsection{Pion and $\rho$-meson bound states}
\label{subsec:4.2}

Numerical solutions of Eq. (\ref{eq:4.27}) in RPA and TDA 
for pion (Eq. (\ref{eq:4.18})) and $\rho$ meson (Eq. (\ref{eq:4.19})) 
interaction kernels are obtained variationally 
with a set of Gaussian test functions,
\begin{eqnarray} 
\psi_i &=& \sum_{n=1,N_h}\langle n|i\rangle \phi_n
\nonumber\\
\langle n|i\rangle &=& {\rm e}^{-k^2n/2\beta^2}
\,,\label{eq:4.34}\end{eqnarray} 
where $\psi_i$ is an eigenfunction in the momentum space $k$, discretized
by $i=1,...,N$, and $\beta$ is a variational parameter, chosen to give the minimum
energy expectation value. In this way we avoid the direct diagonalization 
of Hamiltonian in the momentum space, $|i\rangle$, 
where the interaction kernel is infrared divergent,
such as Coulomb $\sim 1/k^2$ and confining $\sim 1/k^4$ potentials 
at $k\sim 0$. Instead we calculate Hamiltonian matrix elements between
Gaussian functions, and obtain a regular Hamiltonian matrix in 
the $|n\rangle$ space, which is solved as an eigenproblem.  
Typically one choses $N_h\ll N$ to achieve convergence and stability 
of a result (large $N$ for numerical integration of interaction kernels 
Eq. (\ref{eq:4.18}) and Eq. (\ref{eq:4.19}), 
and small $N_h$ to give several lowest eigenvalues from discrete spectrum). 
Such procedure is important when calculating pion mass 
spectrum in the chiral limit, with the current quark mass $m=0$ 
(since there is no regulator in pion denominator as $k\rightarrow 0$).

\begin{table}[ht]
$$
\begin{tabular}{|c|ccc|ccc|} \hline 
   & &TDA, (MeV)& &&RPA, (MeV)&   \\ \hline
 conf. & $504$&$1364$&$2115$ & $222$&$1416$&$2298$ \\ \hline
 conf.$+$Coul. & $608$&$1514$&$2249$ & $427$&$1521$&$2309$ \\ \hline
 conf.$+$Coul.$+$gen.& $513$&$1411$&$2161$ & $180$&$1413$&$2218$   \\ \hline 
\end{tabular}
$$
\caption{Pion spectrum for the ground, first and second exited states
in the TDA and RPA approaches with confining, confining$+$Coulomb and 
confining$+$Coulomb$+$generated potentials taken. Chiral limit
$m=0$ ($\alpha_s=0.4, \sigma=0.18 GeV^2, \Lambda=10 GeV$).}
\label{tab.1}
\end{table}

\begin{table}[ht]
$$
\begin{tabular}{|c|ccc|ccc|} \hline 
 m, (MeV)  && TDA, (MeV)  & && RPA, (MeV)& \\ \hline
 $150$ & $1038$&$1926$&$2936$ & $1037$&$1986$&$3077$ \\ \hline
 $100$ & $885$&$1762$&$2697$ & $868$&$1811$&$2826$   \\ \hline 
  $50$ & $716$&$1590$&$2431$ & $660$&$1626$&$2537$   \\ \hline 
  $10$ & $553$&$1446$&$2212$ & $366$&$1460$&$2283$   \\ \hline 
   $5$ & $532$&$1428$&$2186$ & $293$&$1437$&$2250$   \\ \hline 
   $0$ & $513$&$1411$&$2161$ & $180$&$1413$&$2218$   \\ \hline 
\end{tabular}
$$
\caption{Pion spectrum for the ground, first and second exited states
in the TDA and RPA approaches for different current masses of constituents.
Confining$+$Coulomb$+$generated potentials are taken (the same parameters
as in the table $1$).}
\label{tab.2}
\end{table}

\begin{table}[ht]
$$
\begin{tabular}{|c|ccc|ccc|} \hline 
   & &TDA, (MeV)& &&RPA, (MeV)&   \\ \hline
 conf. & $659$&$1484$&$2258$ & $642$&$1482$&$2256$ \\ \hline
 conf.$+$Coul. & $750$&$1678$&$2515$ & $732$&$1676$&$2514$ \\ \hline
 conf.$+$Coul.$+$gen.& $718$&$1592$&$2377$ & $700$&$1590$&$2376$ \\ \hline 
\end{tabular}
$$
\caption{Spectrum of the $\rho$ meson for the ground, first and second exited 
states in the TDA and RPA approaches with confining, confining$+$Coulomb and 
confining$+$Coulomb$+$generated potentials taken. Chiral limit
$m=0$, the same parameters as in the table $1$.} 
\label{tab.3}
\end{table}

\begin{table}[ht]
$$
\begin{tabular}{|c|ccc|ccc|} \hline 
 m, (MeV)  && TDA, (MeV)  & && RPA, (MeV)& \\ \hline
 $150$ & $1130$&$2086$&$3247$ & $1128$&$2086$&$3247$ \\ \hline
 $100$ & $986$&$1916$&$2990$ & $983$&$1915$&$2990$   \\ \hline 
  $50$ & $839$&$1744$&$2692$ & $833$&$1744$&$2692$   \\ \hline 
  $10$ & $727$&$1616$&$2436$ & $714$&$1615$&$2435$   \\ \hline 
   $5$ & $719$&$1603$&$2406$ & $704$&$1601$&$2405$   \\ \hline 
   $0$ & $718$&$1592$&$2377$ & $700$&$1590$&$2376$   \\ \hline 
\end{tabular}
$$
\caption{Spectrum of the $\rho$ meson for the ground, first and second exited 
states in the TDA and RPA approaches for different current masses of constituents.
Confining$+$Coulomb$+$generated potentials are taken (the same parameters
as in the \mbox{table $1$}).}
\label{tab.4}
\end{table}

RPA/TDA Eq. (\ref{eq:4.27}) has the form $M_{mn}X_n=\lambda_m N_{mn}X_n$ 
in the $|n\rangle$ space, where $N_{mn}=\langle m|n\rangle$,
$X_n$ and $\lambda_n$ are eigenfunctions and eigenvalues. Note 
that the RPA matrix $M_{mn}$ is not hermitian, $A\times B\neq B\times A$
(i.e. it is not a Hamiltonian) and one cannot use algorithms for 
diagonalization of symmetric matrices. In numerical calculations 
of Eq. (\ref{eq:4.27}) we first used the files \texttt{dgeco.f} and 
\texttt{dgesl.f} to find the product $N^{-1}\times M$, 
that reduces eigenvalue problem to $N^{-1}MX=\lambda X$, and then used 
the file \texttt{rg.f} from the SLATEC archive to solve the general matrix 
for eigenstates. Results for the pion and $\rho$-meson masses in RPA
and TDA approaches are presented in Tables \ref{tab.1}-\ref{tab.4}. 
In the chiral limit RPA gives ground state pion masses 
which are significantly lower than those obtained using TDA. 
Including the generated interaction terms
reduces ground state pion mass even more, increasing the mass splitting
between the pion and $\rho$ meson (Tables \ref{tab.1}, \ref{tab.3}).
In the chiral limit we get
\begin{eqnarray}
&& M_{\pi}=180\,MeV\,,\, M_{\rho}=700\,MeV 
\nonumber\\
&& M_{\rho}-M_{\pi}=520\,MeV
\,.\label{eq:4.35}\end{eqnarray} 
Although Coulomb and generated interactions
are both perturbative, they act in different directions: Coulomb [generated]
term increases [decreases] a meson mass. Indeed, this is in an accordance with 
the standard perturbation theory where the leading order perturbative
corrections always shift the energy of a ground state down. 
Effects of using the RPA instead of the TDA including the generated terms 
are not so pronounced for excited states.

The $\pi-\rho$ mass splitting of $520\,MeV$ in the chiral limit
is close enough to the lattice data splitting of $600\,MeV$ \cite{UKQCD}.
However, we are unable to get zero mass pion either in the BCS or
adding the corrections from the leading order flow equations.
Various reasons why the zero mass pion solution cannot appear within 
the BCS approach are summarized in Appendix \ref{app:D}.
The underlying reason of failing to produce the Goldstone boson
seems to be associated with a breakdown of covariance in the BCS model. 
Including higher orders of calculations is necessary in order to approach 
to a covariant result. 

In the chiral limit the $\pi-\rho$ mass splitting, 
$\delta E_{\pi\rho}=M_{\rho}-M_{\pi}$, with all terms contributed
is $\delta E_{\pi\rho}=205\,MeV$ in the TDA and $\delta E_{\pi\rho}=520\,MeV$ 
in the RPA. These values should be compared
to $\delta E_{\pi\rho}=155\,MeV$ in the TDA and $\delta E_{\pi\rho}=420\,MeV$ 
in the RPA, calculated with static confining potential alone. Flow equations 
improve the $\pi-\rho$ mass splitting by $32\%$ in the TDA and by $24\%$ 
in the RPA.

\begin{figure}[!htb]
\begin{center}
\input{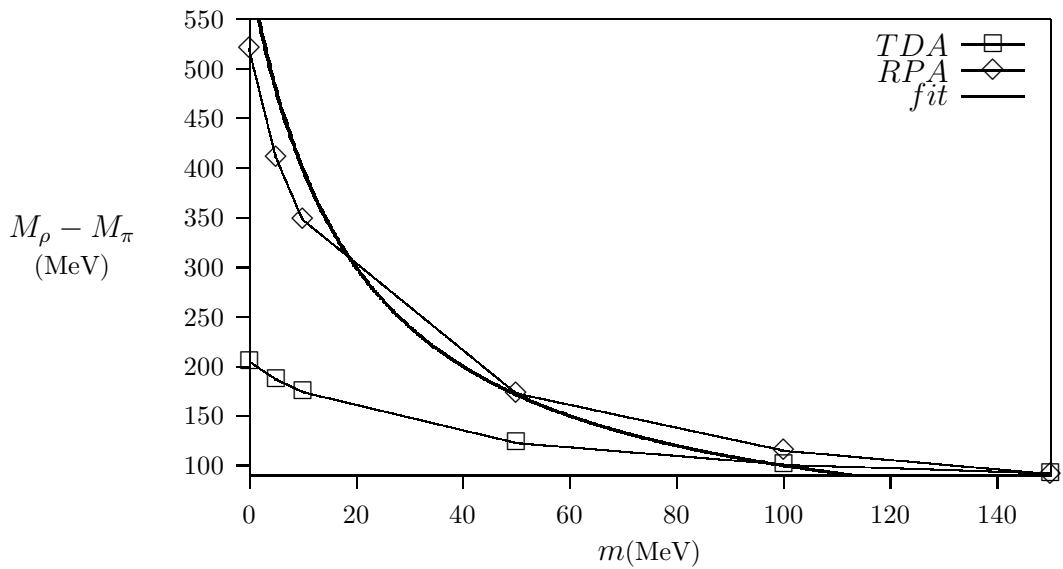}
\end{center}
\caption{$\pi-\rho$ mass splitting for the ground state 
in TDA and RPA approaches.
Chiral limit, $m=0$, and confining$+$Coulomb$+$generated interactions
are taken (same parameters as in table\ref{tab.1}).
Fit function is 12000/(m+20).}
\label{Fig.1}
\end{figure}

\begin{figure}[!htb]
\begin{center}
\input{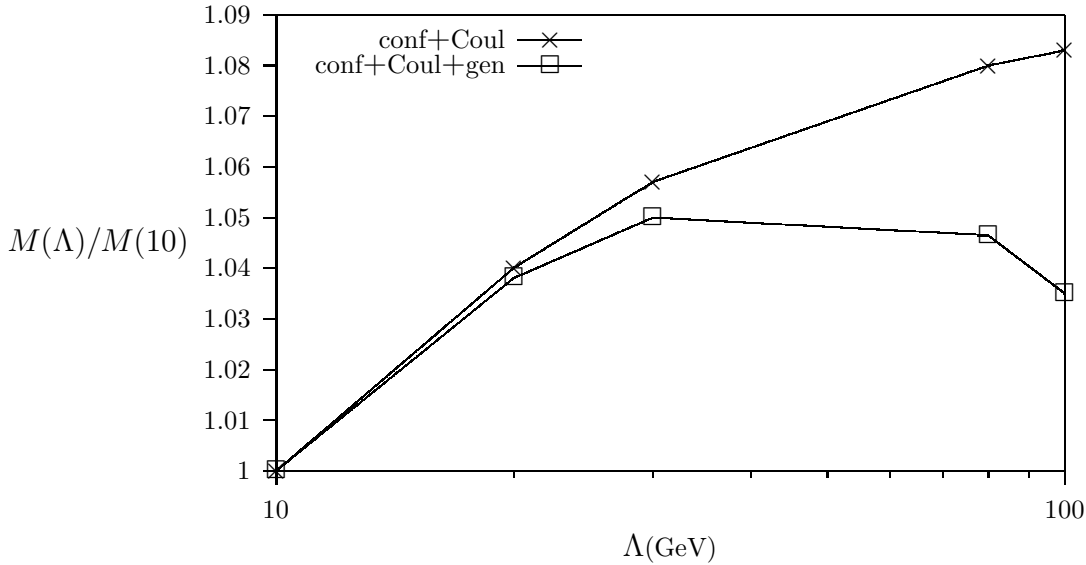}
\end{center}
\caption{Cut-off dependence of the pion mass
in the chiral limit $M(\Lambda)$, normalized to the pion mass
at $\Lambda=10GeV$, $M(10)$, (same parameters as in table\ref{tab.1}).
Line with crosses represents RPA solution with confinement$+$Coulomb,
the line with boxes stays for RPA solution when generated
potentials are added.}
\label{Fig.2}
\end{figure}

For nonzero current quark masses,
the main contribution to meson masses is coming from the solutions of 
the gap equation, while the contributions from bare mass kinetic terms 
are negligible (Tables \ref{tab.2}, \ref{tab.4}).
This supports an idea of the constituent quark model, where interactions
give rise to massive constituents which build a meson mass 
in a valence approximation. Since a numerical solution of the gap equation
carries its mass dependence, it is difficult to analyze analytically
the dependence of the meson spectrum, in particular of the 
$\pi-\rho$ mass splitting, on the current mass.  

Numerically obtained dependence of the $\pi-\rho$ mass splitting,
$M_{\rho}-M_{\pi}$, as a function of the bare mass of one of the quarks 
is shown in Figure \ref{Fig.1}. We find $1/m_{const}$ behavior, where
$m_{const}$ is a constituent quark mass, which is valid for heavy quarks
and continues to be valid for lighter constituent quarks.
This fall-off is more rapid in the RPA than in the TDA. 
From the RPA fit function, a constituent quark mass can be approximated,
uniformly for heavy and light quarks, as $m_{const}=m+20\,(MeV)$,
with the bare quark mass $m$. The $1/m_{const}$ behavior
is characteristic for the hyperfine interaction. However, we reproduce this
behavior taking into account both spin dependent interactions and 
vacuum effects of the dynamical chiral symmetry breaking. 
We do not separate the effects of the hyperfine interactions and the chiral 
symmetry breaking, since after Bogoluibov-Valatin transformation 
all quark interactions are calculated in a chiral noninvariant framework.

We reproduce $1/m_{const}$ behavior throughout the whole range of current 
quark masses. One may conjecture that the hyperfine interaction dominates 
for heavy quarks, and 
the confining potential and the chiral symmetry breaking  
are dominant for light quarks. However, we do not see a change in regime. 
We thus conclude that $1/m_{const}$ behavior of the $\pi-\rho$ mass splitting, 
attributed to the hyperfine interaction, is actually due to both hyperfine and 
the chiral symmetry breaking effects, which is consistent with 
the lattice calculations \cite{UKQCD} and experiment \cite{Groom}.
 
Formally, the static interactions of the form
$\psi^{\dagger}T\psi V_{L+C}\psi^{\dagger}T\psi$, which include 
the sum of confining and Coulomb potentials and 
initiate the breaking of the chiral symmetry, is referred as the 
chiral symmetry breaking interactions, and the generated terms, 
having the structure 
$\psi^{\dagger}T\mbf{\alpha}\psi\psi^{\dagger}T\mbf{\alpha}\psi$, 
are referred as the hyperfine interactions.
 
We find in the chiral limit that in the TDA roughly $30\%$ of the $\pi-\rho$ 
mass splitting is due to the presence of the hyperfine interaction and 
the rest $70\%$ is due to the chiral symmetry breaking. 
In the RPA this ratio is $40\%$ for 
the hyperfine and $60\%$ for the chiral symmetry breaking. However,
the numerical value of this ratio depends on the details of the confining
interaction. In our calculations we used the linear rising potential
generally accepted in the quark model phenomenology with the string tension 
$\sigma=0.18 GeV^2$ predicted by the lattice studies.
It is important to have both terms to reproduce a correct mass splitting.

\begin{figure}[!htb]
\begin{center}
\input{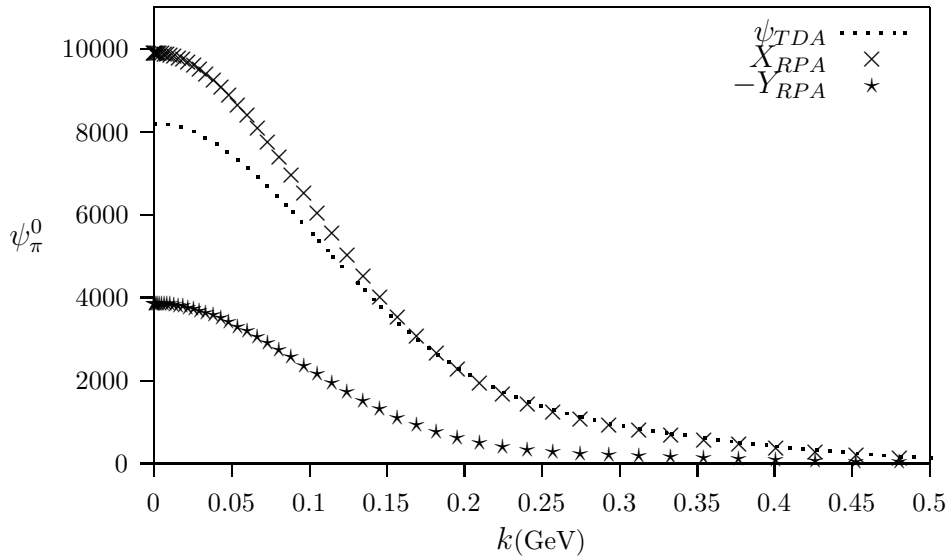}
\end{center}
\caption{Ground state pion wave function in TDA and RPA approaches.
Chiral limit, $m=0$, and confining$+$Coulomb$+$generated interactions
are taken  
($\alpha_s=0.4, \sigma=0.18 GeV^2, \Lambda=10 GeV$).}
\label{Fig.3}
\end{figure}

\begin{figure}[!htb]
\begin{center}
\input{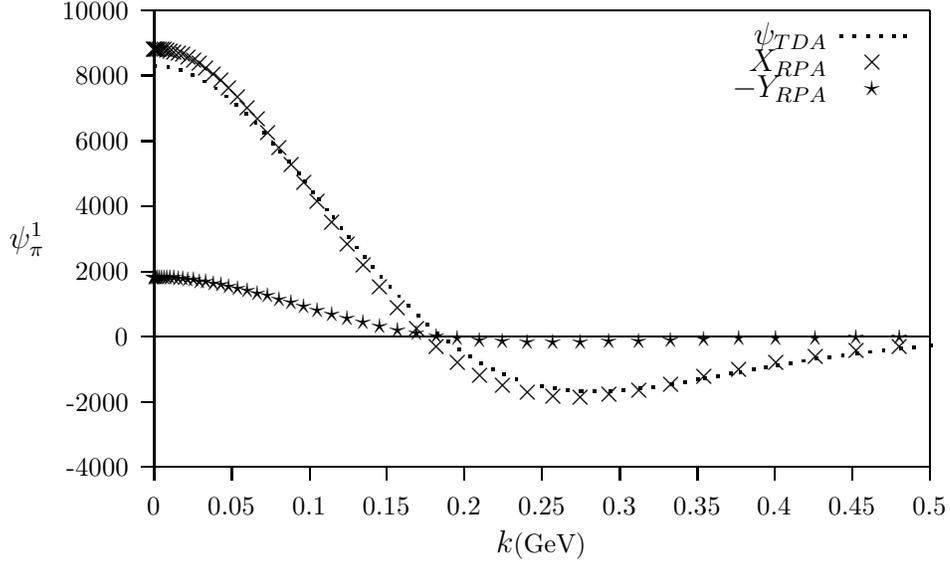}
\end{center}
\caption{Pion wave function for the first exited state in TDA and RPA 
approaches. Chiral limit, $m=0$, and confining$+$Coulomb$+$generated 
interactions are taken into account (same parameters as in Fig.\ref{Fig.1}).}
\label{Fig.4}
\end{figure}

\begin{figure}[!htb]
\begin{center}
\input{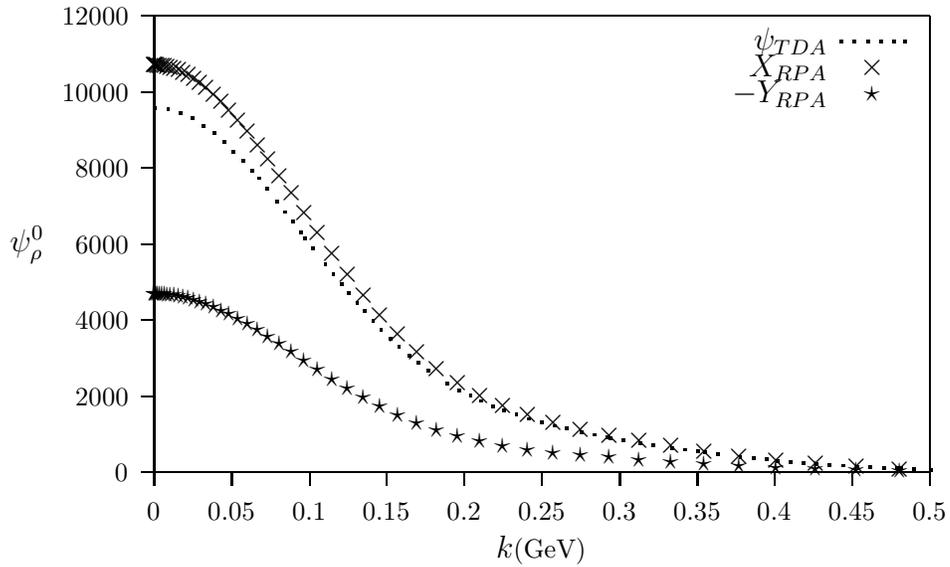}
\end{center}
\caption{Ground state $\rho$ meson wave function in TDA and RPA approaches.
Chiral limit, $m=0$, and confining$+$Coulomb$+$generated interactions
are taken into account (same parameters as in Fig.\ref{Fig.1}).}
\label{Fig.5}
\end{figure}

The dependence of the pion mass in the chiral limit on the cut-off parameter,
$M(\Lambda)$, is shown in Fig.\ref{Fig.2}. The RPA solution, with only 
confining$+$Coulomb potential included, grows unlimited (due to Coulomb),
while adding the generated term stabilizes $M(\Lambda)$, which saturates
roughly at $M(10)$. Stable result confirms that the TDA/RPA equations
are completely renormalized when the generated terms by flow equations 
are included.

The pion wave functions for the ground and first excited states
are depicted in Figs.\ref{Fig.3}, \ref{Fig.4}. The ground state $\rho$-meson
wave function is also shown in Fig.\ref{Fig.5}. The difference between 
the TDA wave function, $\psi_{TDA}$, and the creation component of the RPA 
wave function, $X_{RPA}$, is most significant for the pion ground state, 
while it is smaller for the pion excited states and for the $\rho$-meson
ground state. We conclude, as expected, that the pion ground state is most 
sensitive to the improvement from the RPA over the TDA.

\begin{table}[ht]
$$
\begin{tabular}{|c|c|c|} \hline 
   & TDA, (MeV)&RPA, (MeV)   \\ \hline
 conf. & $20$ & $45$  \\ \hline

 conf.$+$Coul.$+$gen.& $39$ & $92$   \\ \hline 
\end{tabular}
$$
\caption{Pion decay constants in the TDA and RPA approaches with 
confining and confining$+$Coulomb$+$generated potentials taken.} 
\label{tab.5}
\end{table}

Using the TDA and RPA wave functions, we obtain the pion
decay constants given in the table \ref{tab.5}.
We find higher values of $f_{\pi}$ in the RPA than in the TDA.
Taking the confining potential we obtain $f_{\pi}=45\, MeV$ in the RPA,
that is higher than given in Ref. \cite{LeYaouanc} $f_{\pi}=20\,MeV$.
Adding the dynamical terms improves the pion decay constant,
$f_{\pi}=92\,MeV$, that should be compared with the experimental
value $f_{\pi}=96\,MeV$. The Gell'Mann-Oakes-Renner 
relation, Eq. (\ref{eq:4.32}), is satisfied within $6\,\%$, showing 
how close our wave functions are to the exact Hamiltonian eigenfunctions.
This may explain why we obtain a realistic pion decay constant,
$92\, MeV$, in the RPA, although we obtained low value of quark condensate 
$-(155\,MeV)^3$ in subsection \ref{subsec:4.1}.

\section{Summary and Conclusions}
\label{sec:5}

Starting with the leading order QCD Hamiltonian in the Coulomb gauge
and introducing phenomenological potential to account for the nonperturbative
effects, we performed a sequence of transformations of the Hamiltonian:
BV transformation to a massive quasiparticle basis and integral transformation
achieved by flow equations. After BV transformation the phenomenological
potential becomes strong, saturating strong QCD interactions, and describes
the BCS solution, which involves dynamical breaking of the chiral symmetry. 
Residual interactions (weak in the quasiparticle basis) are 
treated perturbatively by flow equations and provide perturbative corrections
to the BCS solution. In this way we have utilized the scheme of the BCS model,
where nonperturbative features such as dynamical chiral symmetry breaking
and massive quasiparticle modes are explicitly present, and included 
perturbatively dynamical interactions in the BCS framework.

Including dynamical interactions by flow equations, we find the correct
ultraviolet behavior of Hamiltonian solutions. Namely, for the first time
the renormalized gap and Bethe-Salpeter equations which are finite in both 
UV and IR regions are obtained. Moreover, no additional UV renormalization
is required in the chiral limit.

Adding dynamical interactions only slightly enlarges the quark dynamical mass;
$m_0(0)=70\, MeV$ in the chiral limit. However, the chiral
condensate has been improved by $68\%$ using flow equations and equals
$\langle\bar{\psi}\psi\rangle_0=-(155\, MeV)^3$. This value is still low 
compared to the predicted lattice value 
$\langle\bar{\psi}\psi\rangle_0=-(250\, MeV)^3$, but such feature of 
lower value is common for all Hamiltonain methods. 

The cut-off dependence of the quark condensate at nonzero current masses
has been obtained which provides the product 
$m(\Lambda)\langle\bar{\psi}\psi\rangle_m(\Lambda)$
invariant with respect to $\Lambda$. The chiral condensate
$\langle\bar{\psi}\psi\rangle_0$ is renormalized by including
the renormalization group running of the strong coupling constant
$\alpha_s(\mbf{q}^2)$.

Bethe-Salpeter equation is solved in the TDA and RPA,
which approximate the pion as a valence $q\bar{q}$ pair
in the TDA and allowing in addition to a $q\bar{q}$ creation for 
a $q\bar{q}$ annihilation operator in the RPA. The pion ground state mass
is most sensitive to the improvement over the TDA approach from the RPA. 
As expected, the pion ground state is obtained lower in the RPA 
than in the TDA.
However, we are unale to get zero mass pion either in the BCS or
adding the leading order corrections from flow equations. An underlying
reason of failing to produce the Goldstone boson might be a breakdown
of covariance in the BCS model, which can be cured by including higher
orders of calculations. Indeed, even in including the leading order 
dynamical interactions in the BCS solution shifts the RPA pion mass
of the ground state down, as expected from the standard perturbation
theory. We obtain for the RPA pion and $\rho$-meson ground states
$M_{\pi}=180\, MeV$ and $M_{\rho}=700\, MeV$, respectively, 
in the chiral limit. 

In the chiral limit the $\pi-\rho$ mass splitting with all terms contributed
is $\delta M_{\pi\rho}=205\, MeV$ in the TDA and $\delta M_{\pi\rho}=520\, MeV$
in the RPA. These values should be compared to 
$\delta M_{\pi\rho}=155\, MeV$ in the TDA and $\delta M_{\pi\rho}=420\, MeV$
in the RPA, calculated with the static confining potential alone.
Flow equations improve the $\pi-\rho$ mass splitting by $32\%$ in the TDA
and by $24\%$ in the RPA.

One may conjecture that the hyperfine interaction dominates for heavy quark masses, 
and the dynamical chiral symmetry breaking is dominant for light quarks. 
However, we cannot
separate these two effects, distinguishing different regimes. By switching
one contribution or another, we find in the chiral limit, that roughly
$30\%$ of the $\pi-\rho$ mass splitting in the TDA is due to the presence 
of the hyperfine interaction and the rest $70\%$ is due to the chiral symmetry 
breaking. In the RPA, this ratio is $40\%$ for the hyperfine and 
$60\%$ for the chiral symmetry breaking.
However, the numerical value of this ratio depends on the details of 
the confining interaction. In our calculations we have used the linear rising
potential generally accepted in the quark model phenomenology with the string tension
$\sigma=0.18\, GeV^2$ predicted by the lattice studies.

It is crucial that the flow equations generate the dynamical spin-dependent hyperfine
interactions which depend on all in- and out-going quark
momenta as well as the momentum of propagating dynamical gluon
and include the momentum dependent solutions of the gap equation. 
In the NJL type model as shown in the second paper of 
Ref. \cite{SzczepaniakSwanson1}, the static hyperfine interaction 
contribute a dominant part to the $\pi-\rho$ mass splitting, 
indicating that this model is close to the nonrelativistic CQM.
However, our model indicates that dynamical interactions in the BCS framework
need to be included using flow equations to improve the CQM. Thus, 
the flow equations seem to provide a working tool to incorporate the corrections 
toward a covariant result.

\vspace{1cm}

{\bf Acknowledgments.}\thinspace The author (E. G.) would like to thank Chueng-Ryong Ji
for reading the manuscript and providing grammatical corrections.

\appendix
\section{Complete QCD motivated Hamiltonian}
\label{app:A}

Here we specify, in second quantized form, the full QCD motivated
Hamiltonian in the Coulomb gauge expanded in the basis of Eq.(\ref{eq:2.6}).
We ignore the pure gluon non-abelian terms, since they do not contribute
to the quark sector.
One-body operators and condensate terms arise from normal ordering
with respect to the trial vacuum state $|\Omega\rangle$.
The upper index over a Hamiltonian operator (e.g. $K^{(0)}$,$V^{(1)}$)
denotes the order (power) in coupling constant.

We specify the $\gamma$-matrices, that are used in this work,
\begin{eqnarray}
\gamma_i &=& \left(
  \begin{array}{c c}
         0 & \sigma_i \\
         -\sigma_i& 0
  \end{array} \right)\,,\,
 \beta=\gamma_0 = \left(
  \begin{array}{c c}
         1 & 0 \\
         0& -1
  \end{array} \right)\,,\,
  \alpha_i=\beta\gamma_i = \left(
  \begin{array}{c c}
         0 & \sigma_i \\
         \sigma_i& 0
  \end{array} \right)\,,\,
 \gamma_5 = \left(
  \begin{array}{c c}
         0 & 1 \\
         1 & 0
  \end{array} \right)\,,
\nonumber\\
\label{eq:a1}\end{eqnarray}
and the spinors are defined in Eq.(\ref{eq:2.7}).   
 
{\it Free quark and gluon part:} Eq.(\ref{eq:2.2}) includes 
the quark $K_q$ and gluon $K_g$ kinetic energies, $K=K_q+K_g$,
and the corresponding condensate terms, $O=O_q+O_g$.
The quark kinetic term reads
\begin{eqnarray}
K_q &=& \sum_{s}\int\frac{d\mbf{k}}{(2\pi)^3}
\label{eq:a2}\\
&& \left(
u_s^{\dagger}(\mbf{k})(\mbf{\alpha}\mbf{k}+\beta m)u_s(\mbf{k})
b_s^{\dagger}(\mbf{k})b_s(\mbf{k}) 
- v_s^{\dagger}(-\mbf{k})(\mbf{\alpha}\mbf{k}+\beta m)v_s(-\mbf{k})
d_s^{\dagger}(-\mbf{k})d_s(-\mbf{k}) \right.
\nonumber\\
&+& \left. u_s^{\dagger}(\mbf{k})(\mbf{\alpha}\mbf{k}+\beta m)v_s(-\mbf{k})
b_s^{\dagger}(\mbf{k})d_s^{\dagger}(-\mbf{k}) 
+ v_s^{\dagger}(-\mbf{k})(\mbf{\alpha}\mbf{k}+\beta m)u_s(\mbf{k})
d_s(-\mbf{k})b_s(\mbf{k}) \right)\,,\nonumber
\end{eqnarray}
which, using the spinors in Eq. (\ref{eq:2.7}) and $\gamma$-matrices
in Eq. (\ref{eq:a1}), is reduced to
\begin{eqnarray}
K_q &=& \sum_{s}\int\frac{d\mbf{k}}{(2\pi)^3}\left(
(kc_{\Phi}(\mbf{k}) + ms_{\Phi}(\mbf{k})) 
[b^{\dagger}_s(\mbf{k})b_s(\mbf{k}) + d^{\dagger}_s(\mbf{k})d_s(\mbf{k})]
\right.
\nonumber\\ 
&+&\left. (ks_{\Phi}(\mbf{k}) - mc_{\Phi}(\mbf{k})) 
[b^{\dagger}_s(\mbf{k})d^{\dagger}_s({-\bf k}) + d_s({-\bf k})b_s(\mbf{k})]
\right)\nonumber\\
&=& \sum_{s}\int\frac{d\mbf{k}}{(2\pi)^3}\sqrt{\mbf{k}^2+m^2}\,
\left(c_{\Theta}(\mbf{k}) 
[b^{\dagger}_s(\mbf{k})b_s(\mbf{k}) + d^{\dagger}_s(\mbf{k})d_s(\mbf{k})]
\right.
\nonumber\\ 
&+&\left. s_{\Theta}(\mbf{k}) 
[b^{\dagger}_s(\mbf{k})d^{\dagger}_s({-\bf k}) + d_s({-\bf k})b_s(\mbf{k})]
\right)
\,,\label{eq:a3}\end{eqnarray}
where $m$ is the bare quark mass, and we have used the connection
between the nonperturbative angle $\Phi$ and the Bogoliubov-Valatin angle
$\Theta$ (See Eq. (\ref{eq:c8}) in Appendix \ref{app:C}). 
In order to separate the zeroth and second order 
contributions we represent the quark kinetic term, using Eq. (\ref{eq:2.8}), as 
\begin{eqnarray}
K_q &=& \sum_{s}\int\frac{d\mbf{k}}{(2\pi)^3}
\sqrt{\mbf{k}^2+M^2(\mbf{k})}\,
[b^{\dagger}_s(\mbf{k})b_s(\mbf{k})+d^{\dagger}_s(\mbf{k})d_s(\mbf{k})]
\nonumber\\
&+& \sum_{s}\int\frac{d\mbf{k}}{(2\pi)^3}
(m-M(\mbf{k}))
\left(s_{\Phi}(\mbf{k}) 
[b^{\dagger}_s(\mbf{k})b_s(\mbf{k}) + d^{\dagger}_s(\mbf{k})d_s(\mbf{k})]
\right.
\nonumber\\ 
&-&\left. c_{\Phi}(\mbf{k}) 
[b^{\dagger}_s(\mbf{k})d^{\dagger}_s({-\bf k}) + d_s({-\bf k})b_s(\mbf{k})]
\right)
\nonumber\\
&=& K_q^{(0)}+ K_q^{(2)}
\,,\label{eq:a4}\end{eqnarray}
where $K_q^{(0)}$ and $K_q^{(2)}$ correspond to the zeroth and second order
quark kinetic energy, respectively. The gluon kinetic energy is given by  
\begin{eqnarray}
 K_g &=& K_g^{(0)} = \sum_a \int\frac{d\mbf{k}}{(2\pi)^3}
 \omega(\mbf{k}) a_i^{a\dagger}(\mbf{k})a_i^a(\mbf{k}) 
\,,\label{eq:a5}\end{eqnarray}
Summing the zeroth order kinetic energy yields
\begin{eqnarray}
K^{(0)} &=& \sum_{s}\int\frac{d\mbf{k}}{(2\pi)^3}E(\mbf{k})
[b^{\dagger}_s(\mbf{k})b_s(\mbf{k}) + d^{\dagger}_s(\mbf{k})d_s(\mbf{k})]
+\sum_a \int\frac{d\mbf{k}}{(2\pi)^3}
 \omega(\mbf{k}) a_i^{a\dagger}(\mbf{k})a_i^a(\mbf{k})\,,
\nonumber\\ \label{eq:a6}
\end{eqnarray}
with the effective quark and free gluon energies, 
$E(\mbf{k})=\sqrt{\mbf{k}^2+M^2(\mbf{k})}$, Eq. (\ref{eq:2.8}), 
and $\omega(\mbf{k})=k$, respectively.
The quark condensate reads
\begin{eqnarray}
O_q &=& N_c\mbf{V} \int\frac{d\mbf{k}}{(2\pi)^3}\sum_{s} 
v_s^{\dagger}(-\mbf{k})(\mbf{\alpha}\mbf{k}+\beta m)v_s(-\mbf{k})
\,,\label{eq:a7}\end{eqnarray}
with the volume $\mbf{V}=(2\pi)^3\delta^{(3)}(0)$.
It is reduced to
\begin{eqnarray}
O_q &=& -4N_c\mbf{V}\int\frac{d\mbf{k}}{(2\pi)^3}\left(
kc_{\Phi}(\mbf{k})+ms_{\Phi}(\mbf{k}) \right)
\nonumber\\
&=& -4N_c\mbf{V}\int\frac{d\mbf{k}}{(2\pi)^3}
\sqrt{\mbf{k}^2+m^2}\,c_{\Theta}(\mbf{k})
\,,\label{eq:a8}\end{eqnarray}
where we have used Eq. (\ref{eq:c8}). 
Separating the zeroth and second order,
one has
\begin{eqnarray}
O_q &=& -4N_c\mbf{V}\int\frac{d\mbf{k}}{(2\pi)^3}
\sqrt{\mbf{k}^2+M^2(\mbf{k})}\,
-4N_c\mbf{V}\int\frac{d\mbf{k}}{(2\pi)^3}
(m-M(\mbf{k}))s_{\Phi}(\mbf{k})
\nonumber\\ 
&=& O_q^{(0)}+O_q^{(2)}
\,.\label{eq:a9}\end{eqnarray}
The gluon condensate is given by
\begin{eqnarray}
 O_g = O_g^{(0)} = (N_c^2-1)\mbf{V}
\int\frac{d\mbf{k}}{(2\pi)^3}\omega(\mbf{k})
\,.\label{eq:a10}\end{eqnarray}
Therefore, the zeroth order condensate term is given by
\begin{eqnarray}
O^{(0)} &=& -4N_c\mbf{V}\int\frac{d\mbf{k}}{(2\pi)^3} 
E(\mbf{k}) + (N_c^2-1)\mbf{V}
\int\frac{d\mbf{k}}{(2\pi)^3}\omega(\mbf{k})
\,,\label{eq:a11}\end{eqnarray}
where the volume is $\mbf{V}=(2\pi)^3\delta^{(3)}(0)$,
and $E(\mbf{k})=\sqrt{\mbf{k}^2+M^2(\mbf{k})}\,$.

{\it Instantaneous interaction:} Eq. (\ref{eq:2.3}) includes
the linear confining and Coulomb interactions in the quark sector
\begin{eqnarray}
V_{inst} &=& \sum_{s_1...s_4}\int 
\left(\prod_{n=1}^{4}\frac{d\mbf{k}_n}{(2\pi)^3}\right)
(2\pi)^3\delta^{(3)}(\mbf{k}_1+\mbf{k}_3-\mbf{k}_2-\mbf{k}_4)
V_{L+C}(\mbf{k}_1,\mbf{k}_2)
\nonumber\\
&\times&\colon [u^{\dagger}_{s_1}(\mbf{k}_1)b^{\dagger}_{s_1}(\mbf{k}_1)
+ v^{\dagger}_{s_1}(-\mbf{k}_1)d_{s_1}(-\mbf{k}_1)]T^a
[u_{s_2}(\mbf{k}_2)b_{s_2}(\mbf{k}_2)
+ v_{s_2}(-\mbf{k}_2)d^{\dagger}_{s_2}(-\mbf{k}_2)]
\nonumber\\
&& [u^{\dagger}_{s_3}(\mbf{k}_3)b^{\dagger}_{s_3}(\mbf{k}_3)
+ v^{\dagger}_{s_3}(-\mbf{k}_3)d_{s_3}(-\mbf{k}_3)]T^a
[u_{s_4}(\mbf{k}_4)b_{s_4}(\mbf{k}_4)
+ v_{s_4}(-\mbf{k}_4)d^{\dagger}_{s_4}(-\mbf{k}_4)]\colon
\nonumber\\
&=& V_{inst}^{(0)}+V_{inst}^{(2)}
\,,\label{eq:a12}\end{eqnarray}
with $V_{L+C}(\mbf{k},\mbf{q})\rightarrow V_{L+C}(\mbf{k}-\mbf{q})$.
Here the linear and Coulomb terms are represented by 
$V_{inst}^{(0)}$ and $V_{inst}^{(2)}$, respectively, and
$V_{L+C}(\mbf{k})$ is defined by
\begin{eqnarray}
C_f V_{L+C}(\mbf{k}) &=& 2\pi C_f \frac{\alpha_s}{\mbf{k}^2}
+4\pi\frac{\sigma}{\mbf{k}^4}
\,,\label{eq:a13}\end{eqnarray}
with the fundamental Casimir operator $C_f=T^aT^a=(N_c^2-1)/2N_c=4/3$.
Terms arising from normal ordering are the one-body operator
(self-energy) and the condensate term. Self-energy is written as 
\begin{eqnarray}
\Sigma_{inst} &=& \sum_s\int\frac{d\mbf{k}d\mbf{q}}{(2\pi)^6}
C_f V_{L+C}(\mbf{k},\mbf{q})
\label{eq:a14} \\
&& \left(
b^{\dagger}_s(\mbf{k})b_s(\mbf{k})\sum_{s^{\prime}}
[(u^{\dagger}_s(\mbf{k})u_{s^{\prime}}(\mbf{q}))
(u^{\dagger}_{s^{\prime}}(\mbf{q})u_s(\mbf{k}))
-(u^{\dagger}_s(\mbf{k})v_{s^{\prime}}(-\mbf{q}))
(v^{\dagger}_{s^{\prime}}(-\mbf{q})u_s(\mbf{k}))] \right.
\nonumber\\
&-&\left. d^{\dagger}_s(-\mbf{k})d_s(-\mbf{k})\sum_{s^{\prime}}
[(v^{\dagger}_s(-\mbf{k})u_{s^{\prime}}(\mbf{q}))
(u^{\dagger}_{s^{\prime}}(\mbf{q})v_s(-\mbf{k}))
-(v^{\dagger}_s(-\mbf{k})v_{s^{\prime}}(-\mbf{q}))
(v^{\dagger}_{s^{\prime}}(-\mbf{q})v_s(-\mbf{k}))] \right.
\nonumber\\
&+&\left. d_s(-\mbf{k})b_s(\mbf{k})\sum_{s^{\prime}}
[(v^{\dagger}_s(-\mbf{k})u_{s^{\prime}}(\mbf{q}))
(u^{\dagger}_{s^{\prime}}(\mbf{q})u_s(\mbf{k}))
-(v^{\dagger}_s(-\mbf{k})v_{s^{\prime}}(-\mbf{q}))
(v^{\dagger}_{s^{\prime}}(-\mbf{q})u_s(\mbf{k}))] \right.
\nonumber\\
&+&\left. b^{\dagger}_s(\mbf{k})d^{\dagger}_s(-\mbf{k})
\sum_{s^{\prime}}
[(u^{\dagger}_s(\mbf{k})u_{s^{\prime}}(\mbf{q}))
(u^{\dagger}_{s^{\prime}}(\mbf{q})v_s(-\mbf{k}))
-(u^{\dagger}_s(\mbf{k})v_{s^{\prime}}(-\mbf{q}))
(v^{\dagger}_{s^{\prime}}(-\mbf{q})v_s(-\mbf{k}))] \right)
\nonumber
\,,\end{eqnarray}
which, using Eq. (\ref{eq:2.7}), is reduced to
\begin{eqnarray}
\Sigma_{inst} &=& \sum_s\int\frac{d\mbf{k}d\mbf{q}}{(2\pi)^6}
C_f V_{L+C}(\mbf{k},\mbf{q})
[b^{\dagger}_s(\mbf{k})b_s(\mbf{k}) + d^{\dagger}_s(-\mbf{k})d_s(-\mbf{k})]
\nonumber\\
&& \frac{1}{2} \left(
(1+s(\mbf{k})s(\mbf{q})+c(\mbf{k})c(\mbf{q})\mbf{\hat{k}}\cdot\mbf{\hat{q}})
- (1-s(\mbf{k})s(\mbf{q})-c(\mbf{k})c(\mbf{q})\mbf{\hat{k}}\cdot\mbf{\hat{q}})
\right)
\nonumber\\
&+& \sum_s\int\frac{d\mbf{k}d\mbf{q}}{(2\pi)^6}
C_f V_{L+C}(\mbf{k},\mbf{q})
[b^{\dagger}_s(\mbf{k})d^{\dagger}_s({-\bf k}) + d_s({-\bf k})b_s(\mbf{k})]
\nonumber\\
&& \frac{1}{2}\left(
2(-c(\mbf{k})s(\mbf{q})+s(\mbf{k})c(\mbf{q})\mbf{\hat{k}}\cdot\mbf{\hat{q}})
\right)
\nonumber\\
&=& \sum_s\int\frac{d\mbf{k}d\mbf{q}}{(2\pi)^6}\left(
s(\mbf{k})s(\mbf{q})+c(\mbf{k})c(\mbf{q})\mbf{\hat{k}}\cdot\mbf{\hat{q}}
\right)C_f V_{L+C}(\mbf{k},\mbf{q})
\nonumber\\ 
&\times& 
[b^{\dagger}_s(\mbf{k})b_s(\mbf{k}) + d^{\dagger}_s(-\mbf{k})d_s(-\mbf{k})]
\nonumber\\
&+& \sum_s\int\frac{d\mbf{k}d\mbf{q}}{(2\pi)^6}\left(
-c(\mbf{k})s(\mbf{q})+s(\mbf{k})c(\mbf{q})\mbf{\hat{k}}\cdot\mbf{\hat{q}}
\right)C_f V_{L+C}(\mbf{k},\mbf{q})
\nonumber\\
&\times&
[b^{\dagger}_s(\mbf{k})d^{\dagger}_s({-\bf k}) + d_s({-\bf k})b_s(\mbf{k})]
\,.\label{eq:a15}\end{eqnarray}
The correction to quark condensate is given by
\begin{eqnarray}
O_{inst} &=& N_c\int\frac{d\mbf{k}d\mbf{q}}{(2\pi)^6}
\sum_{s,s^{\prime}}
[(v^{\dagger}_s(-\mbf{k})u_{s^{\prime}}(\mbf{q}))
(u^{\dagger}_{s^{\prime}}(\mbf{q})v_s(-\mbf{k}))]
C_f V_{L+C}(\mbf{k},\mbf{q})
\,.\label{eq:a16}\end{eqnarray}
which simplifies to
\begin{eqnarray}
O_{inst} &=& 2N_c\int\frac{d\mbf{k}d\mbf{q}}{(2\pi)^6}\left(
1-s(\mbf{k})s(\mbf{q})-c(\mbf{k})c(\mbf{q})\mbf{\hat{k}}\cdot\mbf{\hat{q}}
\right) C_f V_{L+C}(\mbf{k},\mbf{q})
\,.\label{eq:a17}\end{eqnarray}

{\it Dynamical interaction:} Eq. (\ref{eq:2.5}) includes to order
$O(g)$ the quark-gluon coupling
\begin{eqnarray}
V_{qg}^{(1)} &=& -g\sum_{s_1,s_2,a}\int\left(\prod_{n=1}^{3}
\frac{d\mbf{k}_n}{(2\pi)^3} \right)(2\pi)^3
\delta^{(3)}(\mbf{k}_1-\mbf{k}_2-\mbf{k}_3)\frac{1}{\sqrt{2\omega(\mbf{k}_3)}}
\nonumber\\
&\times&\colon
[u^{\dagger}_{s_1}(\mbf{k}_1)b^{\dagger}_{s_1}(\mbf{k}_1)
+ v^{\dagger}_{s_1}(-\mbf{k}_1)d_{s_1}(-\mbf{k}_1)]T^a\alpha_i
[u_{s_2}(\mbf{k}_2)b_{s_2}(\mbf{k}_2)
+ v_{s_2}(-\mbf{k}_2)d^{\dagger}_{s_2}(-\mbf{k}_2)]
\nonumber\\
&& [a_i^a(\mbf{k}_3)+a_i^{a\dagger}(-\mbf{k}_3)]\colon
\,,\label{eq:a18}\end{eqnarray}
where the colon denotes normal ordered products, i.e.
all creation operators are on the left and annihilation on the right.
One-body and condensate terms diverge in the UV region, since they contain
field operator products at one point. They are regulated as discussed
in Section \ref{sec:3}.

\section{Second order flow equations}
\label{app:B}

Solving the second oder flow equation for the particle number conserving
part $H_d$, Eq. (\ref{eq:2.19}), we generate three types of Hamiltonian
operators: two-body effective quark interactions, one-body self-energy terms
and condensates. We consider each separately. In what follows,
$s_{\Phi}(\mbf{k})\equiv s(\mbf{k})$ and the same abbreviation is defined
for a cosine.

\subsection{Effective quark interaction}
\label{app:B1}

We calculate an effective quark interaction only in the sectors 
which contribute to the bound state equations: TDA and RPA.
Using expressions for the effective coupling constant, Eq. (\ref{eq:2.14}),
and the generator, Eq. (\ref{eq:2.15}), the following effective interaction
is generated, which contribute to TDA equation,
\begin{eqnarray}
\frac{dV_{gen}(l)}{dl} &=& \sum_{s_1...s_4}\int 
\left(\prod_{n=1}^{4}\frac{d\mbf{k}_n}{(2\pi)^3}\right)
(2\pi)^3\delta^{(3)}(\mbf{k}_1+\mbf{k}_3-\mbf{k}_2-\mbf{k}_4)
\nonumber\\
&\times&\frac{D_{ij}(\mbf{k}_1-\mbf{k}_2)}{2\omega(\mbf{k}_1-\mbf{k}_2)}
(u^{\dagger}_{s_1}(\mbf{k}_1)\alpha_i u_{s_2}(\mbf{k}_2))
(v^{\dagger}_{s_3}(-\mbf{k}_3)\alpha_j v_{s_4}(-\mbf{k}_4))
\nonumber\\
&\times&[\eta_1(\mbf{k}_1,\mbf{k}_2,\mbf{k}_1-\mbf{k}_2)
g_1(\mbf{k}_3,\mbf{k}_4,\mbf{k}_1-\mbf{k}_2)
+\eta_1(\mbf{k}_3,\mbf{k}_4,\mbf{k}_1-\mbf{k}_2)
g_1(\mbf{k}_1,\mbf{k}_2,\mbf{k}_1-\mbf{k}_2)]
\nonumber\\
&\times&\colon
b_{s_1}^{\dagger}(\mbf{k}_1)T^a b_{s_2}(\mbf{k}_2)
d_{s_3}(-\mbf{k}_3)T^a d_{s_4}^{\dagger}(-\mbf{k}_4)\colon 
\nonumber\\
&+& \sum_{s_1...s_4}\int 
\left(\prod_{n=1}^{4}\frac{d\mbf{k}_n}{(2\pi)^3}\right)
(2\pi)^3\delta^{(3)}(\mbf{k}_1+\mbf{k}_3-\mbf{k}_2-\mbf{k}_4)
\nonumber\\
&\times&\frac{D_{ij}(\mbf{k}_1-\mbf{k}_2)}{2\omega(\mbf{k}_1-\mbf{k}_2)}
(v^{\dagger}_{s_1}(-\mbf{k}_1)\alpha_i v_{s_2}(-\mbf{k}_2))
(u^{\dagger}_{s_3}(\mbf{k}_3)\alpha_j u_{s_4}(\mbf{k}_4))
\nonumber\\
&\times&[\eta_1(\mbf{k}_2,\mbf{k}_1,-(\mbf{k}_1-\mbf{k}_2))
g_1(\mbf{k}_4,\mbf{k}_3,-(\mbf{k}_1-\mbf{k}_2))
\nonumber\\
&+&\eta_1(\mbf{k}_4,\mbf{k}_3,-(\mbf{k}_1-\mbf{k}_2))
g_1(\mbf{k}_2,\mbf{k}_1,-(\mbf{k}_1-\mbf{k}_2))]
\nonumber\\
&\times&\colon
d_{s_1}(-\mbf{k}_1)T^a d^{\dagger}_{s_2}(-\mbf{k}_2)
b^{\dagger}_{s_3}(\mbf{k}_3)T^a b_{s_4}(\mbf{k}_4)\colon
\,,\label{eq:b1}\end{eqnarray}
Integrating this flow equation, we obtain the generated interaction
$V_{gen}(l\rightarrow\infty)=V_{gen}$ (the initial value is 
$V_{gen}(l=0)=0$);
\begin{eqnarray}
V_{gen} &=& \sum_{s_1...s_4}\int 
\left(\prod_{n=1}^{4}\frac{d\mbf{k}_n}{(2\pi)^3}\right)
(2\pi)^3\delta^{(3)}(\mbf{k}_1+\mbf{k}_3-\mbf{k}_2-\mbf{k}_4)
\nonumber\\
&\times&\frac{D_{ij}(\mbf{k}_1-\mbf{k}_2)}{2\omega(\mbf{k}_1-\mbf{k}_2)}
g^2 \frac{D_1+D_1^{\prime}}{D_1^2+D_1^{\prime 2}}
(u^{\dagger}_{s_1}(\mbf{k}_1)\alpha_i u_{s_2}(\mbf{k}_2))
(v^{\dagger}_{s_3}(-\mbf{k}_3)\alpha_j v_{s_4}(-\mbf{k}_4))
\nonumber\\
&\times&\colon
b_{s_1}^{\dagger}(\mbf{k}_1)T^a b_{s_2}(\mbf{k}_2)
d_{s_3}(-\mbf{k}_3)T^a d_{s_4}^{\dagger}(-\mbf{k}_4)\colon
\nonumber\\
&+& \sum_{s_1...s_4}\int 
\left(\prod_{n=1}^{4}\frac{d\mbf{k}_n}{(2\pi)^3}\right)
(2\pi)^3\delta^{(3)}(\mbf{k}_1+\mbf{k}_3-\mbf{k}_2-\mbf{k}_4)
\nonumber\\
&\times&\frac{D_{ij}(\mbf{k}_1-\mbf{k}_2)}{2\omega(\mbf{k}_1-\mbf{k}_2)}
g^2 \frac{\tilde{D}_1+\tilde{D}_1^{\prime}}
{\tilde{D}_1^2+\tilde{D}_1^{\prime 2}} 
(v^{\dagger}_{s_1}(-\mbf{k}_1)\alpha_i v_{s_2}(-\mbf{k}_2))
(u^{\dagger}_{s_3}(\mbf{k}_3)\alpha_j u_{s_4}(\mbf{k}_4))
\nonumber\\
&\times&\colon
d_{s_1}(-\mbf{k}_1)T^a d^{\dagger}_{s_2}(-\mbf{k}_2)
b^{\dagger}_{s_3}(\mbf{k}_3)T^a b_{s_4}(\mbf{k}_4)\colon
\,.\label{eq:b2}\end{eqnarray}
We combine both terms, using symmetry property of the polarization sum
$D_{ji}(\mbf{q})=D_{ij}(\mbf{q})$, and obtain
\begin{eqnarray}
V_{gen} &=& \sum_{s_1...s_4}\int 
\left(\prod_{n=1}^{4}\frac{d\mbf{k}_n}{(2\pi)^3}\right)
(2\pi)^3\delta^{(3)}(\mbf{k}_1+\mbf{k}_3-\mbf{k}_2-\mbf{k}_4)
\nonumber\\
&\times&\frac{D_{ij}(\mbf{k}_1-\mbf{k}_2)}{2\omega(\mbf{k}_1-\mbf{k}_2)}
g^2\left( \frac{D_1+D_1^{\prime}}{D_1^2+D_1^{\prime 2}}
+ \frac{\tilde{D}_1+\tilde{D}_1^{\prime}}
{\tilde{D}_1^2+\tilde{D}_1^{\prime 2}} \right)
(u^{\dagger}_{s_1}(\mbf{k}_1)\alpha_i u_{s_2}(\mbf{k}_2))
(v^{\dagger}_{s_3}(-\mbf{k}_3)\alpha_j v_{s_4}(-\mbf{k}_4))
\nonumber\\
&\times&\colon
b_{s_1}^{\dagger}(\mbf{k}_1)T^a b_{s_2}(\mbf{k}_2)
d_{s_3}(-\mbf{k}_3)T^a d_{s_4}^{\dagger}(-\mbf{k}_4)\colon
\,,\label{eq:b3}\end{eqnarray}
where the energy denominators are
\begin{eqnarray}
D_1 &=& E(\mbf{k}_1)-E(\mbf{k}_2)-\omega(\mbf{k}_1-\mbf{k}_2)\,,\, 
D_1^{\prime} = E(\mbf{k}_3)-E(\mbf{k}_4)-\omega(\mbf{k}_1-\mbf{k}_2)
\nonumber\\
\tilde{D}_1 &=& E(\mbf{k}_2)-E(\mbf{k}_1)-\omega(\mbf{k}_1-\mbf{k}_2)\,,\, 
\tilde{D}_1^{\prime} = E(\mbf{k}_4)-E(\mbf{k}_3)-\omega(\mbf{k}_1-\mbf{k}_2)
\,.\label{eq:b4}\end{eqnarray}
An effective interaction, which contribute to RPA, is defined by
\begin{eqnarray}
\frac{dV_{gen}(l)}{dl} &=& \sum_{s_1...s_4}\int 
\left(\prod_{n=1}^{4}\frac{d\mbf{k}_n}{(2\pi)^3}\right)
(2\pi)^3\delta^{(3)}(\mbf{k}_1+\mbf{k}_3-\mbf{k}_2-\mbf{k}_4)
\nonumber\\
&\times&\frac{D_{ij}(\mbf{k}_1-\mbf{k}_2)}{2\omega(\mbf{k}_1-\mbf{k}_2)}
(u^{\dagger}_{s_1}(\mbf{k}_1)\alpha_i v_{s_2}(-\mbf{k}_2))
(u^{\dagger}_{s_3}(\mbf{k}_3)\alpha_j v_{s_4}(-\mbf{k}_4))
\nonumber\\
&\times&[\eta_0(\mbf{k}_4,\mbf{k}_3,\mbf{k}_1-\mbf{k}_2)
g_{1^{\prime}}(\mbf{k}_1-\mbf{k}_2,\mbf{k}_2,\mbf{k}_1)
-\eta_{1^{\prime}}(\mbf{k}_1-\mbf{k}_2,\mbf{k}_2,\mbf{k}_1)
g_0(\mbf{k}_4,\mbf{k}_3,\mbf{k}_1-\mbf{k}_2)]
\nonumber\\
&\times&\colon
b_{s_1}^{\dagger}(\mbf{k}_1)T^a d^{\dagger}_{s_2}(-\mbf{k}_2)
b^{\dagger}_{s_3}(\mbf{k}_3)T^a d_{s_4}^{\dagger}(-\mbf{k}_4)
\colon 
\nonumber\\
&+&\sum_{s_1...s_4}\int 
\left(\prod_{n=1}^{4}\frac{d\mbf{k}_n}{(2\pi)^3}\right)
(2\pi)^3\delta^{(3)}(\mbf{k}_1+\mbf{k}_3-\mbf{k}_2-\mbf{k}_4)
\nonumber\\
&\times&\frac{D_{ij}(\mbf{k}_1-\mbf{k}_2)}{2\omega(\mbf{k}_1-\mbf{k}_2)}
(v^{\dagger}_{s_1}(-\mbf{k}_1)\alpha_i u_{s_2}(\mbf{k}_2))
(v^{\dagger}_{s_3}(-\mbf{k}_3)\alpha_j u_{s_4}(\mbf{k}_4))
\nonumber\\
&\times&[\eta_0(\mbf{k}_1,\mbf{k}_2,\mbf{k}_1-\mbf{k}_2)
g_{1^{\prime}}(\mbf{k}_1-\mbf{k}_2,\mbf{k}_3,\mbf{k}_4)
-\eta_{1^{\prime}}(\mbf{k}_1-\mbf{k}_2,\mbf{k}_3,\mbf{k}_4)
g_0(\mbf{k}_1,\mbf{k}_2,\mbf{k}_1-\mbf{k}_2)]
\nonumber\\
&\times&\colon
d_{s_1}(-\mbf{k}_1)T^a b_{s_2}(\mbf{k}_2)
d_{s_3}(-\mbf{k}_3)T^a b_{s_4}(\mbf{k}_4)
\colon 
\,,\label{eq:b5}\end{eqnarray}
that gives after integration
\begin{eqnarray}
V_{gen} &=& \sum_{s_1...s_4}\int 
\left(\prod_{n=1}^{4}\frac{d\mbf{k}_n}{(2\pi)^3}\right)
(2\pi)^3\delta^{(3)}(\mbf{k}_1+\mbf{k}_3-\mbf{k}_2-\mbf{k}_4)
\nonumber\\
&\times&\frac{D_{ij}(\mbf{k}_1-\mbf{k}_2)}{2\omega(\mbf{k}_1-\mbf{k}_2)}
g^2\frac{D_0-D_{1^{\prime}}}{D_0^2+D_{1^{\prime }}^2}
(u^{\dagger}_{s_1}(\mbf{k}_1)\alpha_i v_{s_2}(-\mbf{k}_2))
(u^{\dagger}_{s_3}(\mbf{k}_3)\alpha_j v_{s_4}(-\mbf{k}_4))
\nonumber\\
&\times&\colon
b_{s_1}^{\dagger}(\mbf{k}_1)T^a d^{\dagger}_{s_2}(-\mbf{k}_2)
b^{\dagger}_{s_3}(\mbf{k}_3)T^a d_{s_4}^{\dagger}(-\mbf{k}_4)
\colon
\nonumber\\
&+&\sum_{s_1...s_4}\int 
\left(\prod_{n=1}^{4}\frac{d\mbf{k}_n}{(2\pi)^3}\right)
(2\pi)^3\delta^{(3)}(\mbf{k}_1+\mbf{k}_3-\mbf{k}_2-\mbf{k}_4)
\nonumber\\
&\times&\frac{D_{ij}(\mbf{k}_1-\mbf{k}_2)}{2\omega(\mbf{k}_1-\mbf{k}_2)}
g^2\frac{\tilde{D}_0-\tilde{D}_{1^{\prime}}}
{\tilde{D}_0^2+\tilde{D}_{1^{\prime }}^2}
(v^{\dagger}_{s_1}(-\mbf{k}_1)\alpha_i u_{s_2}(\mbf{k}_2))
(v^{\dagger}_{s_3}(-\mbf{k}_3)\alpha_j u_{s_4}(\mbf{k}_4))
\nonumber\\
&\times&\colon
d_{s_1}(-\mbf{k}_1)T^a b_{s_2}(\mbf{k}_2)
d_{s_3}(-\mbf{k}_3)T^a b_{s_4}(\mbf{k}_4)
\colon
\,,\label{eq:b6}\end{eqnarray}
where the energy denominators are 
\begin{eqnarray}
D_0 &=&-(E(\mbf{k}_3)+E(\mbf{k}_4)+\omega(\mbf{k}_1-\mbf{k}_2))
\,,\, 
D_{1^{\prime}}=\omega(\mbf{k}_1-\mbf{k}_2)-E(\mbf{k}_1)-E(\mbf{k}_2)
\nonumber\\
\tilde{D}_0&=&-(E(\mbf{k}_1)+E(\mbf{k}_2)+\omega(\mbf{k}_1-\mbf{k}_2))
\,,\, 
\tilde{D}_{1^{\prime}}=\omega(\mbf{k}_1-\mbf{k}_2)-E(\mbf{k}_3)-E(\mbf{k}_4)
\,.\label{eq:b7}\end{eqnarray}
Here the energy denominators carry the same lower indeces $D_i$
as the generators $\eta_i$ and coupling constants $g_i$ 
which they correspond to. One can further simplify the generated
interactions Eqs. (\ref{eq:b3}) and (\ref{eq:b6}).
In the c.m. frame the generated interactions contributing 
to TDA ($X$ component of the RPA wave function), Eq. (\ref{eq:b3}), are   
\begin{eqnarray}
V_{gen} &=& \sum_{\alpha\beta\delta\gamma}\int
\frac{d\mbf{k}d\mbf{q}}{(2\pi)^6}
2 W_1(\mbf{k},\mbf{q}) D_{ij}(\mbf{k}-\mbf{q})
\label{eq:b8}\\ 
&\times&\left( (u^{\dagger}_{\delta}(\mbf{q})\alpha_i u_{\alpha}(\mbf{k}))
(v^{\dagger}_{\beta}(-\mbf{k})\alpha_j v_{\gamma}(-\mbf{q}))
\colon
b_{\delta}^{\dagger}(\mbf{q})T^a b_{\alpha}(\mbf{k})
d_{\beta}(-\mbf{k})T^a d_{\gamma}^{\dagger}(-\mbf{q}) \colon\right.
\nonumber\\
&+& \left. (u^{\dagger}_{\alpha}(\mbf{k})\alpha_i u_{\delta}(\mbf{q}))
(v^{\dagger}_{\gamma}(-\mbf{q})\alpha_j v_{\beta}(-\mbf{k}))
\colon
b_{\alpha}^{\dagger}(\mbf{k})T^a b_{\delta}(\mbf{q})
d_{\gamma}(-\mbf{q})T^a d_{\beta}^{\dagger}(-\mbf{k}) \colon
\right) \nonumber
\,,\end{eqnarray}
and to RPA ($Y$ component of the RPA wave function), Eq. (\ref{eq:b6}), are
\begin{eqnarray}
V_{gen} &=& \sum_{\alpha\beta\delta\gamma}\int
\frac{d\mbf{k}d\mbf{q}}{(2\pi)^6}
2 W_2(\mbf{k},\mbf{q}) D_{ij}(\mbf{k}-\mbf{q})
\label{eq:b9}\\ 
&\times&\left( (v^{\dagger}_{\gamma}(-\mbf{q})\alpha_i u_{\alpha}(\mbf{k}))
(v^{\dagger}_{\beta}(-\mbf{k})\alpha_j u_{\delta}(\mbf{q}))
\colon
d_{\gamma}(-\mbf{q})T^a b_{\alpha}(\mbf{k})
d_{\beta}(-\mbf{k})T^a b_{\delta}(\mbf{q}) \colon\right.
\nonumber\\
&+& \left. (u^{\dagger}_{\delta}(\mbf{q})\alpha_i v_{\beta}(-\mbf{k}))
(u^{\dagger}_{\alpha}(\mbf{k})\alpha_j v_{\gamma}(-\mbf{q}))
\colon
b_{\delta}^{\dagger}(\mbf{q})T^a d_{\beta}^{\dagger}(-\mbf{k})
b_{\alpha}^{\dagger}(\mbf{k})T^a d_{\gamma}^{\dagger}(-\mbf{q}) \colon
\right) \nonumber
\,,\end{eqnarray}
where potential functions are given by
\begin{eqnarray}
C_f W_1(\mbf{k},\mbf{q}) &=& 
-\frac{1}{2}\frac{C_f g^2}
{\omega^2(\mbf{k}-\mbf{q})+(E(\mbf{k})-E(\mbf{q}))^2}
\nonumber\\
C_f W_2(\mbf{k},\mbf{q}) &=& 
-\frac{1}{2}\frac{C_f g^2}
{\omega^2(\mbf{k}-\mbf{q})+(E(\mbf{k})+E(\mbf{q}))^2}
\,.\label{eq:b10}\end{eqnarray}

\subsection{Self-energies}
\label{app:B2}

Two contraction terms from the commutator $[\eta^{(1)},V_{qg}^{(1)}]$
in Eq. (\ref{eq:2.19}) contribute to the self-energy operators.
In the diagonal one-quark sector ($b^{\dagger}b$ and $d^{\dagger}d$)
the flow equations are written as follows  
\begin{eqnarray}
\frac{d\Sigma_{gen}(l)}{dl} &=& \sum_{s}\int 
\frac{d\mbf{k}d\mbf{q}}{(2\pi)^6}
C_f \frac{D_{ij}(\mbf{k}-\mbf{q})}{2\omega(\mbf{k}-\mbf{q})}
\label{eq:b11}
\\
&\times& \left( \sum_{s^{\prime}}
(u^{\dagger}_{s}(\mbf{k})\alpha_i u_{s^{\prime}}(\mbf{q}))
(u^{\dagger}_{s^{\prime}}(\mbf{q})\alpha_j u_{s}(\mbf{k}))
2\eta_1(\mbf{k},\mbf{q},\mbf{k}-\mbf{q})
g_1(\mbf{k},\mbf{q},\mbf{k}-\mbf{q})\right.
\nonumber\\
&-& \left.\sum_{s^{\prime}}
(u^{\dagger}_{s}(\mbf{k})\alpha_i v_{s^{\prime}}(-\mbf{q}))
(v^{\dagger}_{s^{\prime}}(-\mbf{q})\alpha_j u_{s}(\mbf{k}))
2\eta_0(\mbf{k},\mbf{q},\mbf{k}-\mbf{q})
g_0(\mbf{k},\mbf{q},\mbf{k}-\mbf{q})\right)
b_s^{\dagger}(\mbf{k})b_s(\mbf{k})
\nonumber
\,,\end{eqnarray}
since $\eta_0(\mbf{p},\mbf{k},\mbf{q})=\eta_0(\mbf{k},\mbf{p},\mbf{q})$
(the same holds for $g_0$) and
\begin{eqnarray}
\frac{d\Sigma_{gen}(l)}{dl} &=& \sum_{s}\int 
\frac{d\mbf{k}d\mbf{q}}{(2\pi)^6}
C_f \frac{D_{ij}(\mbf{k}-\mbf{q})}{2\omega(\mbf{k}-\mbf{q})}
\label{eq:b12}
\\
&\times& \left( \sum_{s^{\prime}}
(v^{\dagger}_{s}(-\mbf{k})\alpha_i v_{s^{\prime}}(-\mbf{q}))
(v^{\dagger}_{s^{\prime}}(-\mbf{q})\alpha_j v_{s}(-\mbf{k}))
2\eta_1(\mbf{k},\mbf{q},\mbf{k}-\mbf{q})
g_1(\mbf{k},\mbf{q},\mbf{k}-\mbf{q})\right.
\nonumber\\
&-& \left.\sum_{s^{\prime}}
(v^{\dagger}_{s}(-\mbf{k})\alpha_i u_{s^{\prime}}(\mbf{q}))
(u^{\dagger}_{s^{\prime}}(\mbf{q})\alpha_j v_{s}(-\mbf{k}))
2\eta_0(\mbf{k},\mbf{q},\mbf{k}-\mbf{q})
g_0(\mbf{k},\mbf{q},\mbf{k}-\mbf{q})\right)
d_s^{\dagger}(-\mbf{k})d_s(-\mbf{k})
\nonumber
\,.\end{eqnarray}
Integrating these flow equations with the generators and coupling constants
given by Eq. (\ref{eq:2.16}) and Eq. (\ref{eq:2.18}), respectively,
produces the second order self-energy correction,
$\delta\Sigma=\Sigma(l)-\Sigma(l_0=0)$. The self-energy operator,
at the scale $\lambda$ with $l=1/\lambda$, is given by
\begin{eqnarray}
\Sigma_{gen}(\lambda)&=& \sum_{s}\int 
\frac{d\mbf{k}d\mbf{q}}{(2\pi)^6}
C_f g^2 \frac{D_{ij}(\mbf{k}-\mbf{q})}{2\omega(\mbf{k}-\mbf{q})}
\label{eq:b13} \\
&\times& \left( \sum_{s^{\prime}}
(u^{\dagger}_{s}(\mbf{k})\alpha_i u_{s^{\prime}}(\mbf{q}))
(u^{\dagger}_{s^{\prime}}(\mbf{q})\alpha_j u_{s}(\mbf{k}))
\frac{1}{D_1}{\rm e}^{(-2D_1^2/\lambda^2)} \right.
\nonumber\\
&-& \left.\sum_{s^{\prime}}
(u^{\dagger}_{s}(\mbf{k})\alpha_i v_{s^{\prime}}(-\mbf{q}))
(v^{\dagger}_{s^{\prime}}(-\mbf{q})\alpha_j u_{s}(\mbf{k}))
\frac{1}{D_0}{\rm e}^{(-2D_0^2/\lambda^2)} \right)
b_s^{\dagger}(\mbf{k})b_s(\mbf{k})\nonumber
\,,\end{eqnarray}
and
\begin{eqnarray}
\Sigma_{gen}(\lambda)&=& \sum_{s}\int 
\frac{d\mbf{k}d\mbf{q}}{(2\pi)^6}
C_f g^2 \frac{D_{ij}(\mbf{k}-\mbf{q})}{2\omega(\mbf{k}-\mbf{q})}
\label{eq:b14}\\
&\times& \left( \sum_{s^{\prime}}
(v^{\dagger}_{s}(-\mbf{k})\alpha_i v_{s^{\prime}}(-\mbf{q}))
(v^{\dagger}_{s^{\prime}}(-\mbf{q})\alpha_j v_{s}(-\mbf{k}))
\frac{1}{D_1}{\rm e}^{(-2D_1^2/\lambda^2)} \right.
\nonumber\\
&-& \left.\sum_{s^{\prime}}
(v^{\dagger}_{s}(-\mbf{k})\alpha_i u_{s^{\prime}}(\mbf{q}))
(u^{\dagger}_{s^{\prime}}(\mbf{q})\alpha_j v_{s}(-\mbf{k}))
\frac{1}{D_0}{\rm e}^{(-2D_0^2/\lambda^2)} \right)
d_s^{\dagger}(-\mbf{k})d_s(-\mbf{k})\nonumber
\,,\end{eqnarray}
with 
\begin{eqnarray}
D_0 &=& -(E(\mbf{k})+E(\mbf{q})+\omega(\mbf{k}-\mbf{q}))
\nonumber\\
D_1 &=& E(\mbf{k})-E(\mbf{q})-\omega(\mbf{k}-\mbf{q})
\,.\label{eq:b15}\end{eqnarray}
In the off-diagonal one-quark sector ($bd$ and $b^{\dagger}d^{\dagger}$)
the flow equations are given by  
\begin{eqnarray}
\frac{d\Sigma_{gen}(l)}{dl} &=& \sum_{s}\int 
\frac{d\mbf{k}d\mbf{q}}{(2\pi)^6}
C_f \frac{D_{ij}(\mbf{k}-\mbf{q})}{2\omega(\mbf{k}-\mbf{q})}
\label{eq:b16}\\
&\times& \left( \sum_{s^{\prime}}
(v^{\dagger}_{s}(-\mbf{k})\alpha_i v_{s^{\prime}}(-\mbf{q}))
(v^{\dagger}_{s^{\prime}}(-\mbf{q})\alpha_j u_{s}(\mbf{k}))
-\sum_{s^{\prime}}
(v^{\dagger}_{s}(-\mbf{k})\alpha_i u_{s^{\prime}}(\mbf{q}))
(u^{\dagger}_{s^{\prime}}(\mbf{q})\alpha_j u_{s}(\mbf{k})) \right)
\nonumber\\
&\times&
[\eta_0(\mbf{k},\mbf{q},\mbf{k}-\mbf{q})
g_1(\mbf{k},\mbf{q},\mbf{k}-\mbf{q})
+\eta_1(\mbf{k},\mbf{q},\mbf{k}-\mbf{q})
g_0(\mbf{k},\mbf{q},\mbf{k}-\mbf{q})]
b_s(\mbf{k})d_s(-\mbf{k})\nonumber
\,,\end{eqnarray}
and
\begin{eqnarray}
\frac{d\Sigma_{gen}(l)}{dl} &=& \sum_{s}\int 
\frac{d\mbf{k}d\mbf{q}}{(2\pi)^6}
C_f \frac{D_{ij}(\mbf{k}-\mbf{q})}{2\omega(\mbf{k}-\mbf{q})}
\label{eq:b17}\\
&\times& \left( \sum_{s^{\prime}}
(u^{\dagger}_{s}(\mbf{k})\alpha_i u_{s^{\prime}}(\mbf{q}))
(u^{\dagger}_{s^{\prime}}(\mbf{q})\alpha_j v_{s}(-\mbf{k}))
-\sum_{s^{\prime}}
(u^{\dagger}_{s}(\mbf{k})\alpha_i v_{s^{\prime}}(-\mbf{q}))
(v^{\dagger}_{s^{\prime}}(-\mbf{q})\alpha_j v_{s}(-\mbf{k})) \right)
\nonumber\\
&\times&
[\eta_0(\mbf{k},\mbf{q},\mbf{k}-\mbf{q})
g_1(\mbf{k},\mbf{q},\mbf{k}-\mbf{q})
+\eta_1(\mbf{k},\mbf{q},\mbf{k}-\mbf{q})
g_0(\mbf{k},\mbf{q},\mbf{k}-\mbf{q})]
b^{\dagger}_s(\mbf{k})d^{\dagger}_s(-\mbf{k})\nonumber
\,,\end{eqnarray}
where again the symmetry property of $\eta_0$ and $g_0$,
when interchanging any of their two arguments, was used.
The self-energy operator,
at the scale $l=1/\lambda^2$, is given by
\begin{eqnarray}
\Sigma_{gen}(\lambda) &=& \sum_{s}\int 
\frac{d\mbf{k}d\mbf{q}}{(2\pi)^6}
C_f g^2 \frac{D_{ij}(\mbf{k}-\mbf{q})}{2\omega(\mbf{k}-\mbf{q})}
\nonumber\\
&\times& \left( \sum_{s^{\prime}}
(v^{\dagger}_{s}(-\mbf{k})\alpha_i u_{s^{\prime}}(\mbf{q}))
(u^{\dagger}_{s^{\prime}}(\mbf{q})\alpha_j u_{s}(\mbf{k}))
-\sum_{s^{\prime}}
(v^{\dagger}_{s}(-\mbf{k})\alpha_i v_{s^{\prime}}(-\mbf{q}))
(v^{\dagger}_{s^{\prime}}(-\mbf{q})\alpha_j u_{s}(\mbf{k})) \right)
\nonumber\\
&\times&
\frac{D_0+D_1}{D_0^2+D_1^2}{\rm e}^{-(D_0^2+D_1^2)/\lambda^2}
d_s(-\mbf{k})b_s(\mbf{k})
\,,\label{eq:b18}\end{eqnarray}
and
\begin{eqnarray}
\Sigma_{gen}(\lambda) &=& \sum_{s}\int 
\frac{d\mbf{k}d\mbf{q}}{(2\pi)^6}
C_f g^2 \frac{D_{ij}(\mbf{k}-\mbf{q})}{2\omega(\mbf{k}-\mbf{q})}
\nonumber\\
&\times& \left( \sum_{s^{\prime}}
(u^{\dagger}_{s}(\mbf{k})\alpha_i u_{s^{\prime}}(\mbf{q}))
(u^{\dagger}_{s^{\prime}}(\mbf{q})\alpha_j v_{s}(-\mbf{k}))
-\sum_{s^{\prime}}
(u^{\dagger}_{s}(\mbf{k})\alpha_i v_{s^{\prime}}(-\mbf{q}))
(v^{\dagger}_{s^{\prime}}(-\mbf{q})\alpha_j v_{s}(-\mbf{k})) \right)
\nonumber\\
&\times&
\frac{D_0+D_1}{D_0^2+D_1^2}{\rm e}^{-(D_0^2+D_1^2)/\lambda^2}
b^{\dagger}_s(\mbf{k})d^{\dagger}_s(-\mbf{k})
\,,\label{eq:b19}\end{eqnarray}
with the energy denominators defined in Eq. (\ref{eq:b15}).
Using the spinors, Eq. (\ref{eq:2.7}), and the polarization sum, 
Eq. (\ref{eq:2.10}), the self-energy operators are simplified to
\begin{eqnarray}
\Sigma_{gen}(\lambda) &=& \sum_{s}\int 
\frac{d\mbf{k}d\mbf{q}}{(2\pi)^6}\frac{1}{2}
\left( (1+s(\mbf{k})s(\mbf{q})+c(\mbf{k})c(\mbf{q})
\mbf{\hat{k}}\cdot\mbf{\hat{l}}\mbf{\hat{q}}\cdot\mbf{\hat{l}})
\frac{1}{D_0}{\rm e}^{-2D_0^2/\lambda^2}\right.
\nonumber\\
&-& \left.
(1-s(\mbf{k})s(\mbf{q})-c(\mbf{k})c(\mbf{q})
\mbf{\hat{k}}\cdot\mbf{\hat{l}}\mbf{\hat{q}}\cdot\mbf{\hat{l}})
\frac{1}{D_1}{\rm e}^{-2D_1^2/\lambda^2}
\right) (-\frac{C_fg^2}{\omega(\mbf{l})})
\nonumber\\
&\times& (b_s^{\dagger}(\mbf{k})b_s(\mbf{k})
+d_s^{\dagger}(-\mbf{k})d_s(-\mbf{k}))
\nonumber\\
&+& \sum_{s}\int 
\frac{d\mbf{k}d\mbf{q}}{(2\pi)^6}\frac{1}{2}
\left( 2 (-c(\mbf{k})s(\mbf{q})+s(\mbf{k})c(\mbf{q})
\mbf{\hat{k}}\cdot\mbf{\hat{l}}\mbf{\hat{q}}\cdot\mbf{\hat{l}})
\right)
\frac{D_0+D_1}{D_0^2+D_1^2}{\rm e}^{-(D_0^2+D_1^2)/\lambda^2}
(-\frac{C_fg^2}{\omega(\mbf{l})})
\nonumber\\
&\times& (b_s^{\dagger}(\mbf{k})d_s^{\dagger}(-\mbf{k})
+d_s(-\mbf{k})b_s(\mbf{k}))
\,,\label{eq:b20}\end{eqnarray}
with
\begin{eqnarray}
D_0 &=& -(E(\mbf{k})+E(\mbf{q})+\omega(\mbf{l}))
\nonumber\\
D_1 &=& E(\mbf{k})-E(\mbf{q})-\omega(\mbf{l})
\,,\label{eq:b21}\end{eqnarray}
and $\mbf{l}=\mbf{k}-\mbf{q}$.
For large momenta flowing in the loop, one has 
$D_0\sim D_1\sim -(E(\mbf{q})+\omega(\mbf{l}))\sim -2\omega(\mbf{q})=-2|\mbf{q}|$.
In this limit the self-energy operator, Eq. (\ref{eq:b20}), is reduced to
\begin{eqnarray}
\Sigma_{gen}(\lambda) &=& \sum_{s}\int 
\frac{d\mbf{k}d\mbf{q}}{(2\pi)^6}
\left( s(\mbf{k})s(\mbf{q})+c(\mbf{k})c(\mbf{q})
\mbf{\hat{k}}\cdot\mbf{\hat{l}}\mbf{\hat{q}}\cdot\mbf{\hat{l}} \right)
\left(-\frac{C_f g^2}{D\omega(\mbf{l})}\right)
{\rm e}^{-2D^2/\lambda^2}
\nonumber\\
&\times& (b_s^{\dagger}(\mbf{k})b_s(\mbf{k})
+d_s^{\dagger}(-\mbf{k})d_s(-\mbf{k}))
\nonumber\\
&+& \sum_{s}\int 
\frac{d\mbf{k}d\mbf{q}}{(2\pi)^6}
\left( -c(\mbf{k})s(\mbf{q})+s(\mbf{k})c(\mbf{q})
\mbf{\hat{k}}\cdot\mbf{\hat{l}}\mbf{\hat{q}}\cdot\mbf{\hat{l}} 
\right)
\left(-\frac{C_f g^2}{D\omega(\mbf{l})}\right)
{\rm e}^{-2D^2/\lambda^2}
\nonumber\\
&\times& (b_s^{\dagger}(\mbf{k})d_s^{\dagger}(-\mbf{k})
+d_s(-\mbf{k})b_s(\mbf{k}))
\,,\label{eq:b22}\end{eqnarray}
where $\mbf{l}=\mbf{k}-\mbf{q}$, and the energy denominator is
\begin{eqnarray}
D &=& (E(\mbf{q})+\omega(\mbf{l}))
\,.\label{eq:b23}\end{eqnarray}
This can be simplified further
\begin{eqnarray}
\Sigma_{gen}(\lambda) &=& \sum_{s}\int 
\frac{d\mbf{k}d\mbf{q}}{(2\pi)^6}
\left( s(\mbf{k})s(\mbf{q})+c(\mbf{k})c(\mbf{q})
\mbf{\hat{k}}\cdot\mbf{\hat{l}}\mbf{\hat{q}}\cdot\mbf{\hat{l}} \right)
(\frac{C_f g^2}{(E(\mbf{q})+\omega(\mbf{l}))\omega(\mbf{l})})
{\rm e}^{-4q^2/\lambda^2}
\nonumber\\
&\times& (b_s^{\dagger}(\mbf{k})b_s(\mbf{k})
+d_s^{\dagger}(-\mbf{k})d_s(-\mbf{k}))
\nonumber\\
&+& \sum_{s}\int 
\frac{d\mbf{k}d\mbf{q}}{(2\pi)^6}
\left( -c(\mbf{k})s(\mbf{q})+s(\mbf{k})c(\mbf{q})
\mbf{\hat{k}}\cdot\mbf{\hat{l}}\mbf{\hat{q}}\cdot\mbf{\hat{l}} 
\right)
(\frac{C_f g^2}{(E(\mbf{q})+\omega(\mbf{l}))\omega(\mbf{l})})
{\rm e}^{-4q^2/\lambda^2}
\nonumber\\
&\times& (b_s^{\dagger}(\mbf{k})d_s^{\dagger}(-\mbf{k})
+d_s(-\mbf{k})b_s(\mbf{k}))
\,,\label{eq:b24}\end{eqnarray}
where we rescaled the cut-off, $\lambda\rightarrow\sqrt{2}\lambda$,
that does not change the result. We regulate the self-energy operator
arising by normal ordering the instantaneous interactions Eq. (\ref{eq:a15})
\begin{eqnarray}
\Sigma_{inst}(\lambda) &=& 
\sum_s\int\frac{d\mbf{k}d\mbf{q}}{(2\pi)^6}\left(
s(\mbf{k})s(\mbf{q})+c(\mbf{k})c(\mbf{q})\mbf{\hat{k}}\cdot\mbf{\hat{q}}
\right)C_f V_{L+C}(\mbf{k},\mbf{q}){\rm e}^{-q^2/\lambda^2}
\nonumber\\ 
&\times& 
[b^{\dagger}_s(\mbf{k})b_s(\mbf{k}) + d^{\dagger}_s(-\mbf{k})d_s(-\mbf{k})]
\nonumber\\
&+& \sum_s\int\frac{d\mbf{k}d\mbf{q}}{(2\pi)^6}\left(
-c(\mbf{k})s(\mbf{q})+s(\mbf{k})c(\mbf{q})\mbf{\hat{k}}\cdot\mbf{\hat{q}}
\right)C_f V_{L+C}(\mbf{k},\mbf{q}){\rm e}^{-q^2/\lambda^2}
\nonumber\\
&\times&
[b^{\dagger}_s(\mbf{k})d^{\dagger}_s({-\bf k}) + d_s({-\bf k})b_s(\mbf{k})]
\,,\label{eq:b25}\end{eqnarray}
where the same regulating function is chosen to match the energy denominators.
We followed the regularization prescription suggested by Zhang and Harindranath 
\cite{ZhangHarindranath}, where the divergent instantaneous terms arising after 
normal-ordering are regulated using the same cut-off function, i.e. an exponent, 
as in the divergent dynamical terms, while the argument in the exponent 
matches the energy denominator of the interaction.  
The complete self-energy operator is
\begin{eqnarray}
\Sigma(\lambda) &=&\Sigma_{inst}(\lambda)+\Sigma_{gen}(\lambda)
\,.\label{eq:b26}\end{eqnarray}

\subsection{Quark condensate}
\label{app:B3}

The second order flow equation, Eq. (\ref{eq:2.19}), for the condensate term is
\begin{eqnarray}
\frac{dO_{gen}(l)}{dl} &=& N_c\mbf{V}\int 
\frac{d\mbf{k}d\mbf{q}}{(2\pi)^6}
C_f \frac{D_{ij}(\mbf{k}-\mbf{q})}{2\omega(\mbf{k}-\mbf{q})}
\label{eq:b27} \\
&\times& \left( \sum_{s,s^{\prime}}
(v^{\dagger}_{s}(-\mbf{k})\alpha_i u_{s^{\prime}}(\mbf{q}))
(u^{\dagger}_{s^{\prime}}(\mbf{q})\alpha_j v_{s}(-\mbf{k}))
2\eta_0(\mbf{k},\mbf{q},\mbf{k}-\mbf{q})
g_0(\mbf{k},\mbf{q},\mbf{k}-\mbf{q})
\right) \nonumber
\,,\end{eqnarray}
where the volume is $\mbf{V}=(2\pi)^3\delta^{(3)}(0)$.
Integration yields the correction, $\delta O=O(l)-O(l_0=0)$, where $O(l)$
is a condensate for the flow parameter $l$, related to the energy scale $\lambda$
by $l=1/\lambda^2$. The resulting generated condensate term through 
second order is
\begin{eqnarray}
O_{gen}(\lambda) &=& N_c\mbf{V}\int 
\frac{d\mbf{k}d\mbf{q}}{(2\pi)^6}
C_f g^2 \frac{D_{ij}(\mbf{k}-\mbf{q})}{2\omega(\mbf{k}-\mbf{q})}
\nonumber\\
&\times& \left( \sum_{s,s^{\prime}}
(v^{\dagger}_{s}(-\mbf{k})\alpha_i u_{s^{\prime}}(\mbf{q}))
(u^{\dagger}_{s^{\prime}}(\mbf{q})\alpha_j v_{s}(-\mbf{k}))\right)
\frac{1}{D_0}{\rm e}^{-2D_0^2/\lambda^2}
\,,\label{eq:b28}\end{eqnarray}
which, using the spinors Eq. (\ref{eq:2.7}), is reduced 
\begin{eqnarray}
O_{gen}(\lambda) &=& -2 N_c\mbf{V}\int 
\frac{d\mbf{k}d\mbf{q}}{(2\pi)^6}
(1+s(\mbf{k})s(\mbf{q})+c(\mbf{k})c(\mbf{q})
\mbf{\hat{k}}\cdot\mbf{\hat{l}}\mbf{\hat{q}}\cdot\mbf{\hat{l}})
{\rm e}^{-2D_0^2/\lambda^2}\left(-\frac{C_fg^2}{D_0\omega(\mbf{l})}\right)\,,
\nonumber\\
\label{eq:b29}\end{eqnarray}
where $\mbf{l}=\mbf{k}-\mbf{q}$, and the energy denominator is
\begin{eqnarray}
D_0 &=& -(E(\mbf{k})+E(\mbf{q})+\omega(\mbf{l}))
\,.\label{eq:b30}\end{eqnarray}
The regulated condensate terms, generated and instantaneous,
can be summarized 
\begin{eqnarray}
O_{gen}(\lambda) &=& -2 N_c\mbf{V}\int 
\frac{d\mbf{k}d\mbf{q}}{(2\pi)^6}
(1+s(\mbf{k})s(\mbf{q})+c(\mbf{k})c(\mbf{q})
\mbf{\hat{k}}\cdot\mbf{\hat{l}}\mbf{\hat{q}}\cdot\mbf{\hat{l}})
{\rm e}^{-(q+k+l)^2/\lambda^2}
\nonumber\\
&\times& \left(\frac{C_fg^2}{(E(\mbf{k})+E(\mbf{q})+\omega(\mbf{k}-\mbf{q}))
\omega(\mbf{k}-\mbf{q})}\right)
\,,\label{eq:b31}\end{eqnarray}
and, from Eq. (\ref{eq:a17}), 
\begin{eqnarray}
O_{inst}(\lambda) &=& 2 N_c\mbf{V}\int 
\frac{d\mbf{k}d\mbf{q}}{(2\pi)^6}
(1-s(\mbf{k})s(\mbf{q})-c(\mbf{k})c(\mbf{q})
\mbf{\hat{k}}\cdot\mbf{\hat{q}})
{\rm e}^{-(q+k)^2/\lambda^2}
C_f V(\mbf{k},\mbf{q})\,.\nonumber\\
\label{eq:b32}\end{eqnarray}
The regulating procedure is the same as above (See \cite{ZhangHarindranath}). 
Note that in the exponential factors we have used free dispersion relation, 
$E(\mbf{q})\sim\omega(\mbf{q})=|\mbf{q}|$, valid for large cut-off values.
The complete radiative correction to the quark condensate, Eq. (\ref{eq:a8}),
is
\begin{eqnarray}
O(\lambda) &=& O_{inst}(\lambda)+O_{gen}(\lambda)
\,.\label{eq:b33}\end{eqnarray}

\section{Bogoliubov-Valatin transformation and the gap equation 
with double normal ordering}
\label{app:C}

The Bogoliubov-Valatin transformation for the pairing (BCS) model
relates the operators $b,d$ which annihilate bare vacuum
$|0\rangle$ to a new basis set $B,D$ which annihilate vacuum state
$|\Omega\rangle$, containing the quark condensate, 
\begin{eqnarray}
B_s(\mbf{k}) &=& c_{\Theta/2}(\mbf{k})b_s(\mbf{k})
- h(s) s_{\Theta/2}d_s^{\dagger}(-\mbf{k})
\nonumber\\
D_s(-\mbf{k}) &=& c_{\Theta/2}(\mbf{k})d_s(-\mbf{k})
+ h(s) s_{\Theta/2}b_s^{\dagger}(\mbf{k})
\,,\label{eq:c1}\end{eqnarray}
where $\Theta=\Theta(\mbf{k})$ is the Bogoliubov-Valatin
(BCS) angle, and $h(s)$ is the helicity. Similarly the transformed
quasiparticle spinors are
\begin{eqnarray}
U_s(\mbf{k}) &=& c_{\Theta/2}(\mbf{k})u_s(\mbf{k})
- h(s) s_{\Theta/2}v_s(-\mbf{k})
\nonumber\\
V_s(-\mbf{k}) &=& c_{\Theta/2}(\mbf{k})v_s(-\mbf{k})
+ h(s) s_{\Theta/2}u_s(\mbf{k})
\,,\label{eq:c2}\end{eqnarray}
where the bare particle and quasiparticle spinors are defined,
respectively, as
\begin{eqnarray}
 u_s(\mbf{k}) &=& \frac{1}{\sqrt{2}}\left(
  \begin{array}{c}
  \sqrt{1+s_{\alpha}(\mbf{k})} \chi_{s}\\
  \sqrt{1-s_{\alpha}(\mbf{k})}(\mbf{\sigma}\cdot\hat{\mbf{k}})\chi_{s}
  \end{array} \right),
  \nonumber\\
 v_s(-\mbf{k}) &=& \frac{1}{\sqrt{2}}\left(
  \begin{array}{c}
  -\sqrt{1-s_{\alpha}(\mbf{k})}(\mbf{\sigma}\cdot\hat{\mbf{k}})\eta_{-s}\\
  \sqrt{1+s_{\alpha}(\mbf{k})}\eta_{-s}
  \end{array} \right)  
\,,\label{eq:c3}\end{eqnarray}
and
\begin{eqnarray}
 U_s(\mbf{k}) &=& \frac{1}{\sqrt{2}}\left(
  \begin{array}{c}
  \sqrt{1+s_{\Phi}(\mbf{k})} \chi_{s}\\
  \sqrt{1-s_{\Phi}(\mbf{k})}(\mbf{\sigma}\cdot\hat{\mbf{k}})\chi_{s}
  \end{array} \right),
  \nonumber\\
 V_s(-\mbf{k}) &=& \frac{1}{\sqrt{2}}\left(
  \begin{array}{c}
  -\sqrt{1-s_{\Phi}(\mbf{k})}(\mbf{\sigma}\cdot\hat{\mbf{k}})\eta_{-s}\\
  \sqrt{1+s_{\Phi}(\mbf{k})}\eta_{-s}
  \end{array} \right)  
\,.\label{eq:c4}\end{eqnarray}
Here, $\chi_s$ and $\eta_s$ are the standard two-component Pauli spinors
of a particle and an antiparticle, respectively, with
$\eta_{-s}=-i\sigma_2\chi_s$. We include masses in the definition of spinors, 
therefore the perturbative ($\alpha=\alpha(\mbf{k})$) and nonperturbative
($\Phi=\Phi(\mbf{k})$) angles appear in the sine and cosine as follows
\begin{eqnarray}
\sin(\alpha(\mbf{k})) \equiv s_{\alpha}(\mbf{k}) = \frac{m}{\sqrt{k^2+m^2}}\,,\,
\cos(\alpha(\mbf{k})) \equiv c_{\alpha}(\mbf{k}) = \frac{k}{\sqrt{k^2+m^2}}
\,,\label{eq:c5}\end{eqnarray}
and
\begin{eqnarray}
\sin(\Phi(\mbf{k})) \equiv s_{\Phi}(\mbf{k}) = \frac{M(\mbf{k})}{\sqrt{k^2+M^2(\mbf{k})}}\,,\,
\cos(\Phi(\mbf{k})) \equiv c_{\Phi}(\mbf{k}) = \frac{k}{\sqrt{k^2+M^2(\mbf{k})}}
\,,\label{eq:c6}\end{eqnarray}
where $m$ is the bare mass and $M(\mbf{k})$ is the effective masses.
The nonperturbative angle $\Phi$ is related to the Bogoliubov-Valatin
angle $\Theta/2$, given in Eq. (\ref{eq:c1}), and the perturbative angle
$\alpha$, Eq. (\ref{eq:c5}), as 
\begin{eqnarray}
\Phi &=& \alpha + \Theta
\,,\label{eq:c7}\end{eqnarray}
So that the following holds 
\begin{eqnarray}
s_{\Phi}(\mbf{k}) &=& \frac{m}{\sqrt{k^2+m^2}}c_{\Theta}(\mbf{k})
+\frac{k}{\sqrt{k^2+m^2}}s_{\Theta}(\mbf{k})
\nonumber\\
c_{\Phi}(\mbf{k}) &=& \frac{k}{\sqrt{k^2+m^2}}c_{\Theta}(\mbf{k})
-\frac{m}{\sqrt{k^2+m^2}}s_{\Theta}(\mbf{k})
\,.\label{eq:c8}\end{eqnarray}

To obtain the gap equation with double normal ordering we first decompose
the Hamiltonian, Eq. (\ref{eq:2.1}), in the perturbative basis, 
using (anti)quark $b,d$ operators with spinors $u,v$, Eq. (\ref{eq:c3}),
and the perturbative angle $\alpha$, Eq. (\ref{eq:c5}),
and then normal order $H$ with respect to the perturbative vacuum $|0\rangle$.
We obtain the terms summarized in Appendix \ref{app:A}, replacing the angle 
$\Phi$ by $\alpha$. Next we perform the Bogoliubov-Valatin transformation
from the perturbative to the BCS vacuum, $|0\rangle\rightarrow|\Omega\rangle$,
expressing the quark operators $b,d$ through $B,D$, Eq. (\ref{eq:c1}). 
Condensate terms do not change, one-body operators transform as
\begin{eqnarray}
&& (b_{s}^{\dagger}b_{s}+d_{s}^{\dagger}d_{s})\rightarrow
c_{\Theta}(B_{s}^{\dagger}B_{s}+D_{s}^{\dagger}D_{s})
+h(s)s_{\Theta}(B_{s}^{\dagger}D_{s}^{\dagger}+D_{s}B_{s})
\nonumber\\ 
&& (b_{s}^{\dagger}d_{s}^{\dagger}+d_{s}b_{s})\rightarrow
c_{\Theta}(B_{s}^{\dagger}D_{s}^{\dagger}+D_{s}B_{s})
-h(s)s_{\Theta}(B_{s}^{\dagger}B_{s}+D_{s}^{\dagger}D_{s})
\,,\label{eq:c9}\end{eqnarray}
and two-body interactions, expressed only by the angle $\Phi$
using Eq. (\ref{eq:c8}), coincide with the interaction terms 
of Eq. (\ref{eq:a12}). One should be able to combine
the angles $\alpha$ and $\Theta$ into $\Phi$ for the two-body interactions,
unless a mistake is done. Now we normal order the obtained Hamiltonian
with respect to the new vacuum $|\Omega\rangle$ and obtain an additional 
set of terms. In particular, by normal ordering the BV transformed
two-body interactions, we have the same terms as obtained before 
Eq. (\ref{eq:a15}) in one-body sector. These terms depend only on $\Phi$. 
In addition, there are one-body operators obtained by normal-ordering 
in $|0\rangle$ and BV transformed Eq. (\ref{eq:c9}), which depend on $\alpha$ 
and $\Phi$, and also the BV transformed kinetic terms, depending only on $\Phi$.
Combining all the terms, we obtain the gap equation 
\begin{eqnarray}
&& ks_{\Phi}(\mbf{k})-mc_{\Phi}(\mbf{k})
\nonumber\\
&=& \int\frac{d\mbf{q}}{(2\pi)^3} C_f V_{L+C}(\mbf{k},\mbf{q})
\left( c_{\Phi}(\mbf{k})[s_{\Phi}(\mbf{q})-s_{\alpha}(\mbf{q})] 
- s_{\Phi}(\mbf{k})[c_{\Phi}(\mbf{q})-c_{\alpha}(\mbf{q})]
\mbf{\hat{k}}\cdot\mbf{\hat{q}} \right)
\label{eq:c10} \\
&+& \int\frac{d\mbf{q}}{(2\pi)^3} C_f W(\mbf{k},\mbf{q})
\left( c_{\Phi}(\mbf{k})[s_{\Phi}(\mbf{q})-s_{\alpha}(\mbf{q})]
- s_{\Phi}(\mbf{k})[c_{\Phi}(\mbf{q})-c_{\alpha}(\mbf{q})]
\mbf{\hat{k}}\cdot\mbf{\hat{l}}\mbf{\hat{q}}\cdot\mbf{\hat{l}} \right)
\nonumber
\,,\end{eqnarray}
where the generated terms are also added.
One can express this equation in terms of $\alpha$ and $\Theta$ angles,
using Eq. (\ref{eq:c8}), as
\begin{eqnarray}
E(\mbf{k})s_{\Theta}(\mbf{k})&=&
-\int\frac{d\mbf{q}}{(2\pi)^3}
\left( \Sigma^{cc}c_{\Theta}(\mbf{k})[c_{\Theta}(\mbf{q})-1]
+ \Sigma^{cs}c_{\Theta}(\mbf{k})s_{\Theta}(\mbf{q})
\right. \nonumber\\
&+& \left. \Sigma^{sc}s_{\Theta}(\mbf{k})[c_{\Theta}(\mbf{q})-1]
+ \Sigma^{ss}s_{\Theta}(\mbf{k})s_{\Theta}(\mbf{q}) \right)
\,,\label{eq:c11}\end{eqnarray}
where the self-energy terms include the instantaneous and generated
contributions
\begin{eqnarray}
\Sigma &=& \Sigma_{inst}+\Sigma_{gen}
\,.\label{eq:c12}\end{eqnarray}
The instantaneous self-energies are given only in terms of $\alpha$
\begin{eqnarray}
\Sigma^{cc}_{inst} &=& (-c_{\alpha}(\mbf{k})s_{\alpha}(\mbf{q})
+s_{\alpha}(\mbf{k})c_{\alpha}(\mbf{q})\mbf{\hat{k}}\cdot\mbf{\hat{q}})
C_fV_{L+C}(\mbf{k},\mbf{q})
\nonumber\\
\Sigma^{cs}_{inst} &=& (-c_{\alpha}(\mbf{k})c_{\alpha}(\mbf{q})
-s_{\alpha}(\mbf{k})s_{\alpha}(\mbf{q})\mbf{\hat{k}}\cdot\mbf{\hat{q}})
C_fV_{L+C}(\mbf{k},\mbf{q})
\nonumber\\
\Sigma^{sc}_{inst} &=& (s_{\alpha}(\mbf{k})s_{\alpha}(\mbf{q})
+c_{\alpha}(\mbf{k})c_{\alpha}(\mbf{q})\mbf{\hat{k}}\cdot\mbf{\hat{q}})
C_fV_{L+C}(\mbf{k},\mbf{q})
\nonumber\\
\Sigma^{ss}_{inst} &=& (s_{\alpha}(\mbf{k})c_{\alpha}(\mbf{q})
-c_{\alpha}(\mbf{k})s_{\alpha}(\mbf{q})\mbf{\hat{k}}\cdot\mbf{\hat{q}})
C_fV_{L+C}(\mbf{k},\mbf{q})
\,,\label{eq:c13}\end{eqnarray}
and the following changes should be made in these formulas
to obtain the generated self-energies, $\Sigma_{gen}$;
\begin{eqnarray}
\mbf{\hat{k}}\cdot\mbf{\hat{q}}
\rightarrow \mbf{\hat{k}}\cdot\mbf{\hat{l}}\mbf{\hat{q}}\cdot\mbf{\hat{l}}
\,,\label{eq:c14}\end{eqnarray}
and 
\begin{eqnarray}
V_{L+C}(\mbf{k},\mbf{q})\rightarrow W(\mbf{k},\mbf{q})
\,.\label{eq:c15}\end{eqnarray}

The double normal ordered gap equation, Eqs. (\ref{eq:c10}) and (\ref{eq:c11}), 
is UV finite even with the Coulomb potential, 
$V_{L+C}\sim (\mbf{k}-\mbf{q})^{-2}$, alone. 
For large momenta, $\mbf{q}\sim\Lambda$,
$\Phi({\mbf{q}})\rightarrow \alpha({\mbf{q}})$ and 
$\Theta({\mbf{q}})\rightarrow 0$, i.e. $c_{\Theta}(\mbf{q})\rightarrow 1$
and $s_{\Theta}(\mbf{q})\rightarrow 0$, and terms coming from normal ordering
in $|\Omega\rangle$ and in $|0\rangle$ cancel each other.
Indeed, for high UV momenta two vacua $|0\rangle$ and $|\Omega\rangle$
are viewed as the same, giving the same result of normal-ordered terms
but with opposite signs.
However this gap equation is IR infinite for the confining potential,
$V_{L+C}\sim (\mbf{k}-\mbf{q})^{-4}$. Cancelation of the leading IR divergence 
does not happen, as it happened with single normal ordering in $|\Omega\rangle$,
Eqs. (\ref{eq:3.24}) and (\ref{eq:3.25}). 
At $\mbf{k}\sim \mbf{q}$ the r.h.s. of Eq. (\ref{eq:c10}) behaves
\begin{eqnarray}
\int\frac{d\mbf{q}}{(2\pi)^3} C_f V_{L+C}(\mbf{k},\mbf{q})
s_{\Theta}(\mbf{k})\sim \int d\mbf{q}|\mbf{k}-\mbf{q}|^{-4}
\,,\label{eq:c16}\end{eqnarray}
and diverges. We conclude that the prescription of double normal 
ordering leads to divergencies for some potentials and is artificial.

\section{Pion mass in the BCS model and in the method of flow equations}
\label{app:D}

In this appendix we check if pion $0^{++}$ can have zero mass
in BCS and flow equation approaches. RPA bound state equation, 
Eq. (\ref{eq:4.26}), reads
\begin{eqnarray} 
&& \left(A-B\right)(\mbf{k},\mbf{q})\psi_{-}(\mbf{q})=M\psi_{+}(\mbf{k})
\nonumber\\
&& A(\mbf{k},\mbf{q}) = 2\epsilon(\mbf{k})
-d\mbf{q}F_{xx}(\mbf{k},\mbf{q})
\nonumber\\
&& B(\mbf{k},\mbf{q}) = -d\mbf{q}F_{xy}(\mbf{k},\mbf{q})
\,,\label{eq:d1}\end{eqnarray}
where tensors $F$ include the instantaneous $I$ and 
generated $G$ terms, $F=I+G$. 
In the $\pi$ channel, using Eq. (\ref{eq:4.18}) for $F$,
\begin{eqnarray}
I_{xx}(\mbf{k},\mbf{q}) &=&  
C_fV_{L+C}(\mbf{k}\mbf{q})\left[\,1+s(\mbf{k})s(\mbf{q})
+c(\mbf{k})c(\mbf{q})x\,\right]
\label{eq:d2}\\
I_{xy}(\mbf{k},\mbf{q}) &=& 
C_fV_{L+C}(\mbf{k}\mbf{q})\left[\,-1+s(\mbf{k})s(\mbf{q})
+c(\mbf{k})c(\mbf{q})x\,\right]
\nonumber\\
G_{xx}(\mbf{k},\mbf{q}) &=&  
C_f\tilde{W}(\mbf{k}\mbf{q})\left[\,-1+s(\mbf{k})s(\mbf{q})
+c(\mbf{k})c(\mbf{q})\frac{x(k^2+q^2)-(1+x^2)kq}{(\mbf{k}-\mbf{q})^2}\,\right]
\nonumber\\
G_{xy}(\mbf{k},\mbf{q}) &=& 
C_f\tilde{W}(\mbf{k}\mbf{q})\left[\,1+s(\mbf{k})s(\mbf{q})
+c(\mbf{k})c(\mbf{q})\frac{x(k^2+q^2)-(1+x^2)kq}{(\mbf{k}-\mbf{q})^2}\,\right]
\nonumber
\,,\end{eqnarray}   
and, from Eq. (\ref{eq:3.15}) for $\varepsilon$ in the chiral limit $m=0$, 
\begin{eqnarray}
\varepsilon(\mbf{k}) &=& kc(\mbf{k})
+\int d\mbf{q}C_fV_{L+C}(\mbf{k},\mbf{q})
\left[\,s(\mbf{k})s(\mbf{q})+c(\mbf{k})c(\mbf{q})x\,\right]
\nonumber\\
&+&\int d\mbf{q}C_fW(\mbf{k},\mbf{q})
\left[\,s(\mbf{k})s(\mbf{q})+c(\mbf{k})c(\mbf{q})
\frac{x(k^2+q^2)-(1+x^2)kq}{(\mbf{k}-\mbf{q})^2}\,\right]
\nonumber\\
&=& \int d\mbf{q}C_fV_{L+C}(\mbf{k},\mbf{q})
\left[\,\frac{s(\mbf{q})}{s(\mbf{k})}\,\right]
+\int d\mbf{q}C_fW(\mbf{k},\mbf{q})
\left[\,\frac{s(\mbf{q})}{s(\mbf{k})}\,\right]
\,,\label{eq:d3}\end{eqnarray}   
where we have used the gap equation with $m=0$, Eq. (\ref{eq:3.2}), 
\begin{eqnarray}   
k&=&\int d\mbf{q}C_fV_{L+C}\left[\,c(\mbf{k})\frac{s(\mbf{q})}{s(\mbf{k}}
-c(\mbf{q})x\,\right]
\nonumber\\
&+&\int d\mbf{q}W\left[\,c(\mbf{k})\frac{s(\mbf{q})}{s(\mbf{k})}
-c(\mbf{q})\frac{x(k^2+q^2)-(1+x^2)kq}{(\mbf{k}-\mbf{q})^2}\,\right]
\,,\label{eq:d4}\end{eqnarray}
to get $\varepsilon$. 
In the above formulas the instantaneous interaction, Eq. (\ref{eq:2.24}), 
is given by
\begin{eqnarray}
C_fV_{L+C}(\mbf{k},\mbf{q}) &=& \frac{1}{2}\frac{C_fg^2}{(\mbf{k}-\mbf{q})^2}
+\frac{4\pi\sigma}{(\mbf{k}-\mbf{q})^4}
\,,\label{eq:d5}\end{eqnarray}
and the dynamical interactions generated by flow equations,
Eq. (\ref{eq:2.24}) and Eq. (\ref{eq:2.43}), are
\begin{eqnarray}
C_fW(\mbf{k},\mbf{q}) &=& \frac{C_fg^2}
{\omega(\mbf{k}-\mbf{q})[E(\mbf{q})+\omega(\mbf{k}-\mbf{q})]}
\nonumber\\
C_f\tilde{W}(\mbf{k},\mbf{q}) &=&-\frac{C_fg^2}
{E^2(\mbf{q})+\omega^2(\mbf{k}-\mbf{q})}
\,,\label{eq:d6}\end{eqnarray}
where loop momenta in perturbative corrections are large, 
$|\mbf{k}|\ll |\mbf{q}|$ (See Appendix \ref{app:B}). 
It is convenient to represent the generated interactions
Eq. (\ref{eq:d6}) as
\begin{eqnarray}
C_fW(\mbf{k},\mbf{q})&=&C_fU(\mbf{k},\mbf{q})-C_f\delta W(\mbf{k},\mbf{q})
\nonumber\\
-C_f\tilde{W}(\mbf{k},\mbf{q})&=&C_fU(\mbf{k},\mbf{q})
-C_f\delta \tilde{W}(\mbf{k},\mbf{q})
\,.\label{eq:d7}\end{eqnarray}
with 
\begin{eqnarray}
C_fU(\mbf{k},\mbf{q}) &=& \frac{C_fg^2}{\omega^2(\mbf{k}-\mbf{q})}
\nonumber\\
C_f\delta W(\mbf{k},\mbf{q})&=&\frac{C_fg^2}{\omega^2(\mbf{k}-\mbf{q})}
\left[\,\frac{E(\mbf{q})}{E(\mbf{q})+\omega(\mbf{k}-\mbf{q})}\,\right]
\nonumber\\
C_f\delta \tilde{W}(\mbf{k},\mbf{q})&=&\frac{C_fg^2}{\omega^2(\mbf{k}-\mbf{q})}
\left[\,\frac{E^2(\mbf{q})}{E^2(\mbf{q})+\omega^2(\mbf{k}-\mbf{q})}\,\right]
\,.\label{eq:d8}\end{eqnarray}
In BCS, with only instantaneous interaction present and 
$W=\tilde{W}=0$, we substitute Eqs. (\ref{eq:d2}) and (\ref{eq:d3}) 
into Eq. (\ref{eq:d1}) and get 
\begin{eqnarray}
2\int d\mbf{q}C_fV_{L+C}(\mbf{k},\mbf{q})
\left[\,\frac{s(\mbf{q})}{s(\mbf{k})}\,\right]\psi_{-}(\mbf{k})
-2\int d\mbf{q}C_fV_{L+C}(\mbf{k},\mbf{q})
\psi_{-}(\mbf{q}) &=& M\psi_{+}(\mbf{k})\,.\nonumber\\
\label{eq:d9}\end{eqnarray}
As has been found in Ref. \cite{LeYaouanc}, choosing the wave function solution as 
$\psi_{-}(\mbf{k})=s(\mbf{k})$, the l.h.s. of Eq. (\ref{eq:d9}) 
equals zero leading to zero r.h.s. and $M=0$. In other words,
by this choice the kinetic part, $\sim\psi(\mbf{k})$, 
is equal to the potential part (with an opposite sign), $\sim\psi(\mbf{q})$, 
thus interaction reduces pion mass to zero.  
This statement does not depend on particular
form of the instantaneous potential $V$, as long as a solution
of the gap equation exists for $V$. In particular, for
$V\sim k/[q^2(q^2+1)]$, the solution of the gap equation is constant,
\mbox{$tg(\mbf{k})=s(\mbf{k})/c(\mbf{k})\sim\int dq\,s(\mbf{q})/(q^2+1)=const$},
which reduces the bound state equation to
\begin{eqnarray}
2\int d\mbf{q}C_fV_{L+C}(\mbf{k},\mbf{q})
\left[\,\psi_{-}(\mbf{k})-\psi_{-}(\mbf{q})\,\right] 
&=& M\psi_{+}(\mbf{k})
\,,\label{eq:d10}\end{eqnarray}
where, in order to have $M=0$, we demand 
$\psi_{-}(\mbf{k})=\psi_{-}(\mbf{q})$ for any $\mbf{k},\mbf{q}$.
This leads to a constant wave function, $\psi_{-}(\mbf{k})=const$,
which does not satisfy a normalization condition and means  
a pion bound state is not localized, that is false.

Moreover, in the original RPA system of equations, Eq. (\ref{eq:4.23}),
the wave functions 
\begin{eqnarray}
X(\mbf{k})=-Y(\mbf{k})=s(\mbf{k})
\,,\label{eq:d10a}\end{eqnarray}
give zero l.h.s. and $M=0$ in the BCS \cite{LeYaouanc} (the same type of equation 
as Eq. (\ref{eq:d8})). However, this solution
violates the RPA wave function normalization condition, Eq. (\ref{eq:4.16a}),
\begin{eqnarray} 
\int d\mbf{q}\left[\,X^{\ast}(\mbf{q})X(\mbf{q})-Y^{\ast}(\mbf{q})Y(\mbf{q})
\,\right]
=\int d\mbf{q}\psi_{-}^{\ast}(\mbf{q})\psi_{+}(\mbf{q})=1
\,,\label{eq:d10b}\end{eqnarray}
which follows from the commutation relation of the meson creation/annihilation
operators, Eq. (\ref{eq:4.9}). Also, the covariant dispersion law
$E(\mbf{P})=\sqrt{\mbf{P}^2+M^2}$ does not hold in this case, 
which means a breakdown of covariance. As noted by Le Yaouanc {\it et al.}
\cite{LeYaouanc}, the model is not covariant since we have adopted 
an instantaneous interaction.    

Now we calculate the perturbative correction to the BCS solution using
flow equations. We substitute Eqs. (\ref{eq:d2}), (\ref{eq:d3})
and (\ref{eq:d7}) into Eq. (\ref{eq:d1}), 
with both instantaneous and generated interactions present, and get  
\begin{eqnarray}
 &2&\int d\mbf{q}C_fV_{L+C}(\mbf{k},\mbf{q})
\left[\,\frac{s(\mbf{q})}{s(\mbf{k})}\,\right]\psi_{-}(\mbf{k})
+2\int d\mbf{q}C_fU(\mbf{k},\mbf{q})
\left[\,\frac{s(\mbf{q})}{s(\mbf{k})}\,\right]\psi_{-}(\mbf{k})
\nonumber\\
 -&2&\int d\mbf{q}C_fV_{L+C}(\mbf{k},\mbf{q})\psi_{-}(\mbf{q})
-2\int d\mbf{q}C_fU(\mbf{k},\mbf{q})\psi_{-}(\mbf{q})
\label{eq:d11}\\
 +&2&\int d\mbf{q}(-C_f\delta W(\mbf{k},\mbf{q}))
\left[\,\frac{s(\mbf{q})}{s(\mbf{k})}\,\right]\psi_{-}(\mbf{k})
+2\int d\mbf{q}C_f\delta\tilde{W}(\mbf{k},\mbf{q})\psi_{-}(\mbf{q}) 
= M\psi_{+}(\mbf{k})\nonumber
\,.\end{eqnarray}
Choosing $\psi_{-}(\mbf{k})=s(\mbf{k})$ the sum of the first four terms
becomes zero. The rest two terms give the pion mass
\begin{eqnarray}
M &=& 2C_fg^2\int d\mbf{q}s(\mbf{q})\frac{E(\mbf{q})
[E(\mbf{q})-\omega(\mbf{k}-\mbf{q})]}
{\omega(\mbf{k}-\mbf{q})[E(\mbf{q})+\omega(\mbf{k}-\mbf{q})]
[E^2(\mbf{q})+\omega^2(\mbf{k}-\mbf{q})]}
\,,\label{eq:d12}\end{eqnarray}
at small $\mbf{k}$ ($s(\mbf{k})\sim 1$), where 
$E(\mbf{q})=q$ in the chiral limit and 
$\omega(\mbf{k}-\mbf{q})=|\mbf{k}-\mbf{q}|$. In Eq. (\ref{eq:d11})
all terms are positive except for the difference, which 
is negative after angle average in the leading order; i.e. 
\begin{eqnarray}
\langle E(\mbf{q})-\omega(\mbf{k}-\mbf{q})\rangle_{x}
=\langle kx-\frac{k^2}{2q}(1-x^2)+O(k^3) \rangle_{x}=
-\frac{k^2}{3q}< 0
\,,\label{eq:d13}\end{eqnarray}
where $x=\hat{\mbf{k}}\cdot\hat{\mbf{q}}$. 
Therefore the pion mass is negative  
\begin{eqnarray}
M=-k^2\frac{C_fg^2}{(4\pi^2)6}\int dq\frac{s(\mbf{q})}{q^2}<0
\,.\label{eq:d14}\end{eqnarray}
This cannot hold in reality.
This false result can be understood because the leading order 
perturbative correction generally lowers a mass of the ground state.
Particular feature here is that the dynamical kinetic term
(self energy), $W$, differs from the dynamical potential term,
$\tilde{W}$ in the two-quark channel, Eq. (\ref{eq:d7}). 
This insures a nonzero (negative) correction. Though the flow equation 
correction has been estimated at small
$\mbf{k}$ and it is proportional to $k^2$, it cannot be neglected
(since the integral diverges at small $q$).
Therefore, if the BCS mass is equal to zero which, as discussed before,
violates several principles, then the flow equations shift this mass
to a negative value. This indicates that the statement about zero pion mass 
in the BCS appears to be wrong. We conclude that the BCS does not 
give the zero mass solution for the pion ground state.  

Having shown that the BCS solution Eq. (\ref{eq:d10a}) violates 
the normalization condition Eq. (\ref{eq:d10b}) and is unphysical, 
Eq. (\ref{eq:d14}),   
we estimate the $\pi$ meson mass when $|X|\neq |Y|$ in the BCS model.  
From Eq. (\ref{eq:4.26}), $(A+B)\psi_{+}=M\psi_{-}$,
\begin{eqnarray}
M &=& \int d\mbf{k}d\mbf{q}\psi_{+}^{\ast}(\mbf{k})
(A+B)(\mbf{k},\mbf{q})\psi_{+}(\mbf{q})
\,,\label{eq:d15}\end{eqnarray}
where the normalization condition, Eq. (\ref{eq:d10b}),
has been used. 
Using Eq. (\ref{eq:d1}) for $A$ and $B$ and including
only the instantaneous part, the mass is given by
\begin{eqnarray}
M &=& 2\int d\mbf{k}\psi_{+}^{\ast}(\mbf{k})\psi_{+}(\mbf{k})
\varepsilon(\mbf{k})-\int d\mbf{k}d\mbf{q}\psi_{+}^{\ast}(\mbf{k})
\left[\,I_{xx}(\mbf{k},\mbf{q})+I_{xy}(\mbf{k},\mbf{q})\,\right]
\psi_{+}(\mbf{q})\,.\nonumber\\
\label{eq:d16}\end{eqnarray}
For the $\pi$ meson the instantaneous kinetic term, Eq. (\ref{eq:d3}),  
can be expressed through the instantaneous potential term, 
Eq. (\ref{eq:d2}), in the chiral limit as 
\begin{eqnarray}
\varepsilon(\mbf{k}) &=& kc(\mbf{k})
+\frac{1}{2}\int d\mbf{q}\left[\,I_{xx}(\mbf{k},\mbf{q})
+I_{xy}(\mbf{k},\mbf{q})\,\right]
\,.\label{eq:d17}\end{eqnarray}
Substituting this representation into Eq. (\ref{eq:d16})
the pion mass reads 
\begin{eqnarray}
M &=& 2\int d\mbf{k}\psi_{+}^{\ast}(\mbf{k})\psi_{+}(\mbf{k})
kc(\mbf{k})\label{eq:d18}\\
&+& \int d\mbf{k}d\mbf{q}\psi_{+}^{\ast}(\mbf{k})
\left[\,I_{xx}(\mbf{k},\mbf{q })+I_{xy}(\mbf{k},\mbf{q})\,\right]
[\psi_{+}(\mbf{k})-\psi_{+}(\mbf{q})]
\nonumber\\
&=& 2\int d\mbf{k}\psi_{+}^{\ast}(\mbf{k})\psi_{+}(\mbf{k})
kc(\mbf{k})\nonumber\\
&+& 2\int d\mbf{k}d\mbf{q}\psi_{+}^{\ast}(\mbf{k})
C_fV_{L+C}(\mbf{k},\mbf{q})
\left[\,s(\mbf{k})s(\mbf{q})+c(\mbf{k})c(\mbf{q})x\,\right]
[\psi_{+}(\mbf{k})-\psi_{+}(\mbf{q})]\,.\nonumber
\end{eqnarray}
Here, the first term is always positive and equals  
\begin{eqnarray}
2\int d\mbf{k}d\mbf{q}\psi_{+}^{\ast}(\mbf{k})\psi_{+}(\mbf{k})
C_fV_{L+C}(\mbf{k},\mbf{q})\left[\,c^2(\mbf{k})\frac{s(\mbf{q})}{s(\mbf{k})}
-c(\mbf{k})c(\mbf{q})x\,\right]>0 
\,.\label{eq:d19}\end{eqnarray}
It is the energy of two bare quarks in the transformed BV vacuum.  
In the second term, the kernel averaged over $x$ is positive,
since the term \mbox{$d\mbf{k}d\mbf{q}\psi^{\ast}(\mbf{k})
\left[\,I_{xx}+I_{xy}\,\right]\psi(\mbf{k})$}
represents an effective energy of two quarks which is positive 
(the sum of two quark self energies).
For any positive operator $Q(\mbf{k},\mbf{q})>0$,
symmetric under interchange of arguments, the following holds;
\begin{eqnarray}
&&0< \int d\mbf{k}d\mbf{q}[\psi(\mbf{k})-\psi(\mbf{q})]^{\ast}
Q(\mbf{k},\mbf{q})[\psi(\mbf{k})-\psi(\mbf{q})]
\nonumber\\
&=&2 \int d\mbf{k}d\mbf{q}\left[\,\psi(\mbf{k})^{\ast}
Q(\mbf{k},\mbf{q})\psi(\mbf{k})
-\psi(\mbf{k})^{\ast}
Q(\mbf{k},\mbf{q})\psi(\mbf{q})\,\right]\nonumber\\
&=& 2\int d\mbf{k}d\mbf{q}\psi(\mbf{k})^{\ast}Q(\mbf{k},\mbf{q})
[\psi(\mbf{k})-\psi(\mbf{q})]   
\,.\label{eq:d20}\end{eqnarray}
Therefore the second term in Eq. (\ref{eq:d18}) is 
also positive, provided a positive RPA mass of the pion in the BCS model.
As shown above, flow equations improve slightly this situation
and shift the $\pi$ meson mass down. However, in the leading order 
we are unable to get exactly zero mass pion even with flow equations.
There might be a possibility to reach zero mass pion by extending
calculations to the higher orders, approching to the covariant result.

\section{RPA for the $S$ and $D$ wave $\rho$ mesons}
\label{app:E}

The RPA equations for the $\rho$ wave function components 
$X(\mbf{k}), Y(\mbf{k})$ for $L=0$ and $L=2$ states are
\begin{eqnarray} 
M_nX^{S}(\mbf{k}) = 2\varepsilon(\mbf{k})X^{S}(\mbf{k})
&-&\int\frac{q^2dqdx}{4\pi^2}I_{xx}^{SS}(\mbf{k},\mbf{q})X^{S}(\mbf{q})
-\int\frac{q^2dqdx}{4\pi^2}I_{xx}^{SD}(\mbf{k},\mbf{q})X^{D}(\mbf{q})
\nonumber\\
&-&\int\frac{q^2dqdx}{4\pi^2}I_{xy}^{SS}(\mbf{k},\mbf{q})Y^{S}(\mbf{q})
-\int\frac{q^2dqdx}{4\pi^2}I_{xy}^{SD}(\mbf{k},\mbf{q})Y^{D}(\mbf{q})
\nonumber\\
&-&\int\frac{q^2dqdx}{4\pi^2}G_{xx}^{SS}(\mbf{k},\mbf{q})X^{S}(\mbf{q})
-\int\frac{q^2dqdx}{4\pi^2}G_{xx}^{SD}(\mbf{k},\mbf{q})X^{D}(\mbf{q})
\nonumber\\
&-&\int\frac{q^2dqdx}{4\pi^2}G_{xy}^{SS}(\mbf{k},\mbf{q})Y^{S}(\mbf{q})
-\int\frac{q^2dqdx}{4\pi^2}G_{xy}^{SD}(\mbf{k},\mbf{q})Y^{D}(\mbf{q})
\nonumber\\
-M_nY^{S}(\mbf{k}) = 2\varepsilon(\mbf{k})Y^{S}(\mbf{k})
&-&\int\frac{q^2dqdx}{4\pi^2}I_{yy}^{SS}(\mbf{k},\mbf{q})Y^{S}(\mbf{q})
-\int\frac{q^2dqdx}{4\pi^2}I_{yy}^{SD}(\mbf{k},\mbf{q})Y^{D}(\mbf{q})
\nonumber\\
&-&\int\frac{q^2dqdx}{4\pi^2}I_{yx}^{SS}(\mbf{k},\mbf{q})X^{S}(\mbf{q})
-\int\frac{q^2dqdx}{4\pi^2}I_{yx}^{SD}(\mbf{k},\mbf{q})X^{D}(\mbf{q})
\nonumber\\
&-&\int\frac{q^2dqdx}{4\pi^2}G_{yy}^{SS}(\mbf{k},\mbf{q})Y^{S}(\mbf{q})
-\int\frac{q^2dqdx}{4\pi^2}G_{yy}^{SD}(\mbf{k},\mbf{q})Y^{D}(\mbf{q})
\nonumber\\
&-&\int\frac{q^2dqdx}{4\pi^2}G_{yx}^{SS}(\mbf{k},\mbf{q})X^{S}(\mbf{q})
-\int\frac{q^2dqdx}{4\pi^2}G_{yx}^{SD}(\mbf{k},\mbf{q})X^{D}(\mbf{q})
\nonumber\\
M_nX^{D}(\mbf{k}) = 2\varepsilon(\mbf{k})X^{D}(\mbf{k})
&-&\int\frac{q^2dqdx}{4\pi^2}I_{xx}^{DD}(\mbf{k},\mbf{q})X^{D}(\mbf{q})
-\int\frac{q^2dqdx}{4\pi^2}I_{xx}^{DS}(\mbf{k},\mbf{q})X^{S}(\mbf{q})
\nonumber\\
&-&\int\frac{q^2dqdx}{4\pi^2}I_{xy}^{DD}(\mbf{k},\mbf{q})Y^{D}(\mbf{q})
-\int\frac{q^2dqdx}{4\pi^2}I_{xy}^{DS}(\mbf{k},\mbf{q})Y^{S}(\mbf{q})
\nonumber\\
&-&\int\frac{q^2dqdx}{4\pi^2}G_{xx}^{DD}(\mbf{k},\mbf{q})X^{D}(\mbf{q})
-\int\frac{q^2dqdx}{4\pi^2}G_{xx}^{DS}(\mbf{k},\mbf{q})X^{S}(\mbf{q})
\nonumber\\
&-&\int\frac{q^2dqdx}{4\pi^2}G_{xy}^{DD}(\mbf{k},\mbf{q})Y^{D}(\mbf{q})
-\int\frac{q^2dqdx}{4\pi^2}G_{xy}^{DS}(\mbf{k},\mbf{q})Y^{S}(\mbf{q})
\nonumber\\
-M_nY^{D}(\mbf{k}) = 2\varepsilon(\mbf{k})Y^{D}(\mbf{k})
&-&\int\frac{q^2dqdx}{4\pi^2}I_{yy}^{DD}(\mbf{k},\mbf{q})Y^{D}(\mbf{q})
-\int\frac{q^2dqdx}{4\pi^2}I_{yy}^{DS}(\mbf{k},\mbf{q})Y^{S}(\mbf{q})
\nonumber\\
&-&\int\frac{q^2dqdx}{4\pi^2}I_{yx}^{DD}(\mbf{k},\mbf{q})X^{D}(\mbf{q})
-\int\frac{q^2dqdx}{4\pi^2}I_{yx}^{DS}(\mbf{k},\mbf{q})X^{S}(\mbf{q})
\nonumber\\
&-&\int\frac{q^2dqdx}{4\pi^2}G_{yy}^{DD}(\mbf{k},\mbf{q})Y^{D}(\mbf{q})
-\int\frac{q^2dqdx}{4\pi^2}G_{yy}^{DS}(\mbf{k},\mbf{q})Y^{S}(\mbf{q})
\nonumber\\
&-&\int\frac{q^2dqdx}{4\pi^2}G_{yx}^{DD}(\mbf{k},\mbf{q})X^{D}(\mbf{q})
-\int\frac{q^2dqdx}{4\pi^2}G_{yx}^{DS}(\mbf{k},\mbf{q})X^{S}(\mbf{q})
\,, \nonumber\\
\label{eq:e1}
\end{eqnarray}
where the instantaneous terms $I$ are
\begin{eqnarray} 
I_{xx}^{SS}(\mbf{k},\mbf{q}) &=& I_{yy}^{SS}(\mbf{k},\mbf{q})=
C_fV_{L+C}(\mbf{k},\mbf{q})\frac{1}{2}\left[\phantom{\frac{1}{1}}
\hspace{-0.3cm}\,
(1+s(\mbf{k}))(1+s(\mbf{q})) \right.\nonumber\\
&+&\left. \frac{1}{3}(1-s(\mbf{k}))(1-s(\mbf{q}))(4x^2-1)
+ 2c(\mbf{k})c(\mbf{q})x\, \right]
\nonumber\\
I_{xx}^{SD}(\mbf{k},\mbf{q}) &=& I_{yy}^{SD}(\mbf{k},\mbf{q})=
C_fV_{L+C}(\mbf{k},\mbf{q})\frac{1}{2}\frac{\sqrt{2}}{3}
(1-s(\mbf{k}))(1-s(\mbf{q}))(x^2-1)
\nonumber\\
I_{xx}^{DS}(\mbf{k},\mbf{q}) &=& I_{yy}^{DS}(\mbf{k},\mbf{q})=
C_fV_{L+C}(\mbf{k},\mbf{q})\frac{1}{2}\frac{\sqrt{2}}{3}
(1-s(\mbf{k}))(1-s(\mbf{q}))(x^2-1)
\nonumber\\
I_{xx}^{DD}(\mbf{k},\mbf{q}) &=& I_{yy}^{DD}(\mbf{k},\mbf{q})=
C_fV_{L+C}(\mbf{k},\mbf{q})\frac{1}{2}\left[\,
\frac{1}{2}(1+s(\mbf{k}))(1+s(\mbf{q}))(3x^2-1)\right.\nonumber\\
&+&\left.\frac{1}{6}(1-s(\mbf{k}))(1-s(\mbf{q}))(x^2+5)
+ 2c(\mbf{k})c(\mbf{q})x\, \right]
\nonumber\\
I_{xy}^{SS}(\mbf{k},\mbf{q}) &=& I_{yx}^{SS}(\mbf{k},\mbf{q})=
C_fV_{L+C}(\mbf{k},\mbf{q})\frac{1}{2}\left[\,
-\frac{1}{3}(1+s(\mbf{k}))(1-s(\mbf{q}))\right.\nonumber\\
&+&\left.\frac{1}{3}(1-s(\mbf{k}))(1+s(\mbf{q})) 
+ \frac{2}{3}c(\mbf{k})c(\mbf{q})x\, \right]
\nonumber\\
I_{xy}^{SD}(\mbf{k},\mbf{q}) &=& I_{yx}^{SD}(\mbf{k},\mbf{q})=
C_fV_{L+C}(\mbf{k},\mbf{q})\frac{1}{2}\left[\,
\frac{\sqrt{2}}{3}(1+s(\mbf{k}))(1-s(\mbf{q}))(3x^2-1)\right.\nonumber\\
&+&\left.\frac{2\sqrt{2}}{3}(1-s(\mbf{k}))(1+s(\mbf{q})) 
- \frac{4\sqrt{2}}{3}c(\mbf{k})c(\mbf{q})x\, \right]
\nonumber\\
I_{xy}^{DS}(\mbf{k},\mbf{q}) &=& I_{yx}^{DS}(\mbf{k},\mbf{q})=
C_fV_{L+C}(\mbf{k},\mbf{q})\frac{1}{2}\left[\,
\frac{2\sqrt{2}}{3}(1+s(\mbf{k}))(1-s(\mbf{q}))\right.\nonumber\\
&+&\left.\frac{\sqrt{2}}{3}(1-s(\mbf{k}))(1+s(\mbf{q}))(3x^2-1)
- \frac{4\sqrt{2}}{3}c(\mbf{k})c(\mbf{q})x\, \right]
\nonumber\\
I_{xy}^{DD}(\mbf{k},\mbf{q}) &=& I_{yx}^{DD}(\mbf{k},\mbf{q})=
C_fV_{L+C}(\mbf{k},\mbf{q})\frac{1}{2}\left[\,
\frac{1}{6}(1+s(\mbf{k}))(1-s(\mbf{q}))(3x^2-1)\right.\nonumber\\
&+&\left.\frac{1}{6}(1-s(\mbf{k}))(1+s(\mbf{q}))(3x^2-1) 
+ \frac{2}{3}c(\mbf{k})c(\mbf{q})x\, \right]
\,,\label{eq:e2}\end{eqnarray}
and the generated terms $G$ are
\begin{eqnarray} 
G_{xx}^{SS}(\mbf{k},\mbf{q}) &=& G_{yy}^{SS}(\mbf{k},\mbf{q})=
C_fW_1(\mbf{k},\mbf{q})\frac{1}{2}\left[\,
\frac{1}{3}(1+s(\mbf{k}))(1-s(\mbf{q}))
\left(1-\frac{2(1-x^2)k^2}{(\mbf{k}-\mbf{q})^2}\right)
\right.\nonumber\\
&&\hspace{-3cm}+\left.\frac{1}{3}(1-s(\mbf{k}))(1+s(\mbf{q}))
\left(1-\frac{2(1-x^2)q^2}{(\mbf{k}-\mbf{q})^2}\right)
-\frac{2}{3}c(\mbf{k})c(\mbf{q})
\left(x+\frac{(1-x^2)kq}{(\mbf{k}-\mbf{q})^2}\right)\, \right]
\nonumber\\
G_{xx}^{SD}(\mbf{k},\mbf{q}) &=& G_{yy}^{SD}(\mbf{k},\mbf{q})=
C_fW_1(\mbf{k},\mbf{q})\frac{1}{2}\left[\,
\frac{\sqrt{2}}{6}(1+s(\mbf{k}))(1-s(\mbf{q}))
\left(3x^2-1-\frac{(1-x^2)k^2}{(\mbf{k}-\mbf{q})^2}\right)\right.\nonumber\\
&&\hspace{-3cm}+\left.\frac{\sqrt{2}}{6}(1-s(\mbf{k}))(1+s(\mbf{q}))
\left(2-\frac{(1-x^2)q^2}{(\mbf{k}-\mbf{q})^2}\right)
+\frac{\sqrt{2}}{3}c(\mbf{k})c(\mbf{q})
\left(-2x+\frac{(1-x^2)kq}{(\mbf{k}-\mbf{q})^2}\right)\, \right]
\nonumber\\
G_{xx}^{DS}(\mbf{k},\mbf{q}) &=& G_{yy}^{DS}(\mbf{k},\mbf{q})=
C_fW_1(\mbf{k},\mbf{q})\frac{1}{2}\left[\,
\frac{\sqrt{2}}{6}(1+s(\mbf{k}))(1-s(\mbf{q}))
\left(2-\frac{(1-x^2)k^2}{(\mbf{k}-\mbf{q})^2}\right)\right.\nonumber\\
&&\hspace{-3cm}+\left.\frac{\sqrt{2}}{3}(1-s(\mbf{k}))(1+s(\mbf{q}))
\left(3x^2-1-\frac{(1-x^2)q^2}{(\mbf{k}-\mbf{q})^2}\right)
+\frac{\sqrt{2}}{3}c(\mbf{k})c(\mbf{q})
\left(-2x+\frac{(1-x^2)kq}{(\mbf{k}-\mbf{q})^2}\right)\, \right]
\nonumber\\
G_{xx}^{DD}(\mbf{k},\mbf{q}) &=& G_{yy}^{DD}(\mbf{k},\mbf{q})
\nonumber\\
&=&C_fW_1(\mbf{k},\mbf{q})\frac{1}{2}\left[\,
\frac{1}{2}(1+s(\mbf{k}))(1-s(\mbf{q}))
\left(\frac{5}{6}(1-3x^2)-\frac{1}{6}(1-x^2)
\frac{k^2+9q^2}{(\mbf{k}-\mbf{q})^2}\right)\right.\nonumber\\
&&\hspace{-3cm}+\left.\frac{1}{2}(1-s(\mbf{k}))(1+s(\mbf{q}))
\left(\frac{5}{6}(1-3x^2)-\frac{1}{6}(1-x^2)
\frac{q^2+9k^2}{(\mbf{k}-\mbf{q})^2}\right)
+\frac{1}{3}c(\mbf{k})c(\mbf{q})
\left(-4x-\frac{(1-x^2)kq}{(\mbf{k}-\mbf{q})^2}\right)\, \right]
\nonumber\\
G_{xy}^{SS}(\mbf{k},\mbf{q}) &=& G_{yx}^{SS}(\mbf{k},\mbf{q})=
C_fW_2(\mbf{k},\mbf{q})\frac{1}{2}\left[\,
\frac{1}{3}(1+s(\mbf{k}))(1+s(\mbf{q}))\right.\nonumber\\
&&\hspace{-3cm}+\left.\frac{1}{3}(1-s(\mbf{k}))(1-s(\mbf{q}))(2x^2-1)
+\frac{2}{3}c(\mbf{k})c(\mbf{q})
\left(x-\frac{(1-x^2)kq}{(\mbf{k}-\mbf{q})^2}\right)\, \right]
\nonumber\\
G_{xy}^{SD}(\mbf{k},\mbf{q}) &=& G_{yx}^{SD}(\mbf{k},\mbf{q})=
C_fW_2(\mbf{k},\mbf{q})\frac{1}{2}\left[\,
\frac{\sqrt{2}}{6}(1+s(\mbf{k}))(1+s(\mbf{q}))
\left(2-\frac{3(1-x^2)q^2}{(\mbf{k}-\mbf{q})^2}\right)\right.\nonumber\\
&&\hspace{-3cm}+\left.\frac{\sqrt{2}}{6}(1-s(\mbf{k}))(1-s(\mbf{q}))
\left(1+x^2-\frac{3(1-x^2)k^2}{(\mbf{k}-\mbf{q})^2}\right)
+\frac{\sqrt{2}}{3}c(\mbf{k})c(\mbf{q})
\left(2x+\frac{(1-x^2)kq}{(\mbf{k}-\mbf{q})^2}\right)\, \right]
\nonumber\\
G_{xy}^{DS}(\mbf{k},\mbf{q}) &=& G_{yx}^{DS}(\mbf{k},\mbf{q})=
C_fW_2(\mbf{k},\mbf{q})\frac{1}{2}\left[\,
\frac{\sqrt{2}}{6}(1+s(\mbf{k}))(1+s(\mbf{q}))
\left(2-\frac{3(1-x^2)k^2}{(\mbf{k}-\mbf{q})^2}\right)\right.\nonumber\\
&&\hspace{-3cm}+\left.\frac{\sqrt{2}}{6}(1-s(\mbf{k}))(1-s(\mbf{q}))
\left(1+x^2-\frac{3(1-x^2)q^2}{(\mbf{k}-\mbf{q})^2}\right)
+\frac{\sqrt{2}}{3}c(\mbf{k})c(\mbf{q})
\left(2x+\frac{(1-x^2)kq}{(\mbf{k}-\mbf{q})^2}\right)\, \right]
\nonumber\\
G_{xy}^{DD}(\mbf{k},\mbf{q}) &=& G_{yx}^{DD}(\mbf{k},\mbf{q})
\nonumber\\
&=& C_fW_2(\mbf{k},\mbf{q})\frac{1}{2}\left[\,
\frac{1}{2}(1+s(\mbf{k}))(1+s(\mbf{q}))
\left(\frac{1}{6}(9x^2-1)
-\frac{1}{2}(1-x^2)\frac{k^2+q^2}{(\mbf{k}-\mbf{q})^2})\right)
\right.\nonumber\\
&&\hspace{-3cm}+\left.\frac{1}{2}(1-s(\mbf{k}))(1-s(\mbf{q}))
\left(\frac{1}{6}(7+x^2)
-\frac{1}{2}(1-x^2)\frac{k^2+q^2}{(\mbf{k}-\mbf{q})^2})\right)
+\frac{1}{3}c(\mbf{k})c(\mbf{q})
\left(4x-\frac{(1-x^2)kq}{(\mbf{k}-\mbf{q})^2}\right)\, \right]
\,,\nonumber\\
\label{eq:e3}\end{eqnarray}

\end{document}